\documentclass[twocolumn,preprint2]{aastex631}
\usepackage{amsmath}
\usepackage{gensymb}

\newcommand{\cii}{[\ion{C}{2}]$158 \mu m$}
\newcommand{\ci}{[\ion{C}{1}]$(1-0)$}
\newcommand{\co}{CO(4$-$3)}
\newcommand{\mjypb}{$\rm mJy\ beam^{-1} $}
\newcommand{\jykmps}{$\rm Jy\ km\ s^{-1} $}
\newcommand{\jypbkmps}{Jy\,beam$^{-1}$\,$\times$\,km\,s$^{-1}$}
\newcommand{\kmps}{$\rm  km\ s^{-1} $}
\newcommand{\lsun}{$\rm L_{\sun}$}
\newcommand{\msun}{$\rm M_{\sun}$}
\newcommand{\msunpyr}{$\rm M_{\sun}\ yr^{-1}$}
\newcommand{\mdust}{$\rm M_{dust}$}
\newcommand{\jone}{Q0050+0051}
\newcommand{\jtwo}{Q0052+0140}
\newcommand{\jthree}{Q0101+0201}
\newcommand{\jfour}{Q0107+0314}
\newcommand{\jfive}{Q1227+2848}
\newcommand{\jsix}{Q1228+3128}
\newcommand{\jseven}{Q1230+3320}
\newcommand{\jeight}{Q1416+2649}
\newcommand{\jnine}{Q2121+0052}
\newcommand{\jten}{Q2123$-$0050}
\newcommand{\paper}{Paper\,{\sc I}}
\shorttitle{SUPERCOLD-CGM survey: II.}
\shortauthors{Li et al.}

\graphicspath{{./}{figures/}}
\begin{document}

\title{The SUPERCOLD-CGM survey: II. \ci{} emission and the physical conditions of cold gas in Enormous Ly$\alpha$ nebulae at $z\,\sim\,2$}

\correspondingauthor{Jianan Li}
\email{jiananl@pku.edu.cn}
\correspondingauthor{Bjorn\,H.\,C. Emonts}
\email{bemonts@nrao.edu}
\correspondingauthor{Zheng Cai}
\email{zcai@mail.tsinghua.edu.cn}
\author[0000-0002-1815-4839 ]{Jianan Li}
\affiliation{Department of Astronomy, Tsinghua University, Beijing 100084, People’s Republic of China}

\author[0000-0003-2983-815X]{Bjorn\,H.\,C. Emonts}
\affil{National Radio Astronomy Observatory, 520 Edgemont Road, Charlottesville, VA 22903}

\author[0000-0001-8467-6478]{Zheng Cai}
\affiliation{Department of Astronomy, Tsinghua University, Beijing 100084, People’s Republic of China}

\author{Jianrui Li}
\affil{Department of Astronomy, Tsinghua University, Beijing 100084, People’s Republic of China}

\author[0000-0003-4956-5742]{Ran Wang}
\affiliation{Kavli Institute for Astronomy and Astrophysics, Peking University, Beijing 100871, China}

\author[0000-0003-2566-2126]{Montserrat Villar-Mart\'{i}n}
\affiliation{Centro de Astrobiolog\'{i}a, CSIC-INTA, Ctra. de Torrej\'{o}n a Ajalvir, km 4, 28850 Torrej\'{o}n de Ardoz, Madrid, Spain}

\author[0000-0002-4770-6137]{Fabrizio Arrigoni Battaia}
\affil{Max-Planck-Institut f\"ur Astrophysik, Karl-Schwarzschild-Str 1, D-85748 Garching bei M\"unchen, Germany}

\author[0000-0001-6251-649X]{Mingyu Li}
\affil{Department of Astronomy, Tsinghua University, Beijing 100084, People’s Republic of China}

\author[0000-0003-0111-8249]{Yunjing Wu}
\affil{Department of Astronomy, Tsinghua University, Beijing 100084, People’s Republic of China}

\author[0000-0001-9163-0064]{Ilsang Yoon}
\affil{National Radio Astronomy Observatory, 520 Edgemont Road, Charlottesville, VA 22903}

\author[0000-0003-1939-5885]{Matthew\,D. Lehnert}
\affil{Universit\'{e} Lyon 1, ENS de Lyon, CNRS UMR5574, Centre de Recherche Astrophysique de Lyon, F-69230 Saint-Genis-Laval, France}

\author[0000-0002-0830-2033]{Kyle Massingill} 
\affil{School of Earth and Space Exploration, Arizona State University, Tempe, AZ 85287, USA}

\author[0000-0003-0167-0981]{Craig Sarazin} 
\affil{Department of Astronomy, University of Virginia, 530 McCormick Road, Charlottesville, VA 22904, USA}

\author[0000-0002-7738-6875]{Jason X Prochaska}
\affiliation{University of California, 1156 High Street, Santa Cruz, CA 95064, USA}
\affiliation{Kavli Institute for the Physics and Mathematics of the Universe,
The University of Tokyo, 5-1-5 Kashiwanoha, Kashiwa, 277-8583, Japan}
\affiliation{Division of Science, National Astronomical Observatory of Japan,
2-21-1 Osawa, Mitaka, Tokyo 181-8588, Japan}

\author[0000-0002-3032-1783]{Mark Lacy}
\affil{National Radio Astronomy Observatory, 520 Edgemont Road, Charlottesville, VA 22903}

\author[0000-0002-8472-836X]{Brian Mason}
\affil{National Radio Astronomy Observatory, 520 Edgemont Road, Charlottesville, VA 22903}



\begin{abstract}
We report ALMA and ACA observations of atomic carbon (\ci{}) and dust continuum in 10 Enormous Ly$\alpha$ Nebulae hosting ultra-luminous Type-I QSOs at $z=2.2-2.5$, as part of the SUrvey of Protocluster ELANe Revealing CO/CI in the Ly$\alpha$ Detected CGM (SUPERCOLD-CGM).
We detect \ci{} and dust in all ten QSOs and five companion galaxies. 
We find that the QSOs and companions have higher gas densities and more intense radiation fields than Luminous Infrared galaxies and high-$z$ main sequence galaxies, with the highest values found in the QSOs.
By comparing molecular gas masses derived from  \ci{}, \co{} and dust continuum, we find that the QSOs and companions display a similar low CO conversion factor of $\alpha_{\rm CO}$\,$\sim$\,0.8 \msun{}${[\rm K\,km/s\,pc^2]}^{-1}$.
After tapering our data to low resolution, the \ci{} flux increases for nine QSOs, hinting at the possibility of \ci{} in the circum-galactic medium (CGM)  on a scale of 16$-$40 kpc. However, the \ci{} sensitivity is too low to confirm this for individual targets, except for a tentative (2.7$\sigma$) CGM detection in \jone{} with M$_{\rm H_2}$\,=\, ($1.0 - 2.8$)$\times 10^{10}$ \msun{}. 
The 3$\sigma$ mass limits of molecular CGM for the remaining QSO fields are ($0.2-1.4$)\,$\times$\,10$^{10}$ \msun{}.
This translates into a baryon fraction of $<$0.4-3$\% $ in the molecular CGM relative to the total baryonic halo mass.
Our sample also includes a radio-detected AGN, \jeight{}, which shows \ci{} and \co{} luminosities an order of magnitude fainter for its far-infrared luminosity than other QSOs in our sample, possibly due to a lower molecular gas mass.

%However, the \ci{} sensitivity is too low to confirm this for individual targets, except for \jone{}, where a tentative (2.7$\sigma$) CGM detection reveals a molecular mass of ($1.0 - 2.8$)$\times 10^{10}$ \msun{} within $\sim$5\arcsec (40 kpc). 
%

%Our study highlights the importance of ALMA + ACA in studying the physical conditions of the interstellar medium (ISM) and CGM in the Early Universe. 

\end{abstract}


\section{Introduction}
Active Galactic Nuclei (AGN) play key roles in the formation and evolution of massive galaxies. It has been suggested in theory and simulations that powerful radiative feedback from the AGN is responsible for ejecting or removing the gas from galaxy halos and as a result, quenching star formation (e.g., \citealt{silk98}; \citealt{diMatteo05}; \citealt{croton06}; \citealt{bower06}; \citealt{hopkins08}; \citealt{martizzi12}; \citealt{dubois13,dubois16}; \citealt{lebrun14}; \citealt{schaye15}; \citealt{sijacki15}; \citealt{oppenheimer20}; \citealt{koudmani21}).   
 Cosmic noon  ($z$\,$\sim$\,2$-$3) is the epoch at which the star formation rate, molecular density, and cosmic AGN activity peak (e.g., \citealt{schmidt95}; \citealt{richards06}; \citealt{aird10}; \citealt{madau14}; \citealt{decarli19}; \citealt{khusanova21}). 
Luminous QSOs at cosmic noon are particularly interesting as both the luminous QSO and the active star formation lead to the injection of a large amount of energy into the interstellar medium (ISM) and circum-galactic medium (CGM). This makes them ideal targets to investigate the interplay between the star-formation/AGN feedback and the ISM/CGM properties. 

The molecular ISM, as the fuel for star formation and the material feeding the growth of supermassive black holes (SMBHs), traces the direct link between the AGN and their hosts.  Observations have shown that the SMBHs and their host galaxies co-evolve across redshifts (e.g., \citealt{kormendy13,kormendy20}; \citealt{mountrichas23}). The cold ISM ($\rm \sim 10 - 100\ K$), which is directly linked to star formation,  has long been probed through emission from molecular carbon monoxide, CO($J \rightarrow J-1$), as well as atomic fine structure (FS) lines, e.g., [C {\sc I}] $^3$P$_{1}$\,$-$\,$^3$P$_{0}$ [hereafter \ci{}] and \cii{} (e.g., \citealt{liu15}; \citealt{rosenberg15}; \citealt{diaz17}; \citealt{jiao17,jiao19,jiao21}; \citealt{herrera-camus18}). 
These lines trace different gas temperatures and densities, as well as the effect of cosmic rays \citep{papadopoulos04,bisbas17}. 
In particular, the total molecular gas mass is traced by the \ci{} and low-lying CO rotational transitions, e.g., $\rm CO(1-0)$. 
The ratios between different (sub)-mm emission lines serve as sensitive diagnostics of the physical conditions of the molecular and atomic gas when compared to radiative transfer models (e.g., \citealt{kaufman06}; \citealt{carrili13}; \citealt{gururajan22,gururajan23}; \citealt{pound23}; \citealt{schimek24}).
The cold molecular emission lines with different critical densities but similar excitation potentials serve as excellent probes for the molecular gas density. For example, the \co{} emission, which is usually observed simultaneously with the \ci{} line due to the close frequency separation, has a higher critical density compared to \ci{} ($8.7 \times 10^{4}$ and 470 $\rm cm^{-3}$ for \co{} and \ci{}, respectively; \citealt{carrili13}). Therefore, the \ci{}/\co{} ratio is an ideal tracer of the molecular hydrogen density.

The CGM fills the volume between the ISM and the intergalactic/intracluster medium.  
It is perhaps the key regulator of galactic gas supply by acting as a reservoir that contributes gas to galaxies or that absorbs gas depleted from galaxies, thereby playing a crucial role in feeding or quenching the star formation of galaxies \citep{tumlinson17}.  
The CGM has been extensively studied in atomic and ionized phases through absorption or emission  (e.g., \citealt{mccarthy90}; \citealt{reuland03}; \citealt{villar03}; \citealt{arrigoni15, arrigoni18}; \citealt{lau16}; \citealt{cai17,cai18,cai19}; \citealt{chen17}; \citealt{tumlinson17}). 
In particular, enormous  Lyman-$\alpha$ (Ly$\alpha$) nebulae (ELANe) have been found around massive QSO hosts that extend to up to a few 100 kpcs (e.g., \citealt{cai19}). 
However, the cold CGM (with temperatures of  $\rm \sim 10 - 100\ K$), which provides the direct link between the CGM and star formation, is less extensively explored. 
The cold CGM emission began to be revealed by recent detections of widespread CO (e.g., \citealt{emonts16}; \citealt{ginolfi17}; \citealt{dannerbauer17}; \citealt{cicone21}; \citealt{li21}; \citealt{jones23}; \citealt{scholtz23}; \citealt{chen24}), \cii{} (e.g., \citealt{cicone15}; \citealt{fujimoto20}; \citealt{ginolfi20}; \citealt{herrera-camus21}; \citealt{debreuck22}; \citealt{meyer22}) and \ci{} (e.g., \citealt{emonts18,emonts23}; \citealt{scholtz23}) emission surrounding galaxies, which extend for tens to hundreds of kpc.

For a systematic study of the role that the cold CGM plays in the evolution of massive galaxies in the Early Universe, we carried out the SUrvey of Protocluster ELANe Revealing CO/CI in the Ly$\alpha$ Detected CGM (SUPERCOLD-CGM). 
We simultaneously target the \co{} and \ci{} lines in a sample of 10 enormous Ly$\alpha$ nebulae around ultra-luminous (i-magnitude brighter than 18.5)  Type-I QSOs at $z\sim 2$ using both the main (12 m) and Morita (ACA 7 m) arrays of the Atacama Large Millimeter/submillimeter Array (ALMA). Including the 7 m array increases the surface brightness sensitivity for detecting extended low-surface brightness emission in the CGM. 
The \co{} emission is detected in all the QSO hosts, with 70$\%$ of our QSO fields having at least one CO companion. 
In particular, extended \co{} emission has been revealed in QSO fields on scales of 15$-$100 kpc for 60$\%$ of our targets, with the majority of them showing extended \co{} emission predominantly in the direction of the companion galaxies. Detailed descriptions of the sample and the \co{} results are reported in \citet[][hereafter \paper]{li23}.
In this work, we present the \ci{} results.

The paper is structured as follows. 
We present the observations and data reduction details in Section \ref{observations}. Observational results are presented in Section \ref{results} where we measure the \ci{} and dust continuum emission properties for the QSOs, companion galaxies, and the CGM.  
In Section \ref{discussions}, we discuss the physical properties of the molecular gas utilizing ratios between different molecular emission lines and the far-infrared (FIR) luminosities for the QSOs, companion galaxies, and CGM. In addition, we measure the molecular gas masses of the QSOs, companion galaxies, and the CGM in Section \ref{discussions}. Finally, we summarize our main conclusions in Section \ref{summary}.
We adopt a flat $\Lambda$CDM cosmology model with  $\rm H_{0} = 70\  km \ s^{-1} \ Mpc^{-1}$, $\rm \Omega_{\Lambda}=0.7$, and $\rm \Omega_{M}=0.3$, so that $1''$ corresponds to $8.25-8.16$ kpc at $z=2.22 - 2.37$ respectively.

\begin{figure*}
\centering
\includegraphics[width=0.3\linewidth]{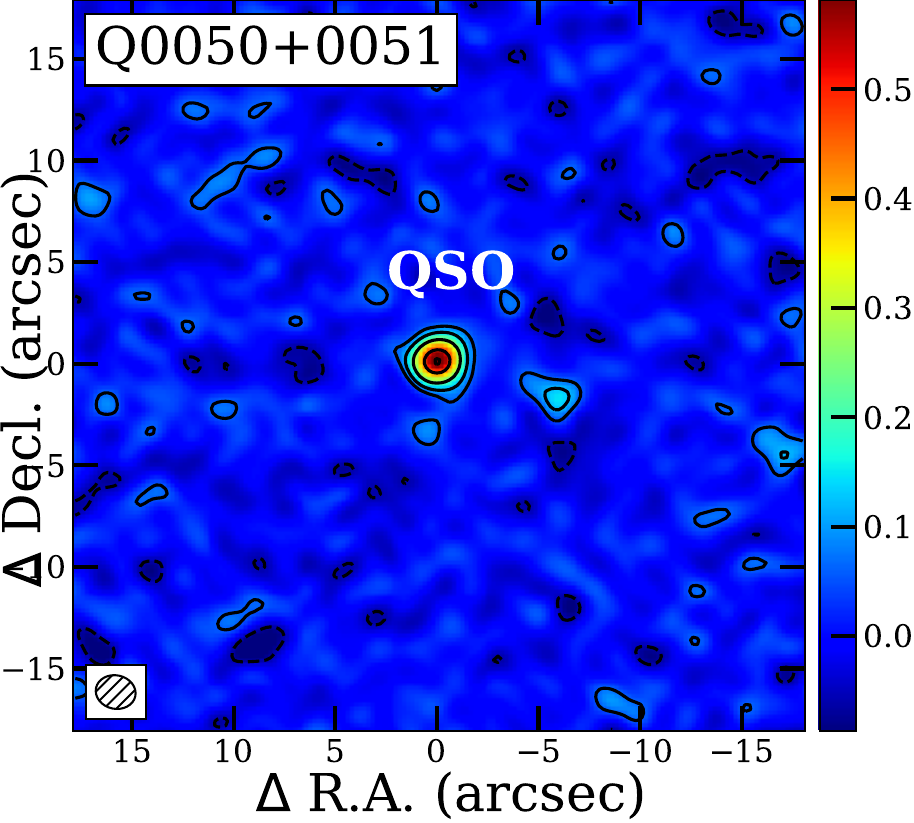}
\includegraphics[width=0.3\linewidth]{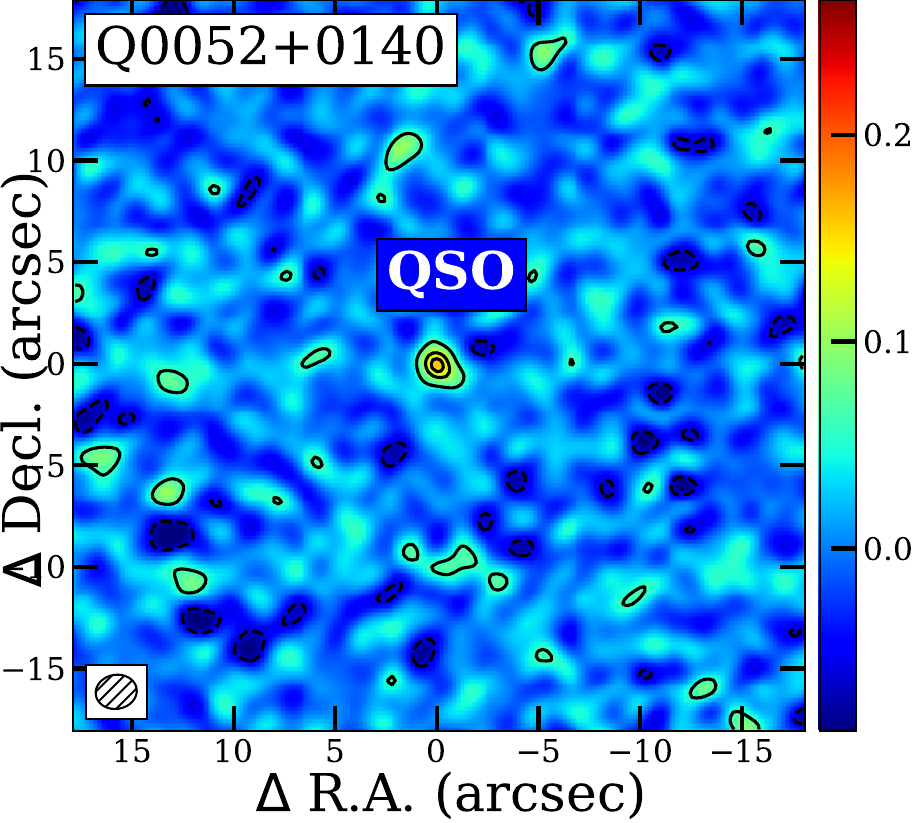}
\includegraphics[width=0.3\linewidth]{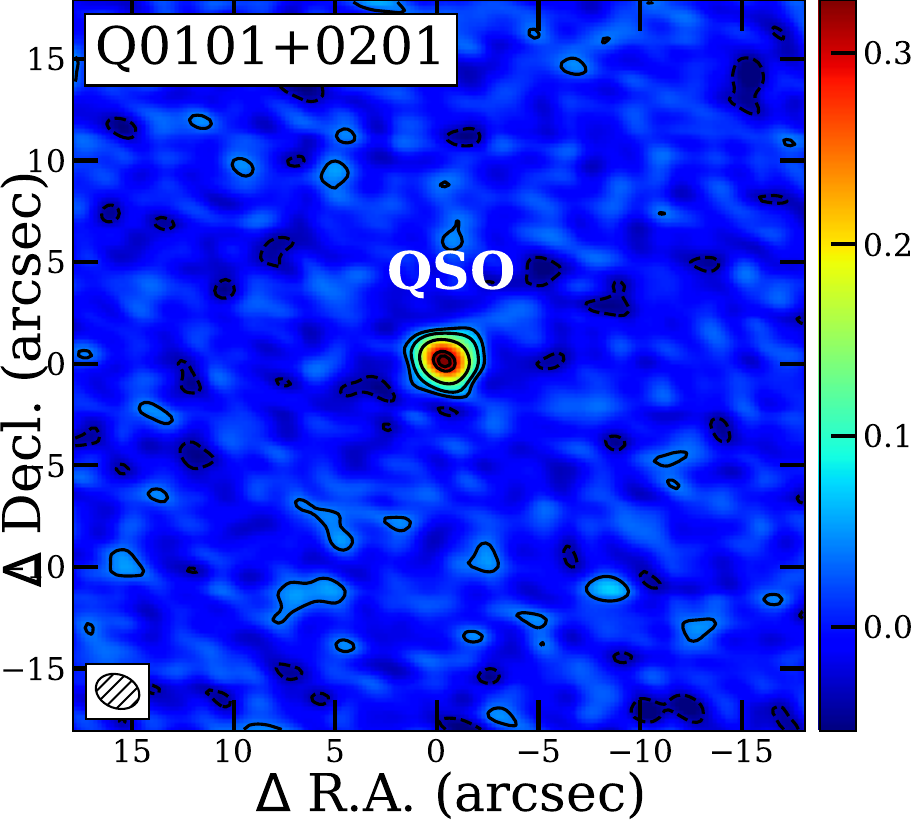}
\includegraphics[width=0.3\linewidth]{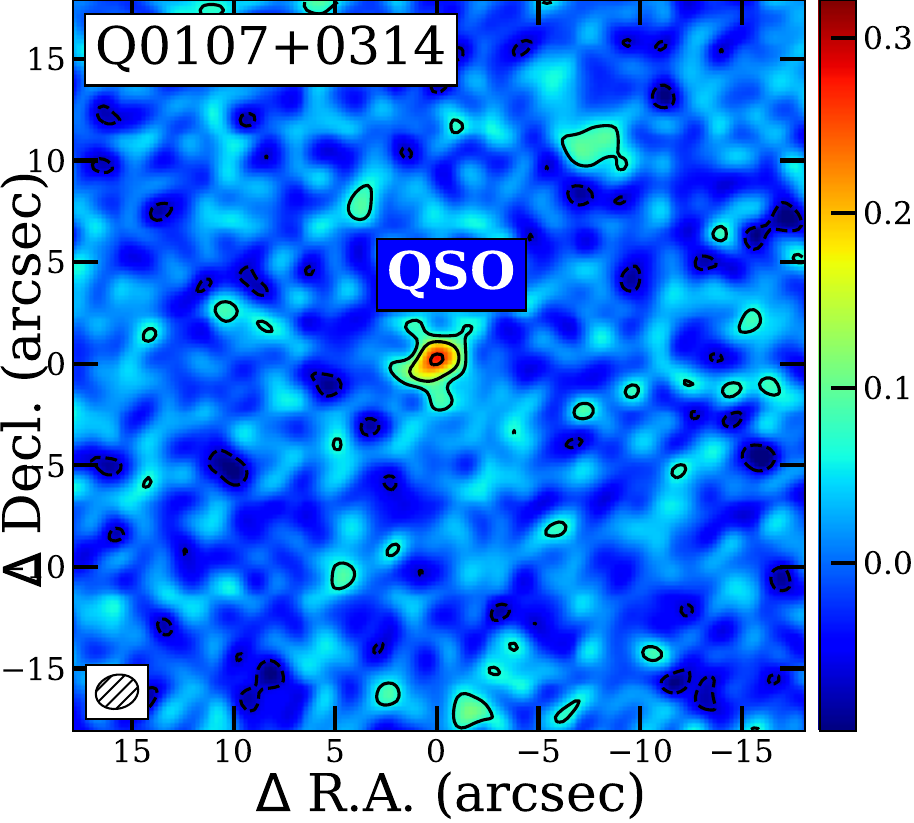}
\includegraphics[width=0.3\linewidth]{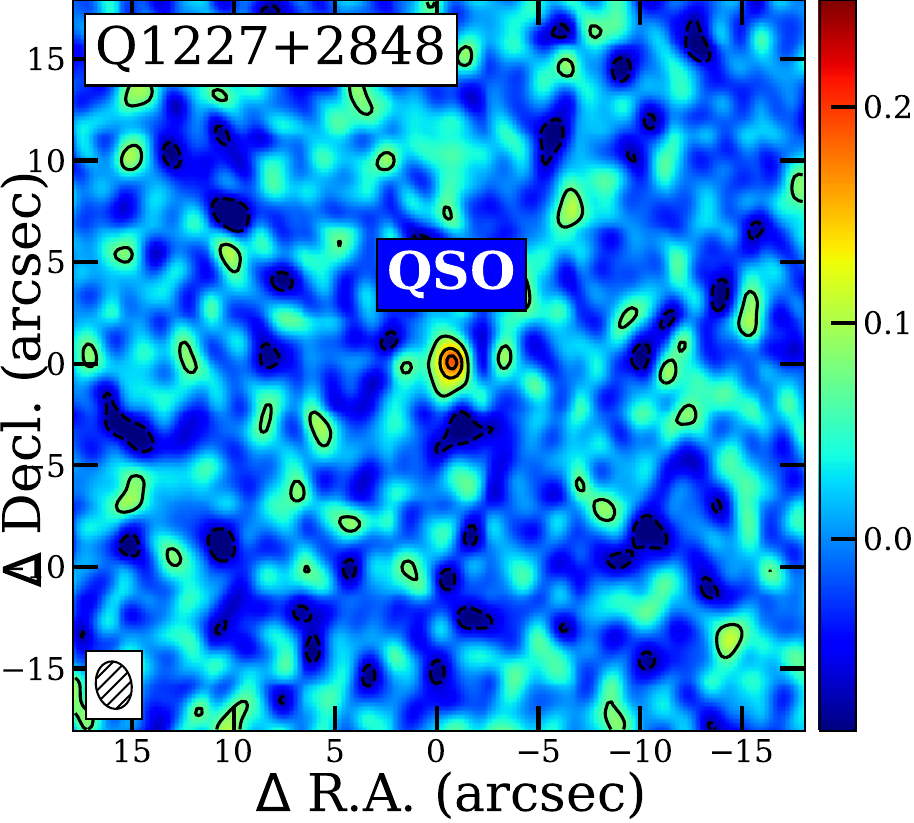}
\includegraphics[width=0.3\linewidth]{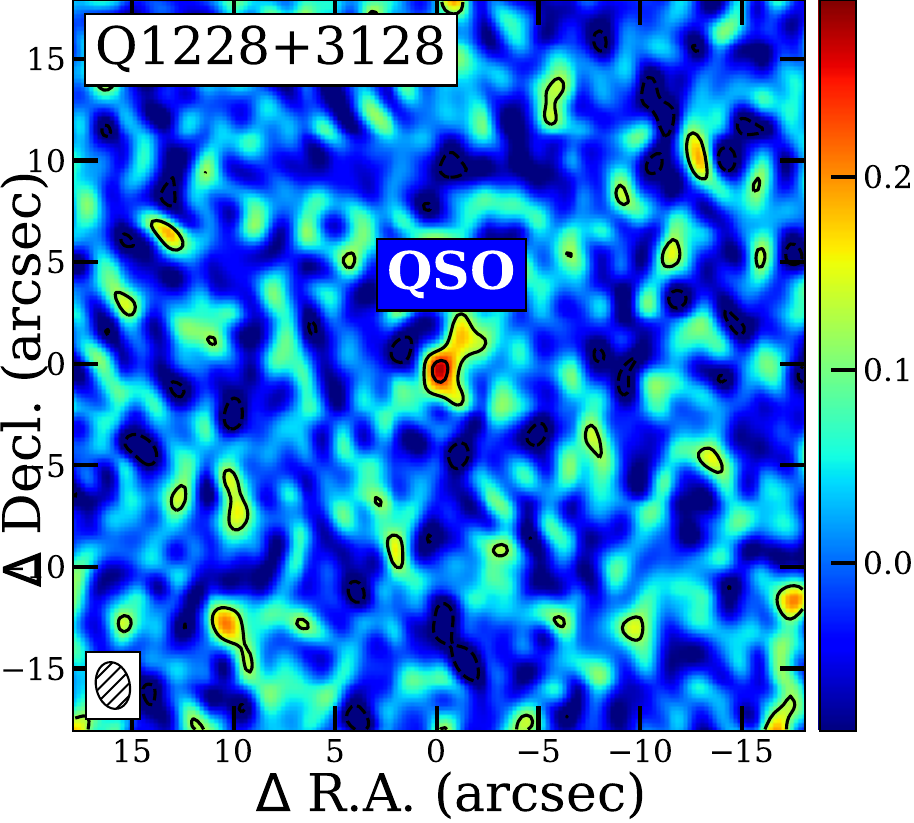}
\includegraphics[width=0.3\linewidth]{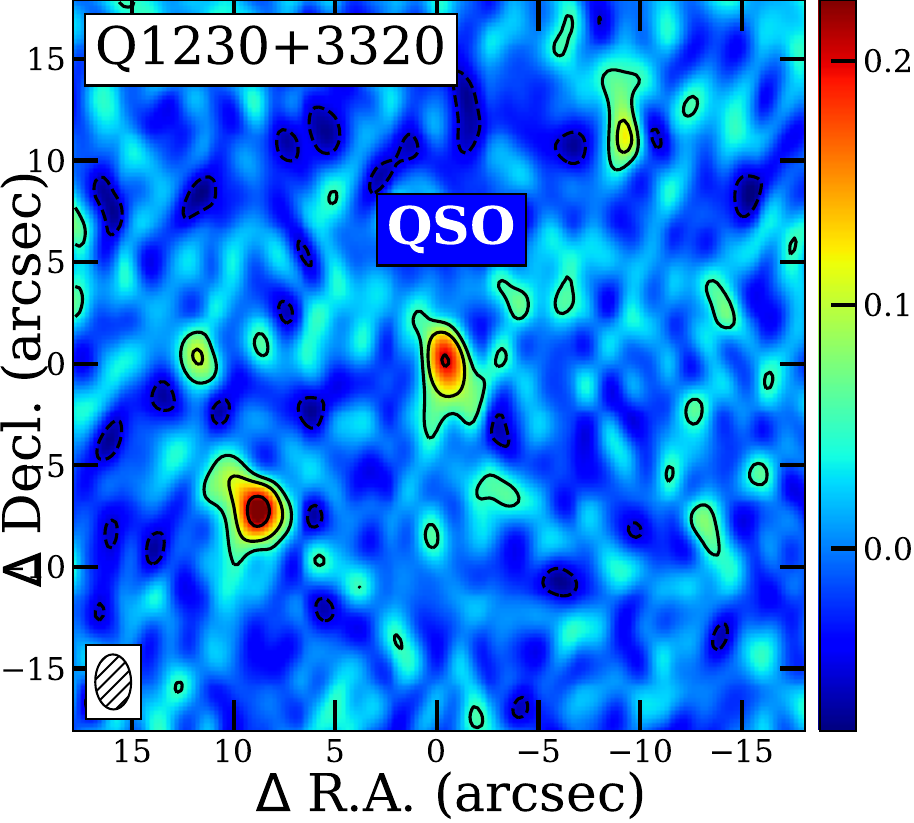}
\includegraphics[width=0.3\linewidth]{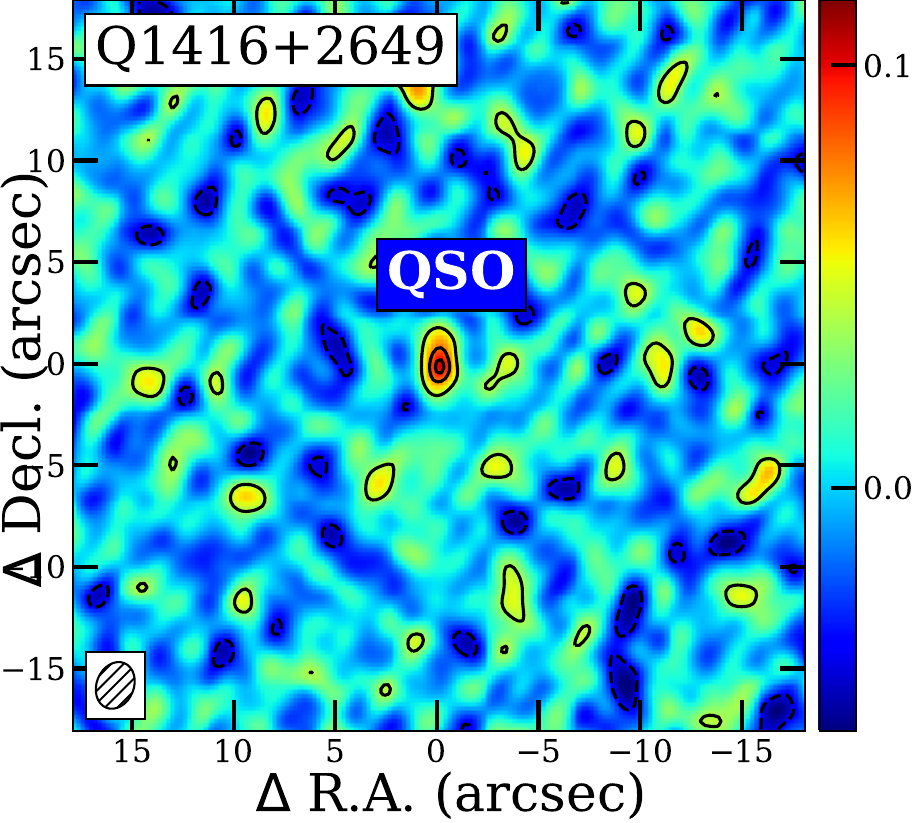}
\includegraphics[width=0.3\linewidth]{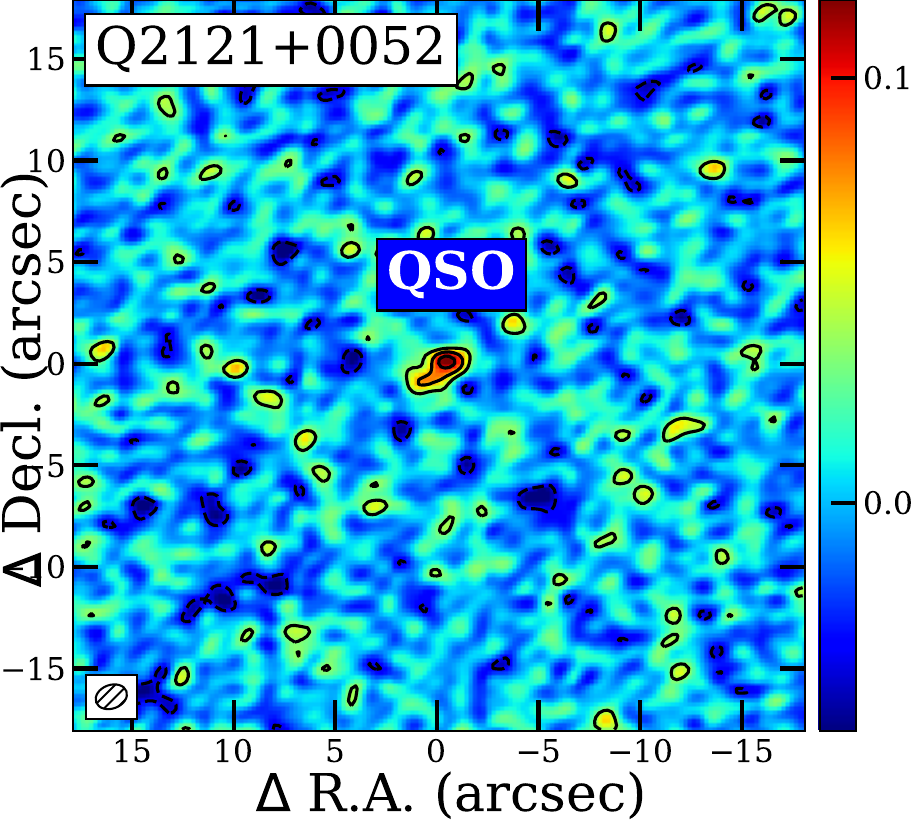}
\includegraphics[width=0.3\linewidth]{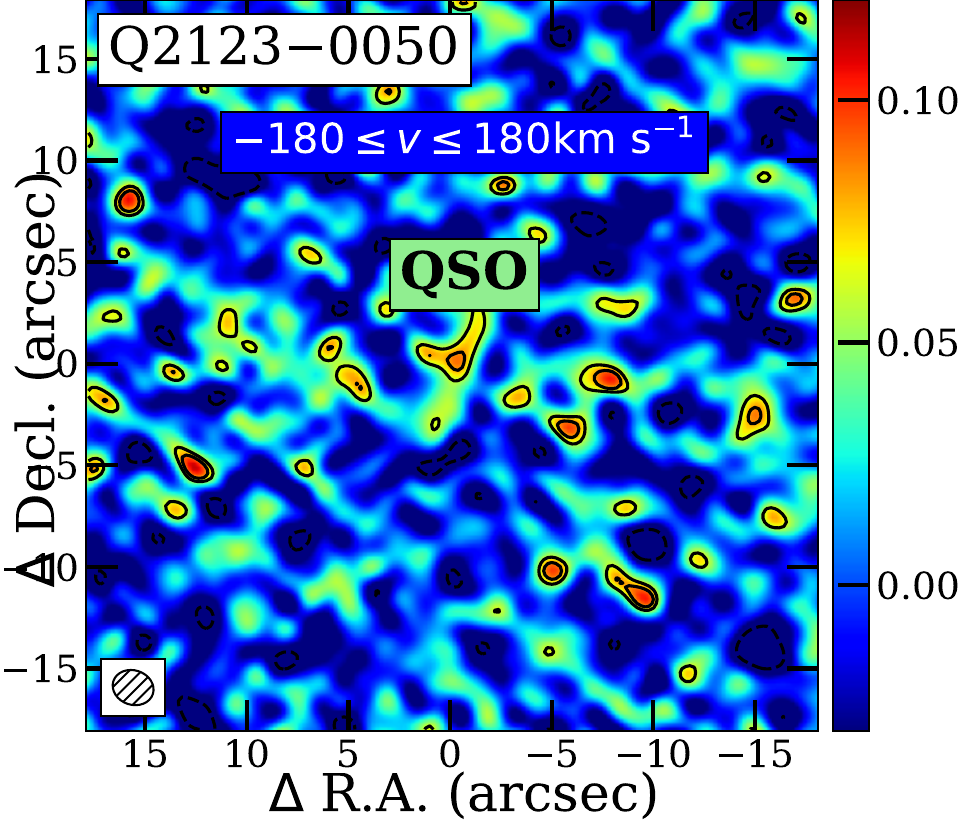}
\caption{\label{intensity_qso} \ci{} intensity maps for the QSOs. Color bars show scale in a unit of \jypbkmps{}.  We show the contours with a factor of 2$\times$ increase, except for the last one which we highlight the peak of the emission. 
Contours denote [-2,2,4,8,16,21]$\times \sigma$ ($\sigma = 0.029$ \jypbkmps{}) for \jone{}, [-2,2,4,5]$\times \sigma$ ($\sigma = 0.029$ \jypbkmps{}) for \jtwo{}, [-2,2,4,8,16,18]$\times \sigma$ ($\sigma = 0.018$ \jypbkmps{}) for \jthree{}, [-2,2,4,8]$\times \sigma$ ($\sigma = 0.032$ \jypbkmps{}) for \jfour{}, [-2,2,4,5]$\times \sigma$ ($\sigma = 0.036$ \jypbkmps{}) for \jfive{}, [-2,2,4]$\times \sigma$ ($\sigma = 0.058$ \jypbkmps{}) for \jsix{}, [-2,2,4,8]$\times \sigma$ ($\sigma = 0.025$ \jypbkmps{}) for \jseven{}, [-2,2,4,5]$\times \sigma$ ($\sigma = 0.019$ \jypbkmps{}) for \jeight{}, [-2,2,4,6]$\times \sigma$ ($\sigma = 0.018$ \jypbkmps{}) for \jnine{}, and [-2,2,3.3]$\times \sigma$ ($\sigma = 0.029$ \jypbkmps{}) for \jten{}. For \jten{}, we also show the velocity range for generating the intensity map. } 
\end{figure*}

\begin{figure*}
\centering
\includegraphics[width=0.32\linewidth]{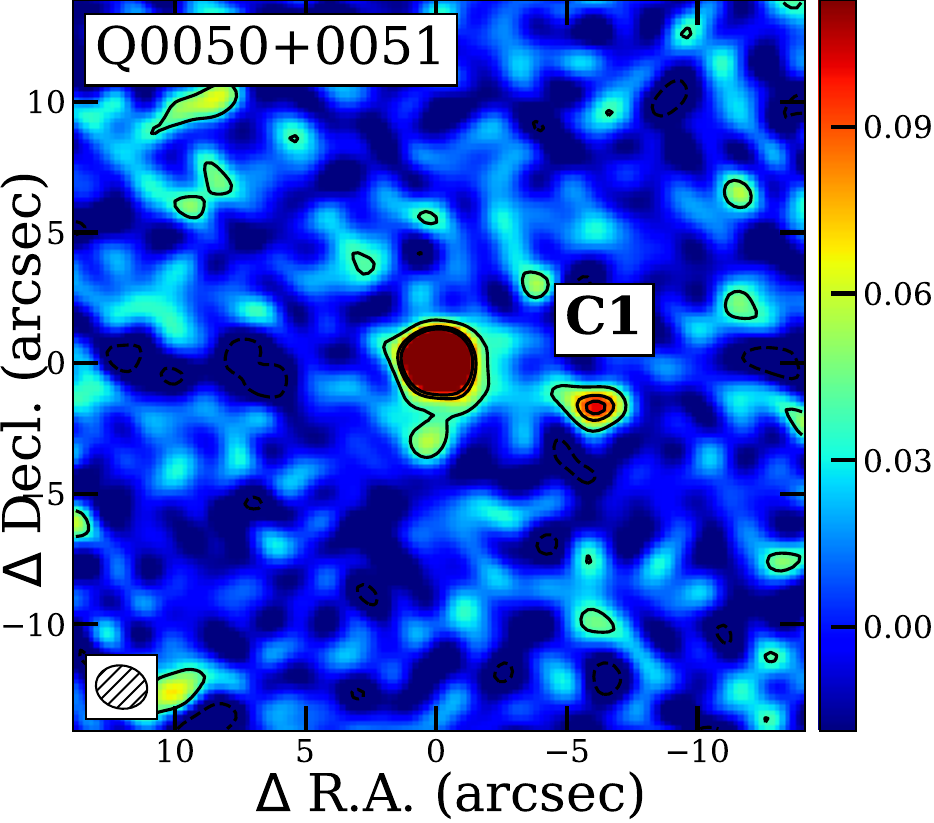}
\includegraphics[width=0.32\linewidth]{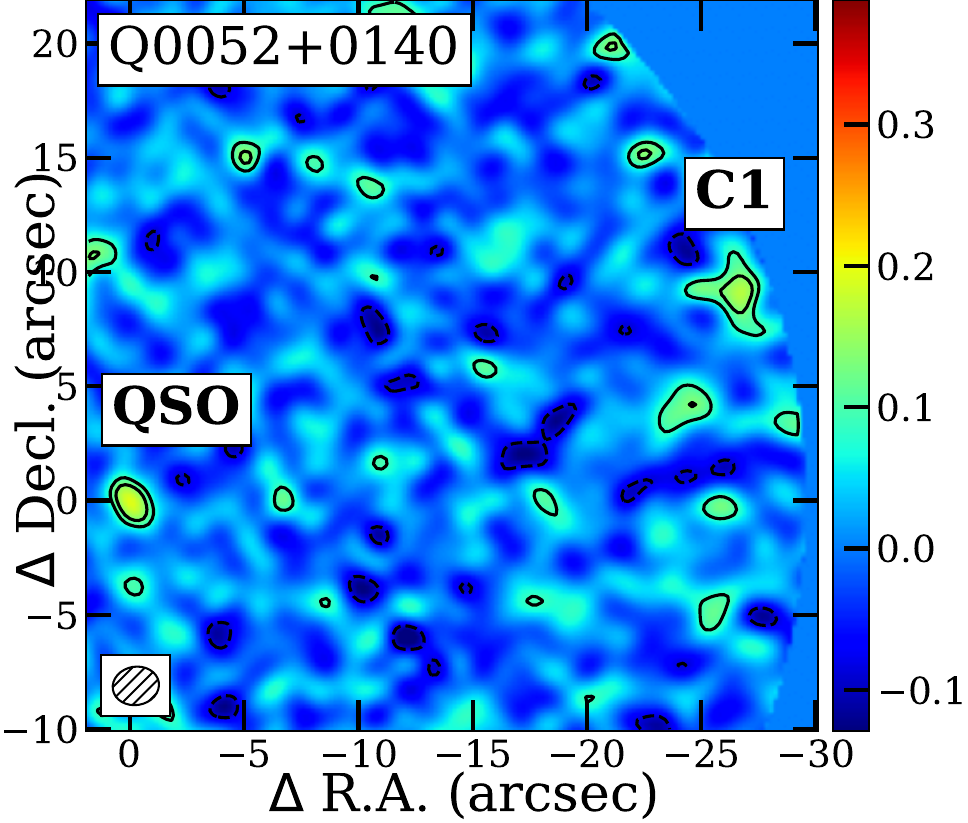}
\includegraphics[width=0.32\linewidth]{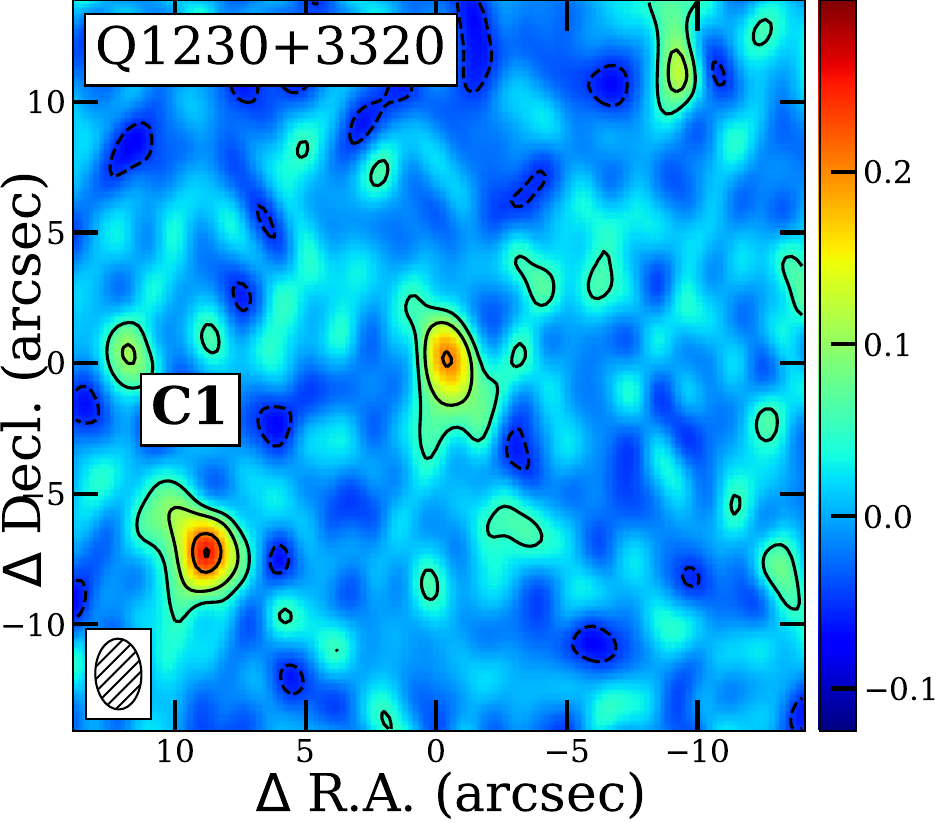}
\includegraphics[width=0.32\linewidth]{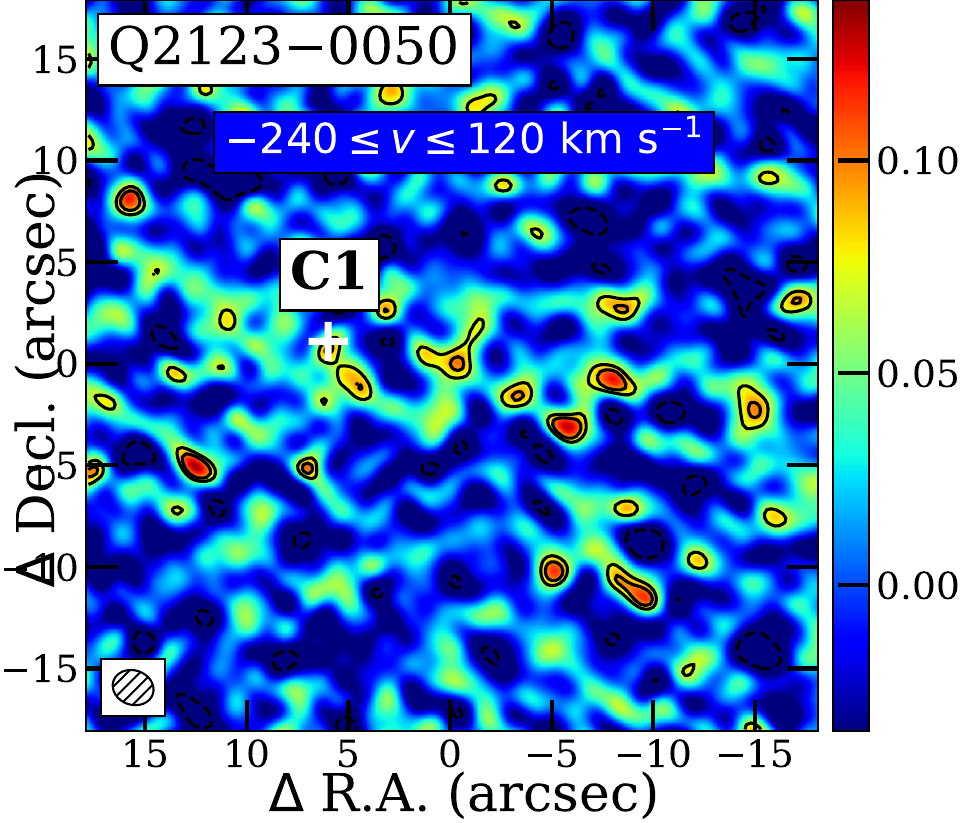}
\includegraphics[width=0.32\linewidth]{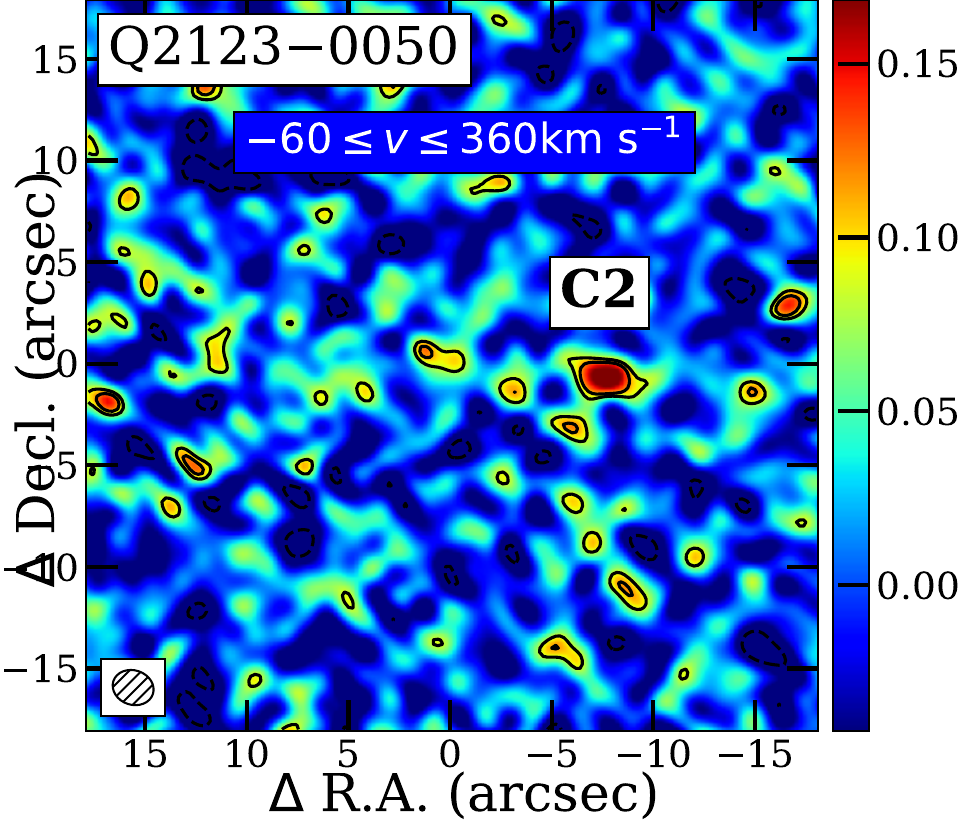}
\caption{\label{intensity_comp} \ci{} intensity maps for the companion galaxies in our QSO fields.  X and Y axis show the positions of the companion galaxies relative to the QSOs.  Colorbars show line flux scale in the unit of \jypbkmps{}.
Contours denote the detections of [-2,2,4,5]$\times \sigma$ (1$\sigma$ = 0.019 \jypbkmps{}) for the companion galaxy C1 in the \jone{} field, [-2,2,3]$\times \sigma$ (1$\sigma$ = 0.043 \jypbkmps{}) for C1 in the \jtwo{} field, [-2,2,4,8,10]$\times \sigma$ (1$\sigma$ = 0.025 \jypbkmps{}) for  C1 in the \jseven{} field, [-2,2,3.3]$\times \sigma$ (1$\sigma$ = 0.030 \jypbkmps{}) for C1 in the \jten{}  field, and [-2,2,4]$\times \sigma$ (1$\sigma$ = 0.041 \jypbkmps{}) for  C2 in the \jten{} field. For the companion galaxies C1 and C2 of \jten{}, we show the velocity ranges for generating the intensity maps. We show the position of the companion galaxy C1 of \jten{} as a white cross. } 
\end{figure*}

\begin{figure*}
\centering
\includegraphics[width=0.49\linewidth]{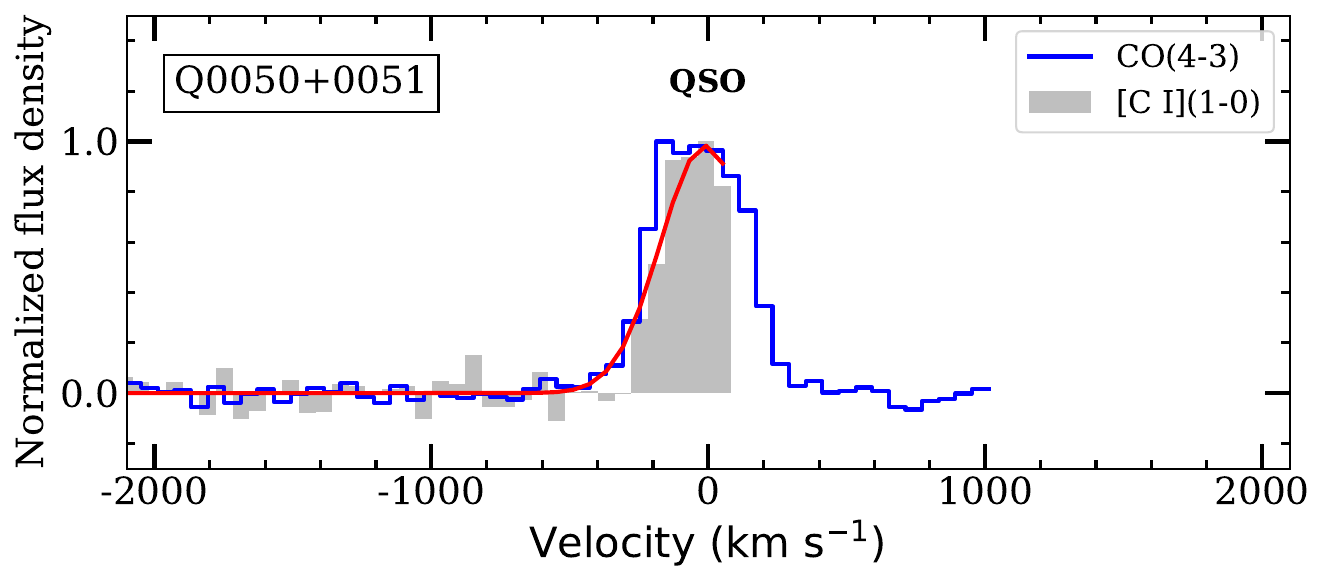}
\includegraphics[width=0.49\linewidth]{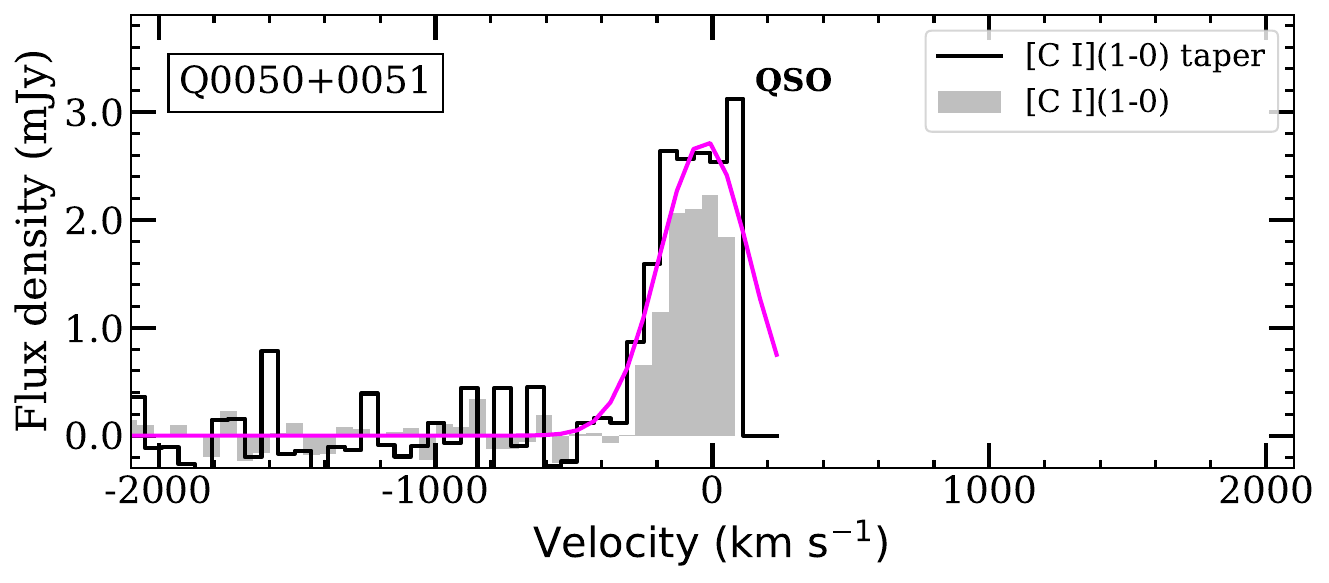}
\includegraphics[width=0.49\linewidth]{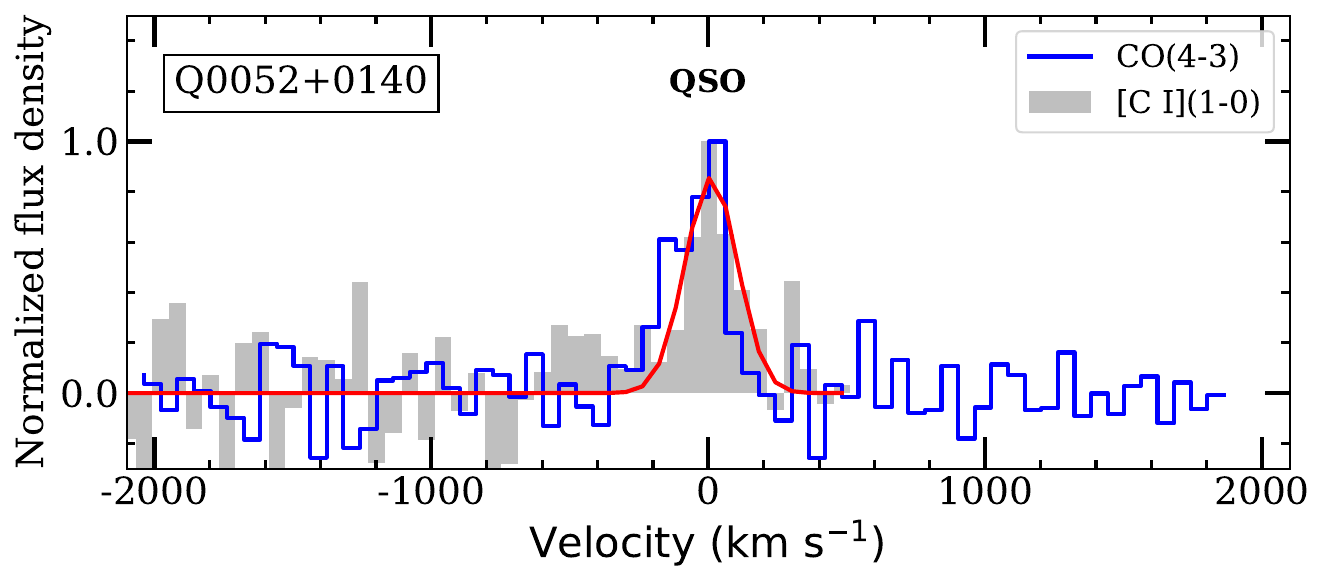}
\includegraphics[width=0.49\linewidth]{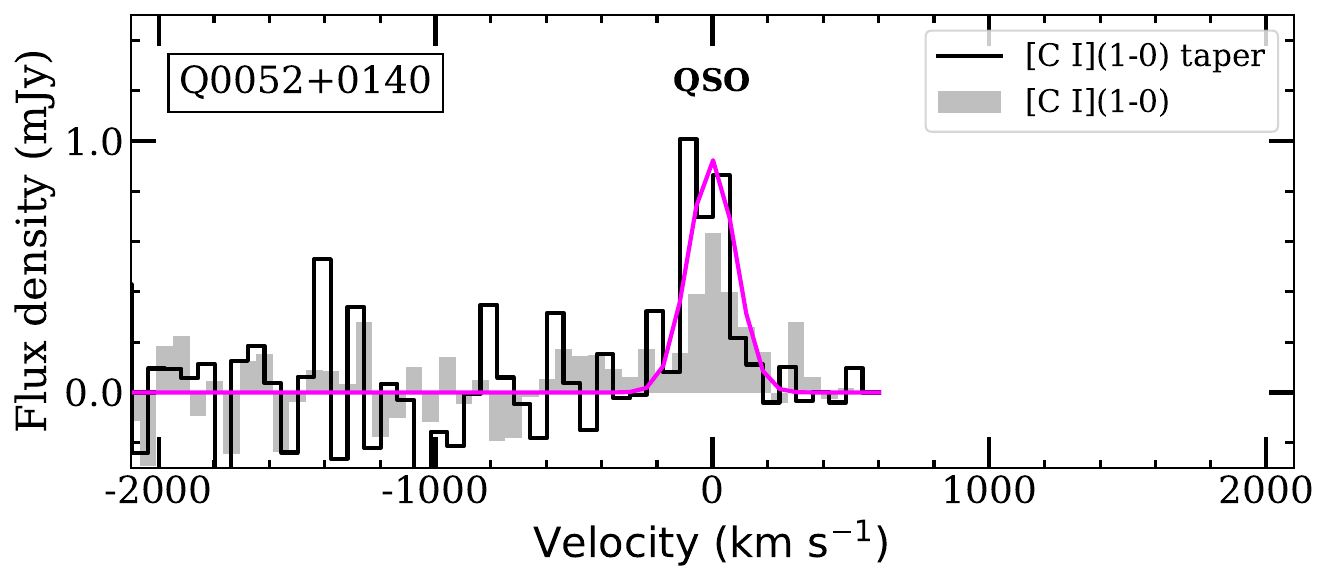}
\includegraphics[width=0.49\linewidth]{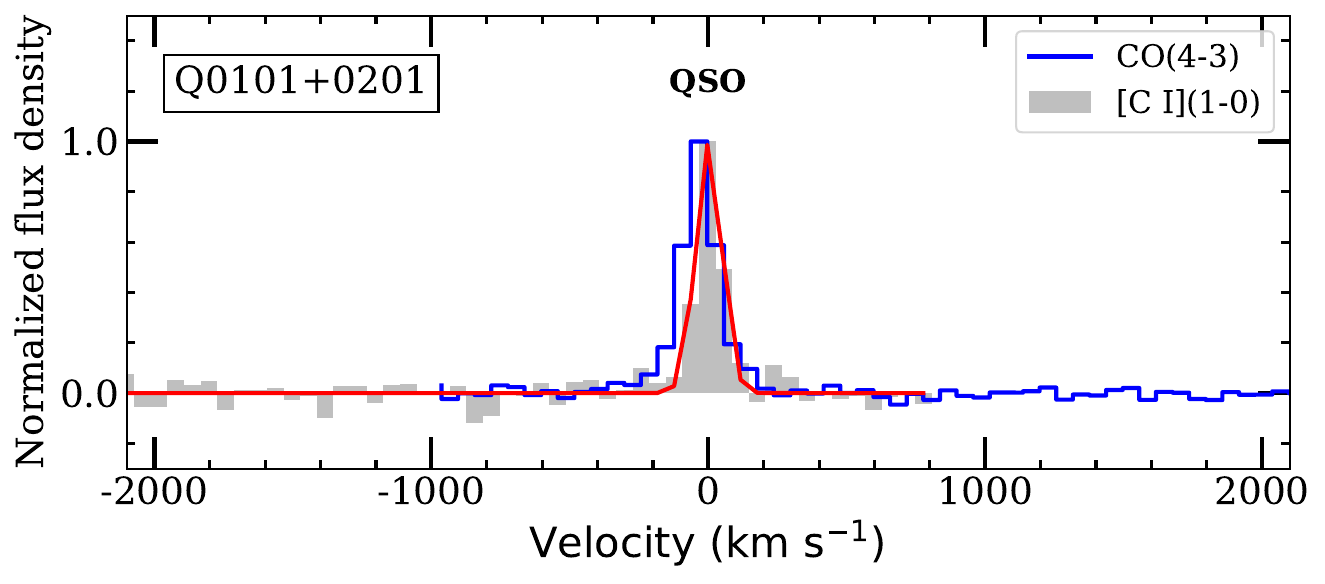}
\includegraphics[width=0.49\linewidth]{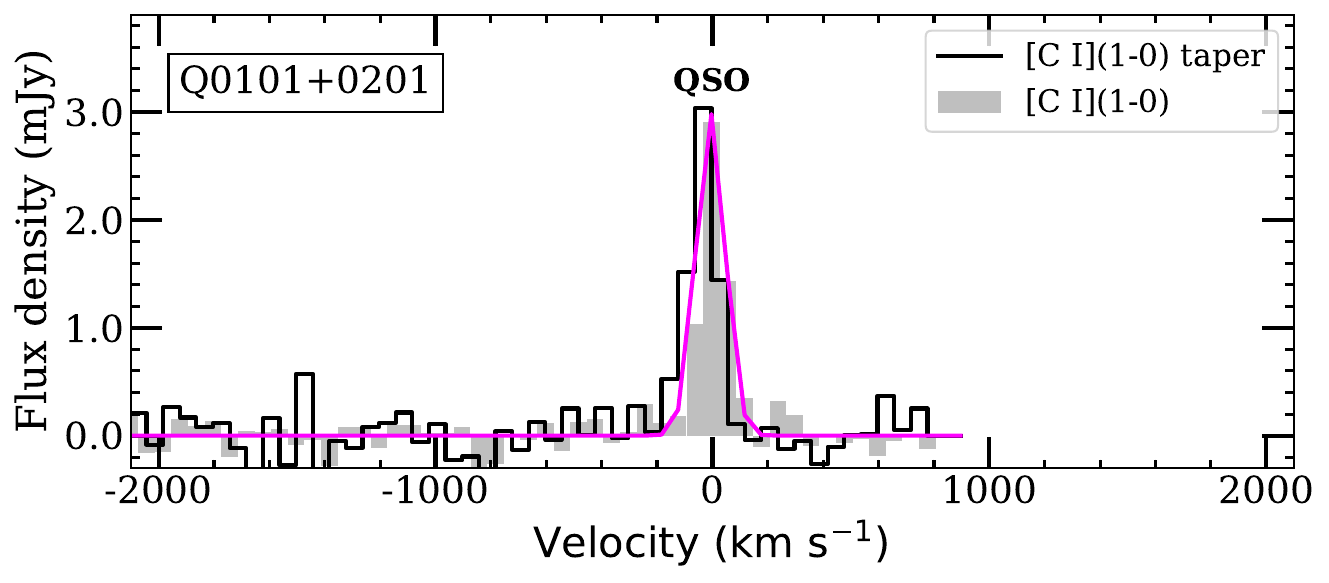}
\includegraphics[width=0.49\linewidth]{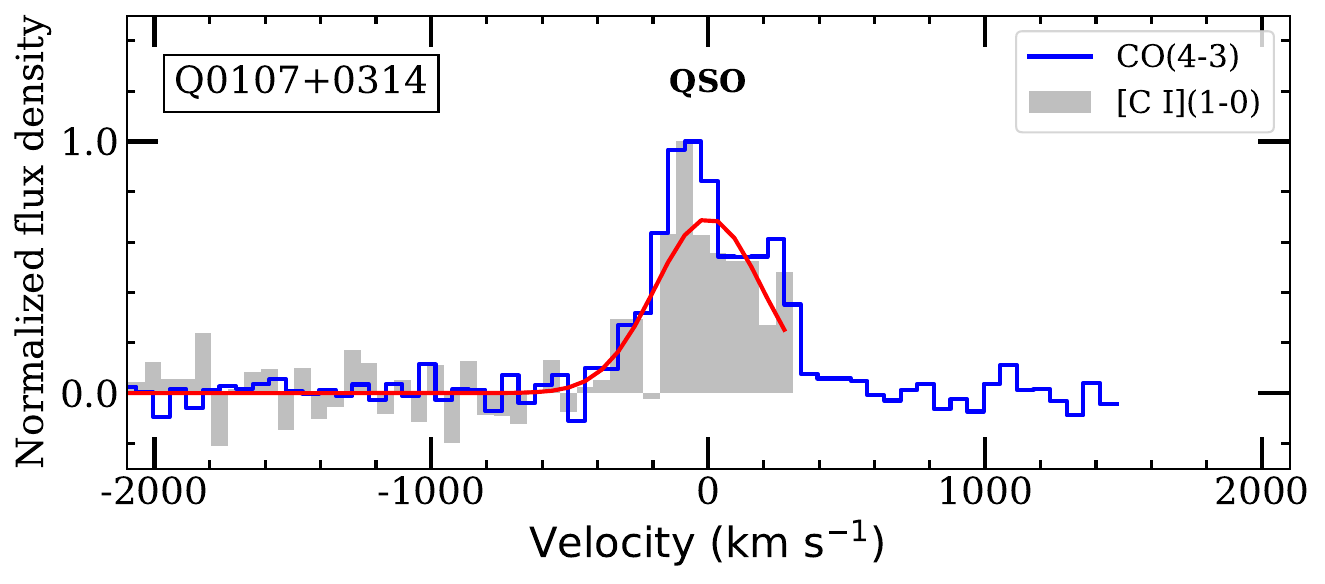}
\includegraphics[width=0.49\linewidth]{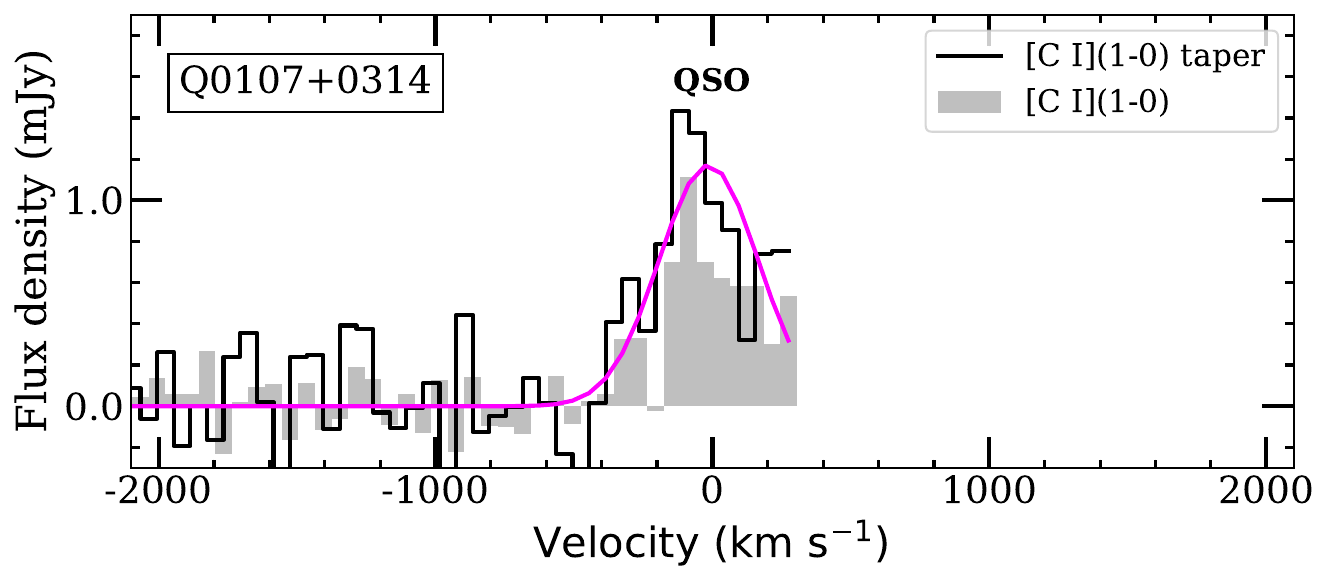}
\includegraphics[width=0.49\linewidth]{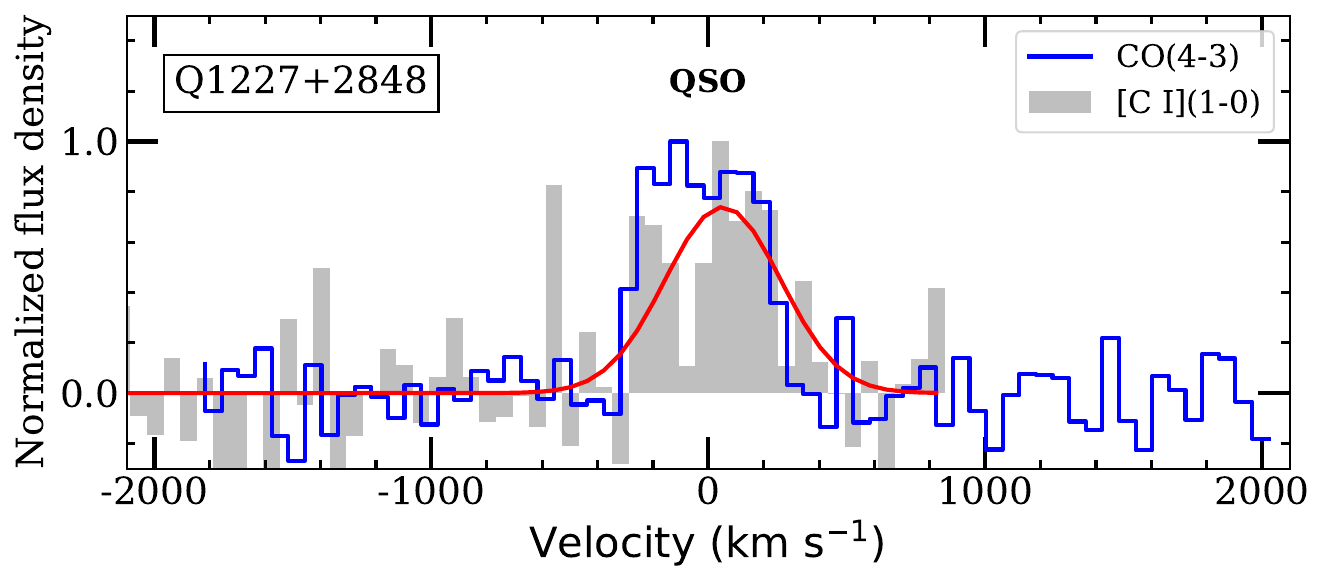}
\includegraphics[width=0.49\linewidth]{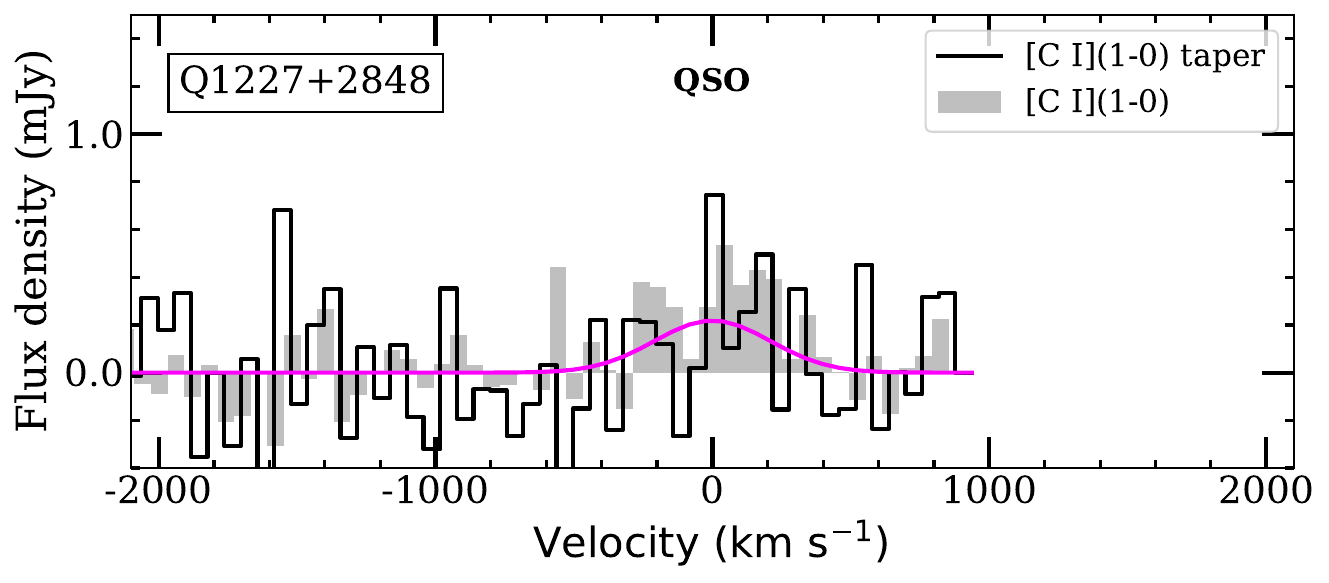}
\caption{\label{spectra1}\ci{} spectra for QSOs. \textbf{Left column:} The \co{} (solid blue lines) and \ci{} (grey histograms) spectra normalized to the peak flux densities. Red solid lines represent Gaussian profile fit to the \ci{} lines. Note that the \ci{} and \co{} are from the 12 m array data of ALMA. \textbf{Right column}: The \ci{} spectra obtained using 12 m array data of ALMA (grey histograms) and 7 m +12 m array data of ALMA after uv-tapering (solid black lines). Gaussian fit to the spectra extracted from the 7 m +12 m array data after uv-tapering are denoted as solid magenta lines. }
\end{figure*}

\begin{figure*}
\addtocounter{figure}{-1}
\centering
\includegraphics[width=0.49\linewidth]{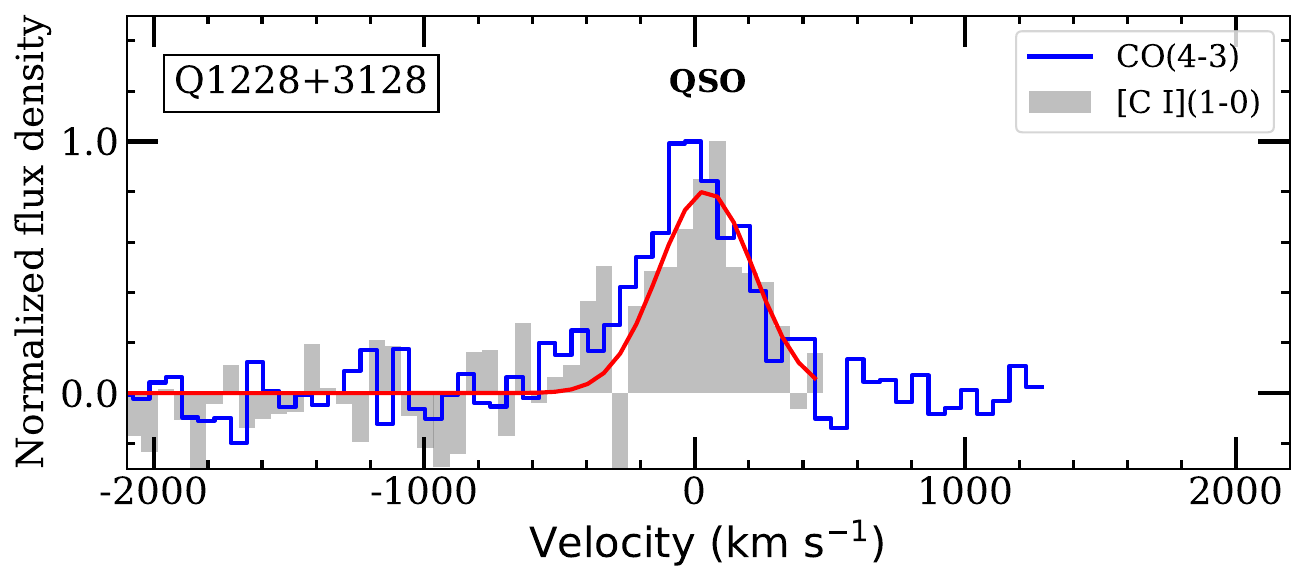}
\includegraphics[width=0.49\linewidth]{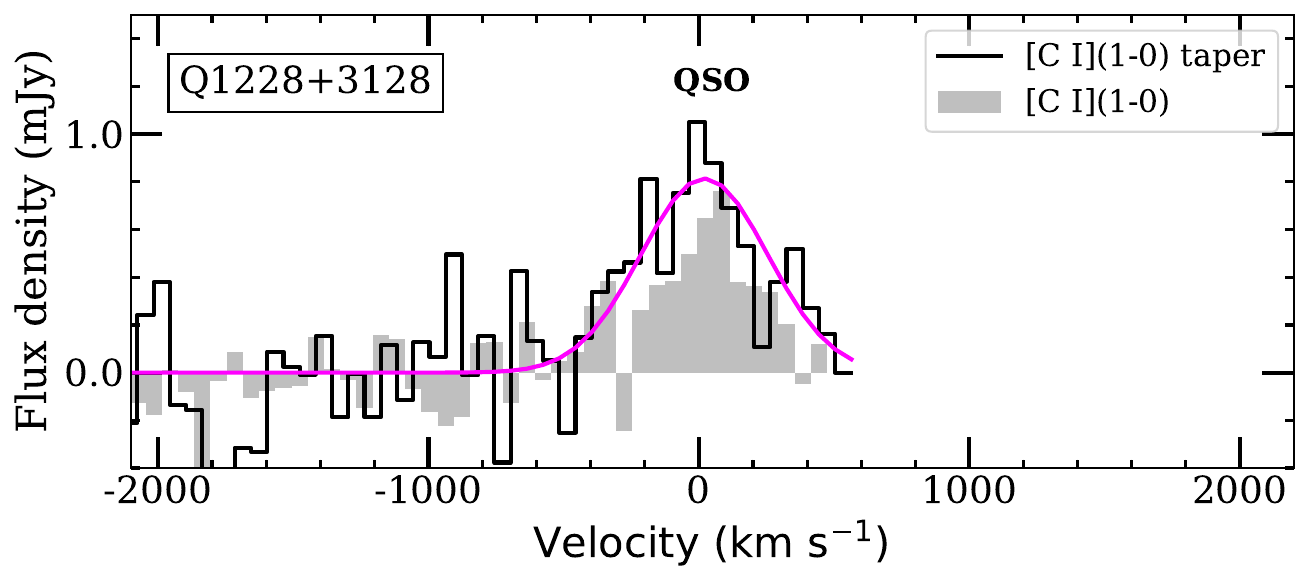}
\includegraphics[width=0.49\linewidth]{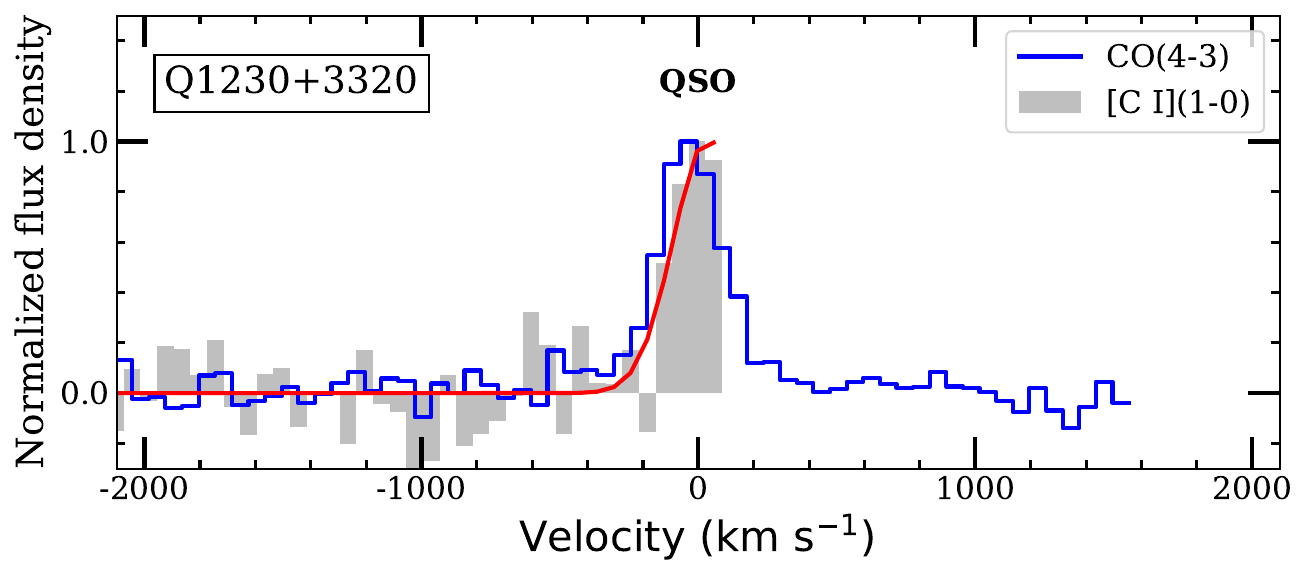}
\includegraphics[width=0.49\linewidth]{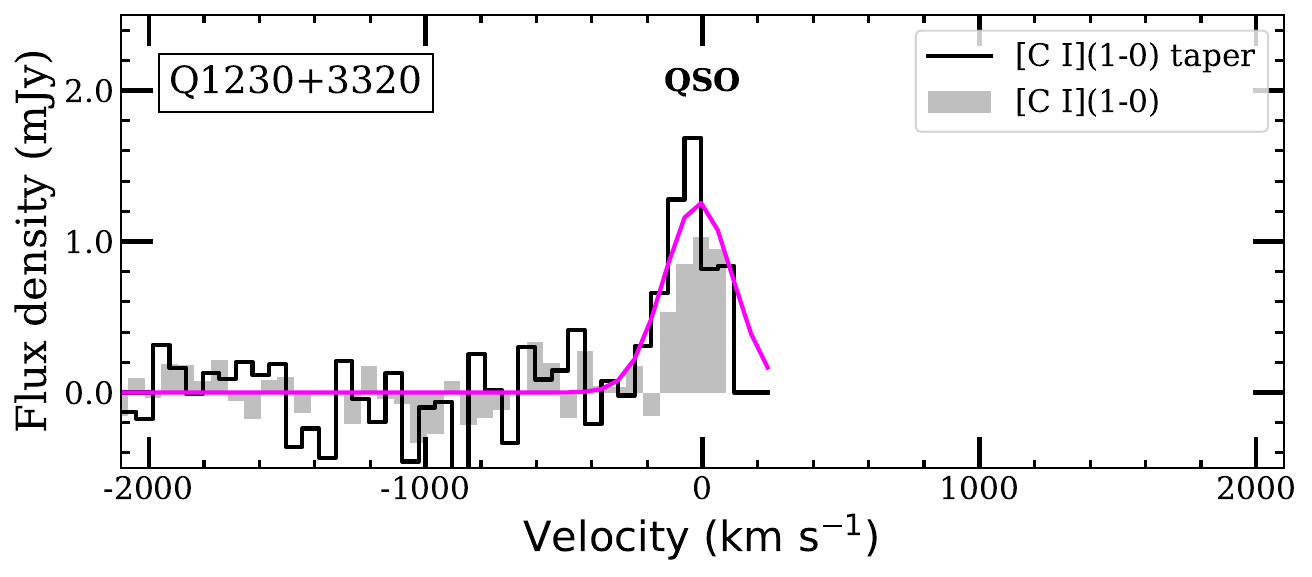}
\includegraphics[width=0.49\linewidth]{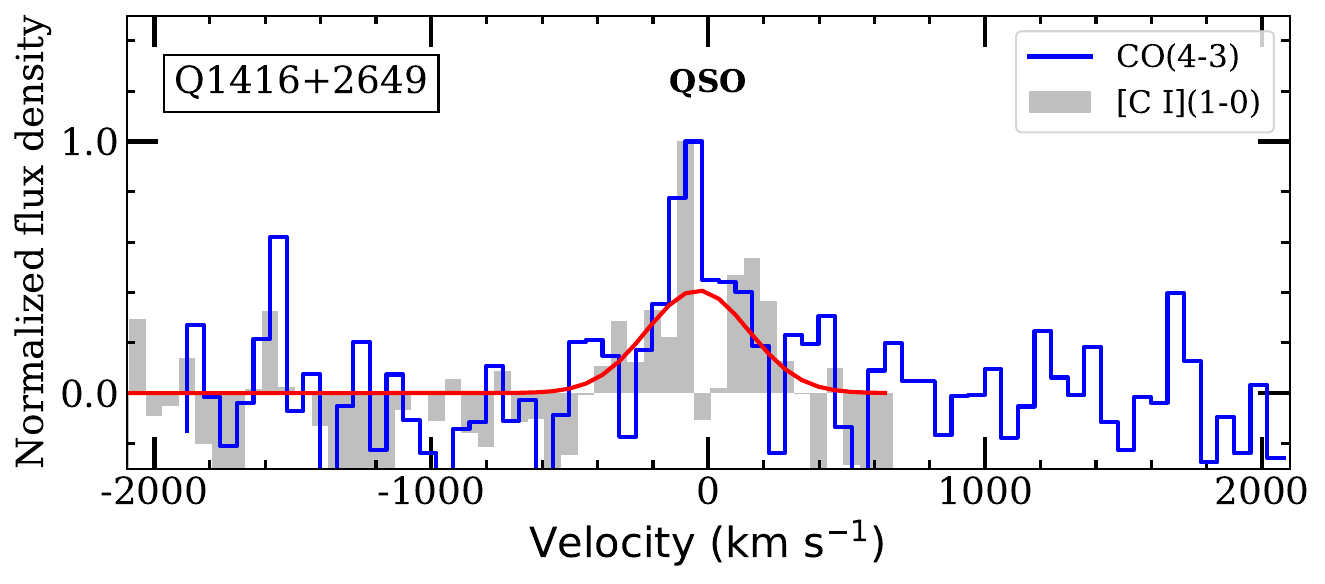}
\includegraphics[width=0.49\linewidth]{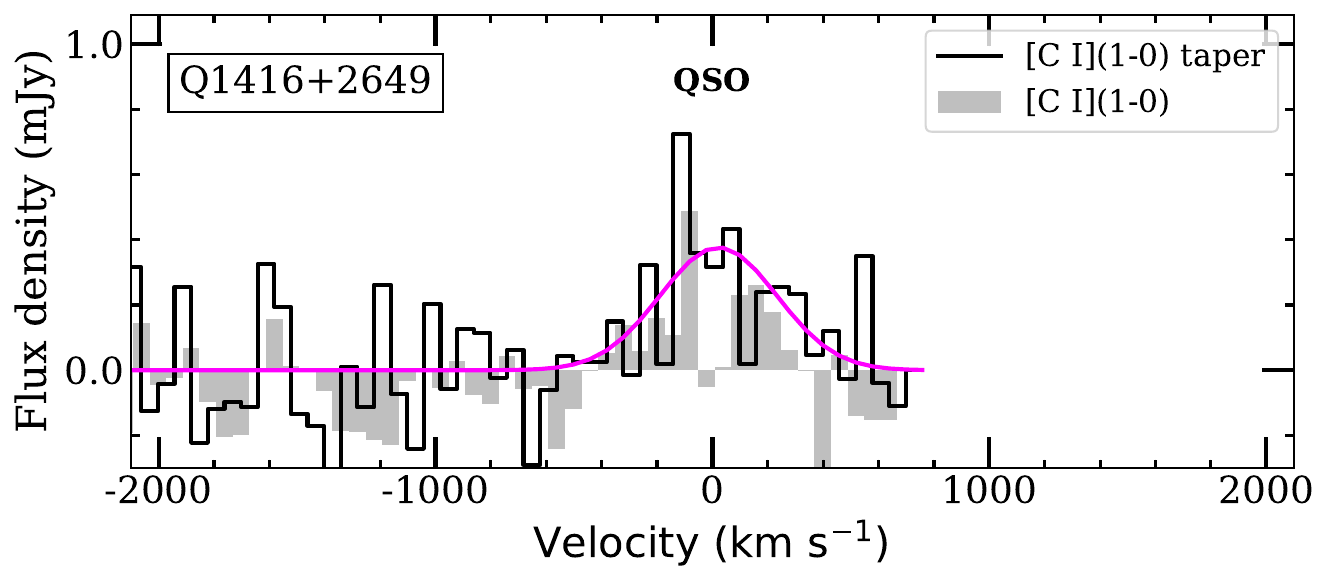}
\includegraphics[width=0.49\linewidth]{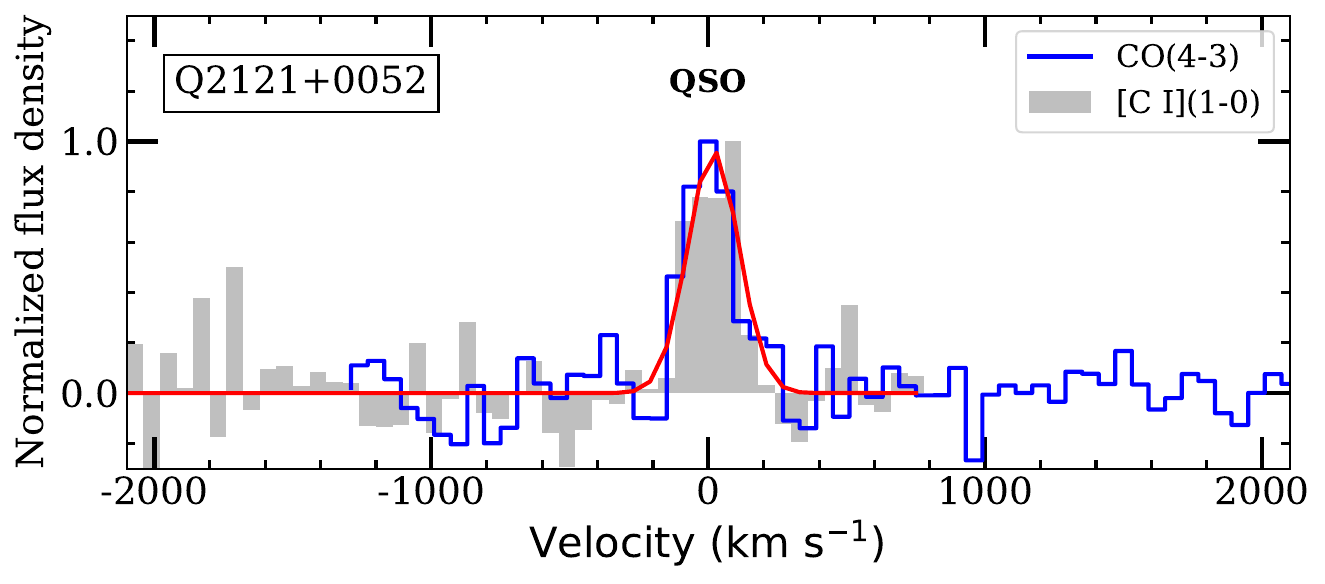}
\includegraphics[width=0.49\linewidth]{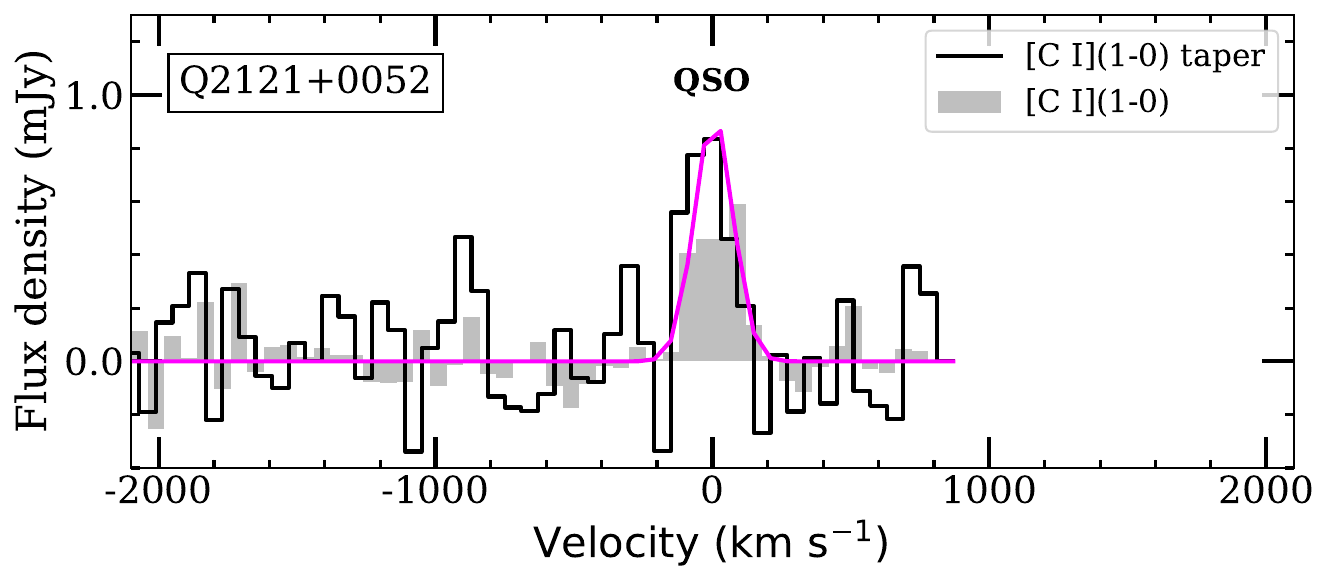}
\includegraphics[width=0.49\linewidth]{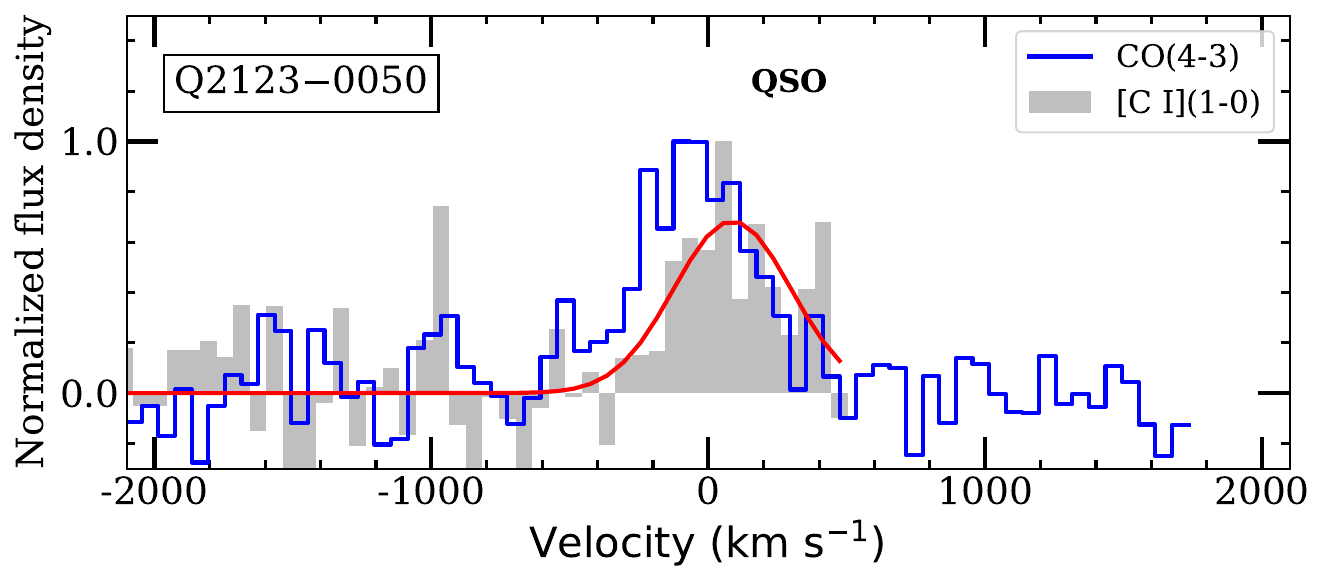}
\includegraphics[width=0.49\linewidth]{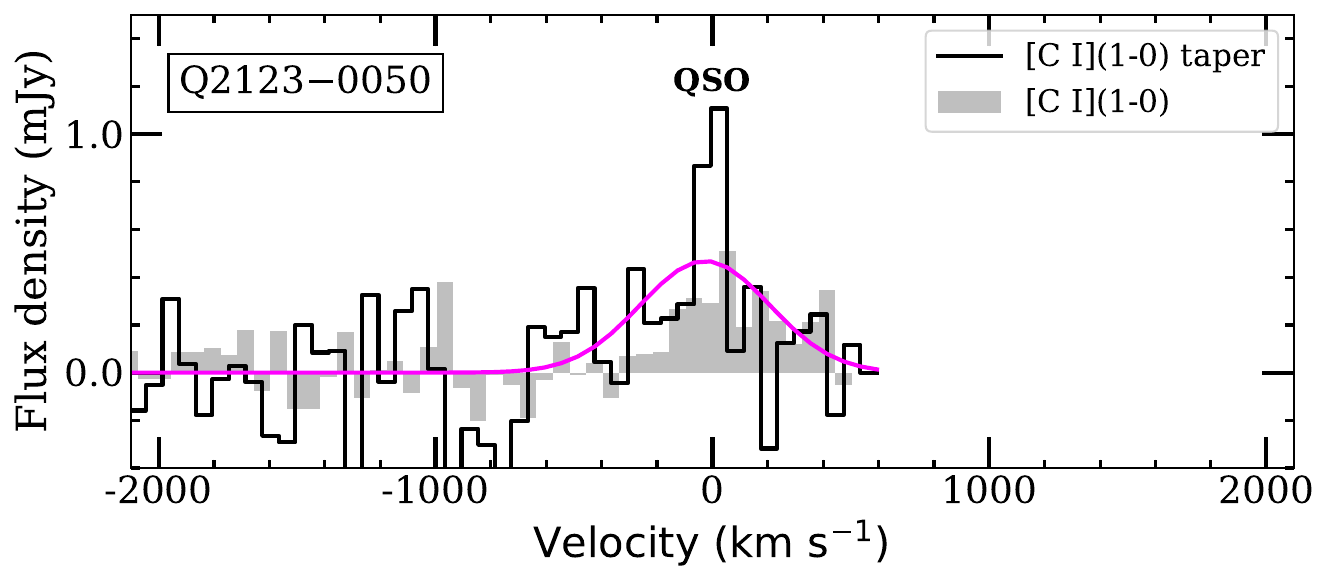}
\caption{Continued...}
\end{figure*}

\begin{figure*}
\centering
\includegraphics[width=0.49\linewidth]{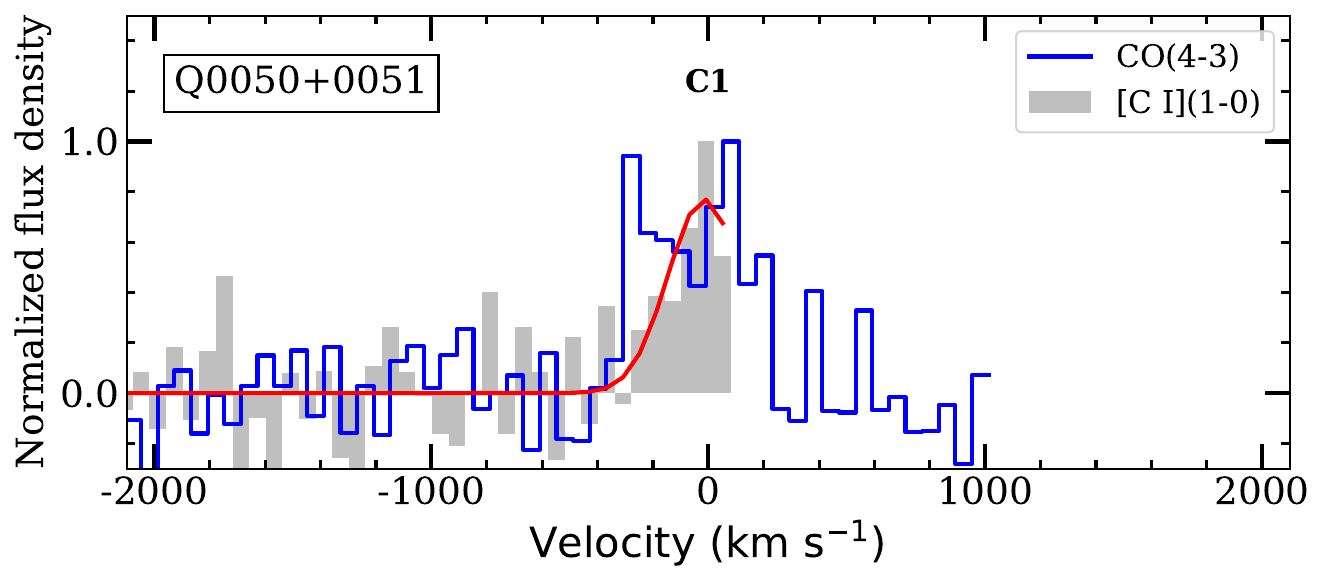}
\includegraphics[width=0.49\linewidth]{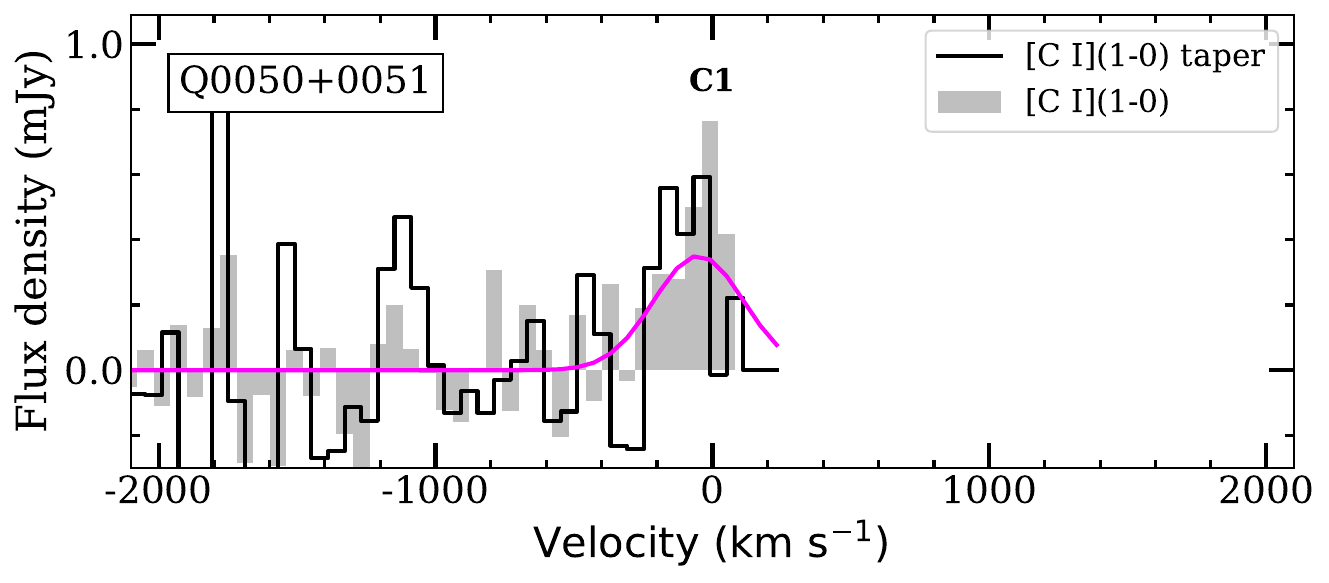}
\includegraphics[width=0.49\linewidth]{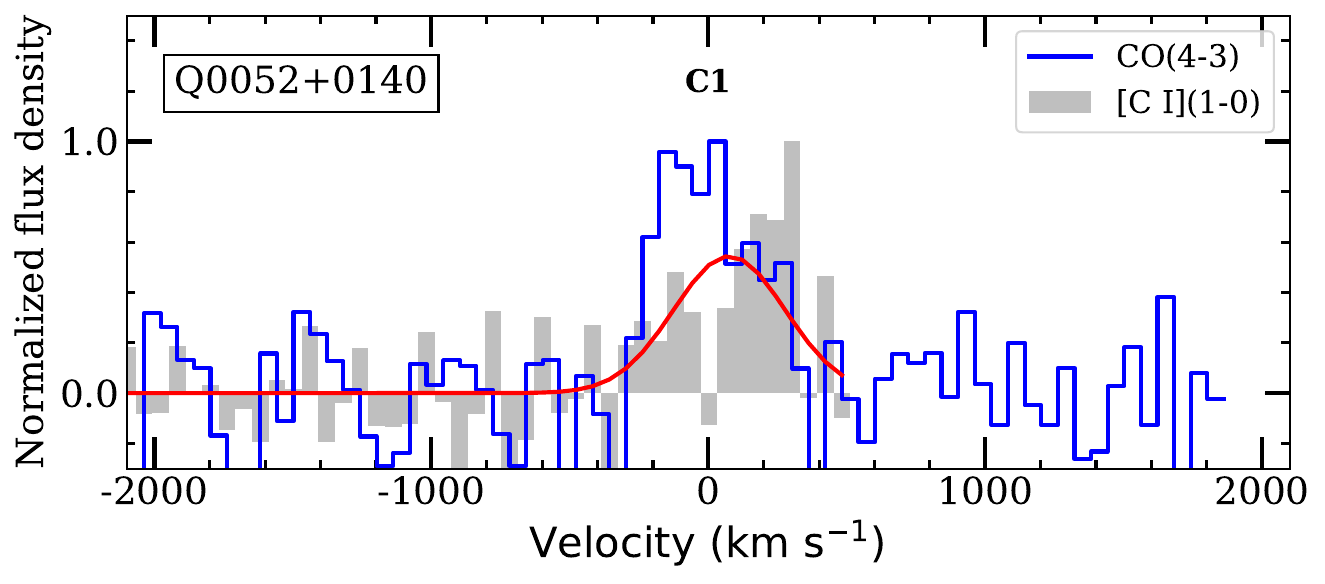}
\includegraphics[width=0.49\linewidth]{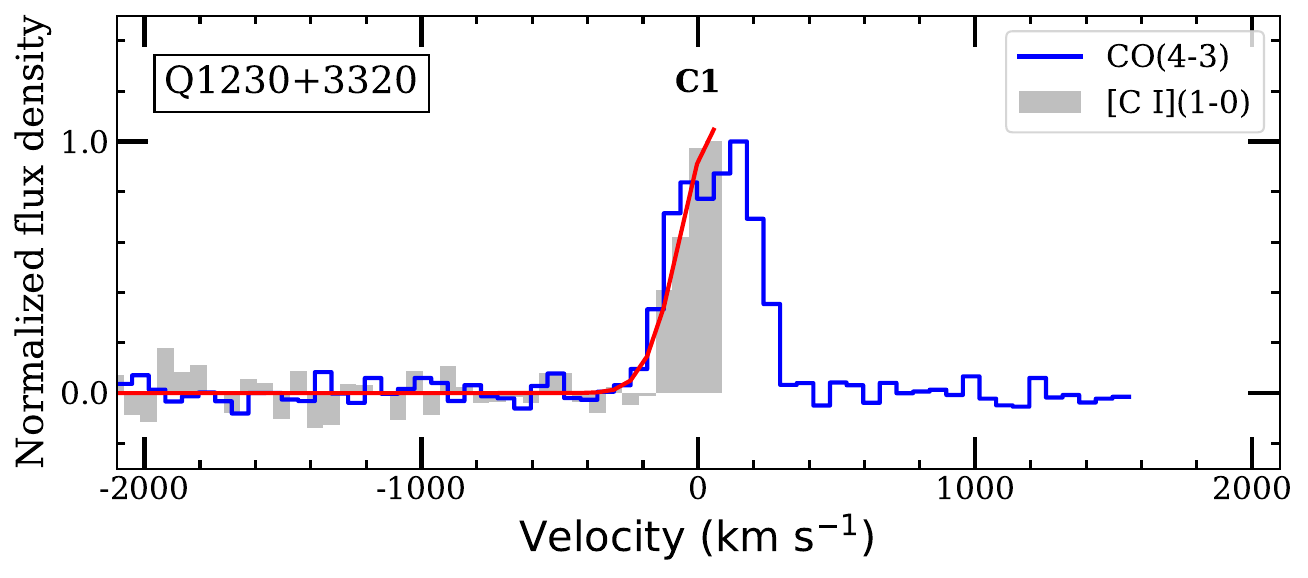}
\includegraphics[width=0.49\linewidth]{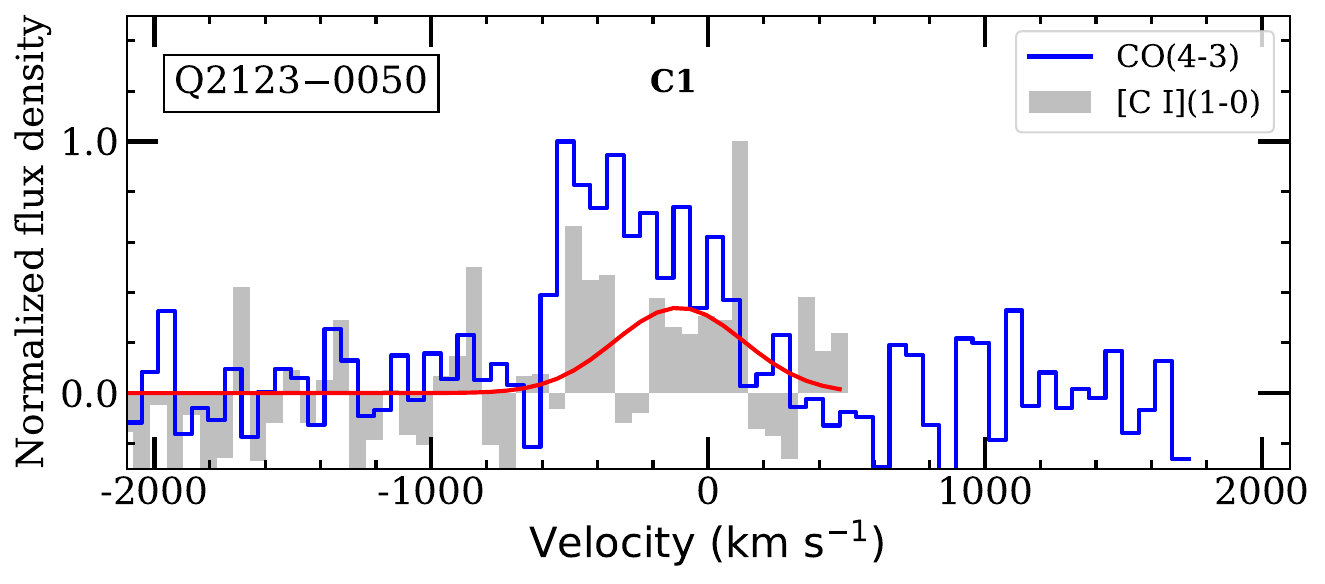}
\includegraphics[width=0.49\linewidth]{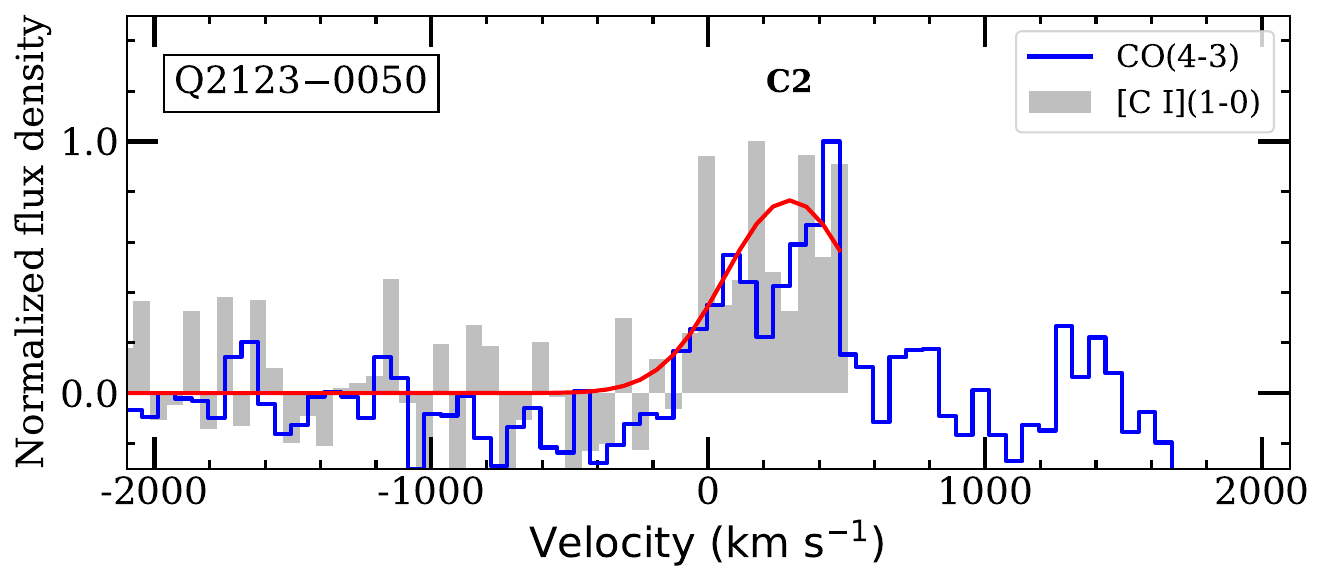}
\caption{Similar to Figure \ref{spectra1} (left column) but showing the \ci{} spectra for the companion galaxies in the QSO fields. \label{spectraco} For the companion galaxy C1 in the \jone{} field, we also show the \ci{} spectrum after uv-tapering. } 
\end{figure*}

% upper limits are calculated on flux_to_luminosity.ipynb

\begin{longrotatetable}
\begin{deluxetable*}{cclcccccccccc}
%\movetableright=10mm
\tabletypesize{\scriptsize}
\setlength{\tabcolsep}{2pt} % Adjust the value as needed
\tablecaption{Measured \ci{} properties for our sample \label{tab-measurements}}
%\tablewidth{0pt}
\tablehead{\colhead{Source} & \colhead{$z_{\rm CO}$/$\rm v_{c}$} &\colhead{component}& \colhead{RA} & \colhead{DEC} & \colhead{ $S_{\nu} d\nu_{[\rm C\ I]}$ }&\colhead{ $\rm FWHM_{[\rm C\ I]}$ }&\colhead{ $S_{\nu} d\nu_{[\rm C\ I]}^{taper}$ }& \colhead{ $\rm FWHM_{[\rm C\ I]}^{taper}$} &\colhead{ $S_{\nu} d\nu_{[\rm C\ I]}^{CGM}$ } &\colhead{$\rm RMS $}&\colhead{$\rm RMS^{taper} $}   &\colhead{R$_{Ly\alpha}$ }\\
\colhead{}&\colhead{}&\colhead{}&\colhead{}&\colhead{}& \colhead{\jykmps{}}&\colhead{\kmps{}}&\colhead{\jykmps{}}&\colhead{\kmps{}}& \colhead{\jykmps{}} &\colhead{mJy beam$^{-1}$} &\colhead{mJy beam$^{-1}$} &\colhead{kpc} 
}  
\decimalcolnumbers
\startdata
\jone{} & 2.2432 &QSO& 00:50:21.22 & +00:51:35.0 &0.89$\pm$0.04 &382& 1.11$\pm$0.07 &382&0.22(2.7$\sigma$)25$\%$ &0.14&0.27&116\\
&-14&C1 (CO emitter$\_\rm A$) &00:50:20.84 &+00:51:33.4&0.31$\pm$0.06& 480 &\nodata&\nodata\\
&323&C2 (CO emitter$\_\rm B$)&00:50:20.14& +00:51:32.4& $\textless$0.07&174&\nodata&\nodata\\ %rms =  
\jtwo{}&2.3101&QSO& 00:52:33.67& +01:40:40.8&0.13$\pm$0.04&222$\pm$49& 0.20$\pm$0.06 & 198$\pm$44&0.07(1.0$\sigma$)54$\%$ &0.14&0.26&127\\
&76&C1(CO emitter$\_\rm A$)&00:52:31.94 &+01:40:50.4&0.67$\pm$0.13&443&\nodata&\nodata\\
\jthree{}&2.4503&QSO&01:01:16.54& +02:01:57.4&0.34$\pm$0.02&110$\pm$5&0.39$\pm$0.04&123$\pm$9&0.05(1.2$\sigma$)15$\%$&0.14&0.24\\
\jfour{}&2.2825&QSO&01:07:36.90& +03:14:59.2&0.37$\pm$0.04&453$\pm$60&0.53$\pm$0.10&785$\pm$254& 0.16(1.4$\sigma$)43$\%$&0.14&0.27&114\\
&-121&C1(CO emitter$\_\rm A$)&01:07:35.76& +03:14:34.9& \nodata &\nodata&\nodata&\nodata  \\
&593&C2(CO emitter$\_\rm B$)&01:07:38.03& +03:14:46.7&\nodata &\nodata &\nodata&\nodata  \\
\jfive{}&2.2653&QSO& 12:27:27.48 & +28:48:47.9& 0.21$\pm$0.04&493$\pm$112&\nodata &\nodata &\nodata&0.15&0.24&$>$164\\
\jsix{}&2.2218& QSO&12:28:24.97 &+31:28:37.7 &0.27$\pm$0.06&414$\pm$65&0.48$\pm$0.10&553$\pm$87&0.21(1.8$\sigma$)78$\%$&0.17&0.30&$>$124\\
&-217&C1(Extended emission)& 12:28:25.42 & +31:28:49.0&$\textless$0.09&313&\nodata&\nodata \\ 
&315&C2(CO emitter$\_\rm A$)&12:28:26.16 &+31:28:51.5 &$\textless$0.22&616&\nodata&\nodata \\
\jseven{}&2.3287&QSO & 12:30:35.47 &+33:20:00.5 &0.32$\pm$0.04&292&0.39$\pm$0.04&292 &0.07(1.2$\sigma$)22$\%$&0.17&0.30& $>$204\\
&81&C1(CO emitter) &12:30:36.20 & +33:19:53.1&0.74$\pm$0.10&349\\
\jeight{}&2.2990 & QSO & 14:16:17.38 & +26:49:06.2 &0.09$\pm$0.03&436$\pm$179&0.20$\pm$0.07& 497$\pm$137&0.11(1.4$\sigma$)122$\%$&0.12&0.21& $>$141\\
&-864&C1(CO emitter$\_\rm A$) & 14:16:17.53 & +26:49:03.4&$\textless$0.03&156 &\nodata&\nodata \\
&-621&C2(CO emitter$\_\rm B$) & 14:16:17.06 &+26:48:58.2 &$\textless$0.08&683&\nodata&\nodata \\
&73&C3(CO emitter$\_\rm C$) & 14:16:17.25 &+26:49:03.9&$\textless$0.03&133&\nodata&\nodata\\
\jnine{}&2.3732&QSO&21:21:59.04& +00:52:24.1 &0.13$\pm$0.03 &218$\pm$39&0.16$\pm$0.05&164$\pm$35&0.03(0.5$\sigma$)23$\%$ &0.12 &0.25&$>$141\\
\jten{}&2.2807&QSO& 21:23:29.46 &$-$00:50:52.9 &0.19$\pm$0.03&498$\pm$109 &0.26$\pm$0.05&487 &0.07(1.0$\sigma$)37$\%$&0.15&0.29&$>$154\\
&-272&C1(CO emitter$\_\rm $blue) &21:23:29.88 &$-$00:50:51.8&0.10$\pm$0.03&549&\nodata&\nodata  \\
&285&C2(CO emitter$\_\rm $red)&21:23:28.98 &$-$00:50:53.4&0.25$\pm$0.03& 520&0.26$\pm$0.05&520\\
\enddata
\tablecomments{ Column 1: source name. Column 2: QSO redshift ($z_{\rm CO}$) measured from the \co{} line and center velocity $\rm v_{c}$ (in km\,s$^{-1}$) of the companion galaxies relative to the QSO redshift. Column 3: \ci{} emitting component in the QSO field, QSO or its companions. For companion galaxies, we also list the names presented in \paper{} in brackets. Columns 4 and 5: RA and DEC for the emitting component. Columns 6 and 7: \ci{} fluxes and line widths in FWHM for the 12 m array data. Uncertainties are based on the uncertainty of the Gaussian fit to the line. 
Width values with no errors correspond to QSOs and companion galaxies for which the \ci{} width was fixed to that of \co{} (see Table 2 in \paper{}).
Columns 8 and 9: \ci{} fluxes and line widths in FWHM for the 7 m + 12 m array data after uv-tapering. Column 10: \ci{} fluxes of the CGM emission estimated from the flux difference between the 7 m + 12 m array data after uv-tapering and the 12 m array data. The SNRs of the CGM emission is in brackets. The percentage of the flux increase after uv-tapering relative to the 12 m array data is also listed. Columns 11 and 12: RMS of the 12 m array data and 7 m + 12 m array data after uv-tapering per 60 \kmps{} channel. Column 13: projected Ly$_{\alpha}$ extent \citep{cai19}.}
\end{deluxetable*}
\end{longrotatetable}

\begin{deluxetable*}{cccccccc}
\tablecaption{Derived properties for spectral lines\label{tab-luminosity}}
\tablewidth{0pt}
\tablehead{\colhead{Source} & \colhead{component} & \colhead{$L_{[\rm C\ I]}$}&\colhead{$L_{\rm CO}$}& $r_{[\rm C\ I]/CO}$ & $M_{\rm H_{2},CO}$ & $M_{\rm H_{2},[C\ I]}$ & $M_{\rm H_{2},dust}$  \\
\colhead{}&\colhead{}&\colhead{$\times 10^{7}$ \lsun{}}&\colhead{$\times 10^{7}$ \lsun{}}&\colhead{} &\colhead{$\times 10^{10}$ \msun{}}& \colhead{$\times 10^{10}$ \msun{}}&\colhead{$\times 10^{10}$ \msun{}}
}
\decimalcolnumbers
\startdata
\jone{} & QSO &4.49 $\pm$ 0.20 &10.21 $\pm$ 0.57& 0.36 $\pm$ 0.03 &3.19 $\pm$ 0.18 &4.23 $\pm$ 0.19&2.86 $\pm$ 0.43 \\
&C1 &1.55 $\pm$ 0.32&1.70 $\pm$ 0.19& 0.75 $\pm$ 0.18 &0.53 $\pm$ 0.06 &1.46 $\pm$ 0.3 & 1.46 $\pm$ 0.3\\
&C2& $\textless$0.34&0.61 $\pm$ 0.09&$\textless$0.46 &0.19 $\pm$ 0.03 &$\textless$0.32 &$<$0.32 \\
\jtwo{}&QSO& 0.69 $\pm$ 0.21&1.24 $\pm$ 0.10& 0.46 $\pm$ 0.14 &0.39 $\pm$ 0.03&0.65 $\pm$ 0.20&0.78 $\pm$ 0.12\\
&C1 & 3.54 $\pm$ 0.67&4.53 $\pm$ 0.25 & 0.64 $\pm$ 0.13 &1.42 $\pm$ 0.08&  3.34 $\pm$ 0.63 & 3.34 $\pm$ 0.63 \\
\jthree{}&QSO&2.00 $\pm$ 0.12&6.94 $\pm$ 0.39& 0.24 $\pm$ 0.02&2.17 $\pm$ 0.12 &1.88 $\pm$ 0.11&2.38 $\pm$ 0.36 \\
\jfour{}&QSO&1.92 $\pm$ 0.31&   5.50 $\pm$ 0.19  & 0.29 $\pm$ 0.05& 1.72 $\pm$ 0.06& 1.81 $\pm$ 0.29&1.42 $\pm$ 0.21\\
&C1 & \nodata& 4.09 $\pm$ 0.78 & \nodata & 1.28 $\pm$ 0.24&\nodata &$<$0.40  \\
&C2 & \nodata &1.22 $\pm$ 0.19  & \nodata & 0.38 $\pm$ 0.06&\nodata&$<$0.45  \\
\jfive{}&QSO&1.08 $\pm$ 0.31&  3.08 $\pm$ 0.19& 0.29 $\pm$ 0.08&0.96 $\pm$ 0.06&1.02 $\pm$ 0.29&1.41 $\pm$ 0.21\\
\jsix{}&QSO&1.34 $\pm$ 0.30&3.81 $\pm$ 0.28&0.29 $\pm$ 0.07&1.19 $\pm$ 0.09&1.26 $\pm$ 0.28&$<$94.85\\
&C1 &$\textless$0.46&  0.65 $\pm$ 0.19&$\textless$0.58& 0.20 $\pm$ 0.06& $\textless$0.43&$<$0.43 \\
&C2 & $\textless$1.09& 1.91 $\pm$ 0.51 & $\textless$0.47&0.60 $\pm$ 0.16& $\textless$1.02&$<$1.02\\
\jseven{}&QSO &1.72 $\pm$ 0.22&5.90 $\pm$ 0.35&0.24 $\pm$ 0.03&1.85 $\pm$ 0.11&1.62 $\pm$ 0.21&1.58 $\pm$ 0.24\\
&C1& 3.98 $\pm$ 0.54 &8.72 $\pm$ 0.45& 0.37 $\pm$ 0.05& 2.73 $\pm$ 0.14&   3.74 $\pm$ 0.51&3.74 $\pm$ 0.51 \\
\jeight{} &QSO&  0.47 $\pm$ 0.20&0.69 $\pm$ 0.10& 0.56 $\pm$ 0.32&0.22 $\pm$ 0.03&0.44 $\pm$ 0.20&4.37 $\pm$ 0.65\\
&C1&$\textless$0.16&  0.49 $\pm$ 0.10& $\textless$0.27 & 0.15 $\pm$ 0.03& $\textless$0.15&$<$0.15\\
&C2 & $\textless$0.43 & 1.92 $\pm$ 0.20 & $\textless$0.18& 0.60 $\pm$ 0.06& $\textless$0.40 &0.40\\
&C3  & $\textless$0.16 & 0.30 $\pm$ 0.10 & $\textless$0.44& 0.09 $\pm$ 0.03&$\textless$0.15&$<$0.15 \\
\jnine{}&QSO&0.72 $\pm$ 0.17&1.46 $\pm$ 0.10& 0.41 $\pm$ 0.10& 0.46 $\pm$ 0.03&0.68 $\pm$ 0.16&0.35 $\pm$ 0.05\\
\jten{}&QSO&0.99 $\pm$ 0.26&2.43 $\pm$ 0.24& 0.33 $\pm$ 0.09 &0.76 $\pm$ 0.08& 0.93 $\pm$ 0.24&0.99 $\pm$ 0.15\\
&C1&0.55 $\pm$ 0.17& 1.90 $\pm$ 0.19& 0.24 $\pm$ 0.08& 0.59 $\pm$ 0.06&0.52 $\pm$ 0.16&$<$0.52 \\
&C2& 1.31 $\pm$ 0.18 &  1.99 $\pm$ 0.19& 0.54 $\pm$ 0.09&0.62 $\pm$ 0.06 &1.23 $\pm$ 0.17&1.23 $\pm$ 0.17\\
\enddata 
\tablecomments{Column 1 and 2: source name and the \ci{} emitting component. Columns 3-5: \ci{} and \co{} luminosities and the $L^{'}_{\rm [C I](1-0)}/L^{'}_{\rm CO(4-3)}$ ratios. Columns: 6-8: molecular gas masses derived using \co{}, \ci{}, and  ALMA continuum. For the QSOs, we adopt $\alpha_{\rm CO}$ = 0.8 \msun{}${[\rm K\ km/s\ pc^2]}^{-1}$, $ R_{\rm 41}$ = 0.87 and $X_{\mathrm{[C\,\scriptscriptstyle{I}\scriptstyle{]}}}$ = $8.4\times 10^{-5}$ to estimate the molecular gas mass using the \co{} and \ci{} lines (see Section \ref{molecualrmass} for explanations of $\alpha_{\rm CO}$, $ R_{\rm 41}$, $X_{\mathrm{[C\,\scriptscriptstyle{I}\scriptstyle{]}}}$, and $\delta_{\rm DGR}$).  For the companion galaxies, we use $\alpha_{\rm CO}$ = 0.8 \msun{}${[\rm K\ km/s\ pc^2]}^{-1}$, $ R_{\rm 41}$ = 0.85, and $X_{\mathrm{[C\,\scriptscriptstyle{I}\scriptstyle{]}}}$ = $8.4\times 10^{-5}$ to derive the molecular gas mass from the \co{} and \ci{} lines. As for the molecular gas mass derived from the ALMA continuum, we adopt $\delta_{\rm DGR} = \frac{1}{50}$ and $T_{\rm dust}$ = 47 K for the QSOs and companions.}
\end{deluxetable*}

\begin{deluxetable*}{cccrrrr}
\tablecaption{Continuum measurements\label{tab-fir}}
\tablewidth{0pt}
\tablehead{\colhead{Source} & \colhead{component} & \colhead{frequency}&\colhead{flux density}&\colhead{$L_{\rm FIR(SF)}$}&  \colhead{SFR} & \colhead{\mdust{}} \\
\colhead{}&\colhead{}& \colhead{GHz}&\colhead{mJy} &\colhead{$\times 10^{12}$ \lsun{}} &\colhead{\msunpyr{}}&\colhead{$\times 10^{8}$\msun{}}\\
}
\decimalcolnumbers
\startdata
\jone{} & QSO & 147.6001&0.27&6.2& 1295&2.86 \\
&C1 & & 0.05&1.2&255 & 2.93 $\pm$ 0.6\\
&C2&& $<$0.05&$<$1.1&$<$229 & $<$0.64\\
\jtwo{}&QSO&144.3370&0.07 &1.7&356&0.78\\
&C1 &  &0.15&3.7&752& 6.67 $\pm$ 1.26  \\
\jthree{}&QSO&138.1957&0.19&5.1&1076&2.38\\
\jfour{}&QSO&145.7495&0.13&3.0&644&1.42 \\
&C1 && $<$0.03 & $<$0.7 & $<$149&$<$0.79 \\
&C2 && $<$0.06  &$<$1.5 & $<$317 &$<$0.90\\
\jfive{}&QSO&146.2051&0.13&3.0&640&1.41\\
\jsix{}&QSO&148.4225&7.71&$<$173.7&$<$36488&$<$94.85\\
& C1&&$<$0.14& $<$3.2 & $<$668&$<$0.87\\
&C2&&$<$0.22 &$<$5.0 & $<$1042&$<$2.05 \\
\jseven{}&QSO &143.7226&0.14&3.4&718&1.58 \\
&C1&&0.19&4.6 &949&7.49 $\pm$ 1.01\\
\jeight{} &QSO&  144.7541 & 0.18(0.39)&4.3(9.3)&908&2.01 \\
&C1 &&$<$0.03 &$<$0.7 &$<$151&$<$0.30 \\ 
&C2 && 0.05 & 1.1& 232&0.80\\
&C3 &&$<$0.03&$<$0.7  &$<$151&$<$0.30 \\
\jnine{}&QSO&141.4330&0.03&0.8& 160&0.35\\
\jten{}&QSO&145.6604&0.09&2.1&447&0.99\\
&C1 &&$<$0.04& $<$1.0&$<$207&$<$1.03 \\
&C2& &0.11&2.5&533 &2.46 $\pm$ 0.33\\
\enddata 
\tablecomments{Columns 1 and 2: source name and the \ci{} emitting component. Columns 3-4: continuum frequencies in the observed frame, and continuum flux densities.  Column 5:  FIR luminosity integrated from 42.5$\mu$ to 122.5 $\mu$m assuming a dust temperature of 47 K and emissivity index of $\beta = 1.6$, which is a measure of the FIR emission from star formation. Columns 6 and 7: Star formation rate and dust mass derived from the dust continuum emission, assuming again a dust temperature of 47 K and emissivity index of $\beta = 1.6$. 
\jeight{} is a QSO with a ``radio-detected" AGN. We list both the observed continuum flux density/FIR luminosity  (in parentheses) and the continuum flux density/FIR luminosity after subtracting the contribution from the radio emission. The SFR and dust mass of \jeight{} are derived based on the continuum emission after subtracting the synchrotron contribution from the radio source. 
} 
\end{deluxetable*}

\section{Observations and data reduction}\label{observations}
We observed the \co{} and \ci{} lines and the dust continuum emission in a sample of 10 ultra-luminous Type-I QSOs at $z\sim 2$, each showing extended Ly$\alpha$ emission on scales of $>$100 kpc, using both the 7 m and 12 m arrays of ALMA.  
Detailed descriptions of the observations are presented in \paper{}, in which we published the \co{} data. 
Including the 7 m array enables us to image large-scale structures that are not well sampled by the 12 m array of ALMA.
In this way, we search for low-surface-brightness molecular gas emission surrounding QSO host galaxies. 
In the current work, we focus on the \ci{} detections in the QSOs, companion galaxies, and the CGM.

We performed data calibration and imaging using the Common Astronomy Software Applications package (CASA) version 6.4.1 \citep{casa22}.
We calibrated the data using the ALMA calibration pipeline \citep{hunter23} version 2022.2.0.68 by running the standard pipeline calibration script that was included with the archival ALMA data. 
We use all line-free channels to image the continuum emission. 
For the spectral lines, we first subtract the continuum emission using the UVCONTSUB task in CASA assuming a first-order polynomial for the continuum using line-free channels, and then use the continuum subtracted data to generate the spectral line datacubes. 
We employed two imaging methods for our sample. First, to spatially resolve the \ci{} emission in the QSOs and their companions, we imaged the 12 m array data using the TCLEAN task in CASA with natural weighting to maximize the signal-to-noise ratio (SNR). Second, to search for widespread CGM emission, we combined the  data from the 7 m and 12 m arrays and used the ``uv-tapering" technique to manually downweight baselines with lengths longer than 18.75 k$\lambda$,\footnote{ Specifying a taper of 18.75k$\lambda$ in CASA task {\it tclean} would correspond to applying a Gaussian with FWHM\,=\,4.8$^{\prime\prime}$ in the image plane.} and image the data using natural weighting to maximize the SNR. 
This will lower the spatial resolution but enhance the surface brightness sensitivity, which is required to trace the  faint, widespread CGM emission. The final spatial resolution was 1\farcs6$-$2\farcs7 $\times$ 1\farcs2$-$1\farcs8  with PA = $-80^{\circ}$---$78^{\circ}$ for the 12 m array data and 5\farcs2$-$5\farcs4 $\times$ 4\farcs9$-$5\farcs1  with PA = $-90^{\circ}$---$84^{\circ}$ for the 7 m + 12 m array data after uv-tapering.
 The \ci{} beam sizes are listed in Table \ref{tab:appendix}.
We binned the \ci{} line in all of our targets to 60 \kmps{} channel width, and the resulting rms was between 0.12 \mjypb{} and 0.17 \mjypb{} per binned channel for the 12 m array data and between 0.21 \mjypb{} and 0.30 \mjypb{} for the 7 m + 12 m array data after uv-tapering (Table\,\ref{tab-measurements}).

\section{Results}\label{results}

We report robust detections of the \ci{} emission line in nine of our QSOs at SNR of $3.3 - 22.3 \sigma$ and a tentative detection of the \ci{} line in \jseven{} at $3.0  \sigma$. Figure \ref{intensity_qso} shows the intensity maps generated from the 12 m array data of the QSOs. Among the 13 companion galaxies previously identified as \co{} emitters in seven of our QSO fields (\paper), 11 are covered in our frequency setup, and five of them (in four QSO fields) are detected in \ci{} emission as well. The intensity maps of these companion galaxies are shown in Figure \ref{intensity_comp}. 

Nine QSOs show flux increase after uv-tapering {\bf (Figure\,\ref{spectra1})}, although the flux increase is only significant in \jone{} at 2.7$\sigma$ level. While we will describe this result in detail in Section\,\ref{cgm-result}, it shows that we need to take the tapered data into account in our analysis of the QSOs. For the QSO \jfive{}, the \ci{} line is detected in the 12 m array data but not detected in the tapered data due to the low SNR. For the companion galaxies, we do not find any indication of flux increase after uv-tapering.

Figure \ref{spectra1} shows the \ci\ spectra of the QSOs, while Figure\,\ref{spectraco} shows the spectra of the companions. The spectra were extracted against the peak of the \ci\ signal in the QSOs and companion galaxies.  The \ci\ emission in the QSOs and their companions appeared spatially unresolved (see Appendix for details). Therefore, this method of extracting a spectrum against the peak of the emission gives the total \ci\ luminosity in the central beam, and also allows for a fair comparison between the full-resolution 12 m and the tapered 12 + 7 m data. Due to the uncertainties in redshift determined from previous Ly$\alpha$ emission, and to secure the full frequency coverage for the \co{} line when observing \co{} and \ci{} lines at the same time, the proposed spectral setup only fully covers the \ci{} emission in seven of our QSOs. For the remaining three QSOs \jone{}, \jfour{}, and \jseven{}, we only obtained part of the \ci{} spectra. 

{In Figures\,\ref{spectra1} and \ref{spectraco},} we use a Gaussian profile to fit the spectrum for each of the QSOs and companions where a full \ci{} spectrum is obtained. We adopt a Gaussian profile with the line center and width fixed to the values derived from the \co{} line to fit each of the QSOs and companions where \ci{} spectra are partially covered. The measured \ci{} widths and fluxes are shown in Table \ref{tab-measurements}. For the majority of the \ci{} spectra detected in the QSOs and companion galaxies, we find no significant deviation from a Gaussian profile. The derived \ci{} widths are in agreement with those found for the \co{} line. We derive the 3$\sigma$ \ci{} flux upper limits for the companion galaxies where the \ci{} line is undetected using $\rm 3\times RMS_{channel} \times \sqrt{\rm \delta \nu \times FWHM}$. Here $\rm \delta \nu$ is the channel width, which is 60 \kmps{} for all our observations, and FWHM is the line width expected for the \ci{} line, for which we adopt the value measured from the \co{} line (\paper). 

Table \ref{tab-measurements} also gives the line luminosities and molecular gas mass. At high redshift, two luminosities are frequently used \citep{carrili13}. One is $L_{\mathrm{\rm line}}$ in the unit of \lsun{}:
\begin{equation}
 L_{\mathrm{\rm line}}  = 1.04 \times 10^{-3}  S_{\rm line}\Delta \nu D_{L}^{2} \nu_{\rm obs}\ [\rm L_\odot]
\end{equation}
where $z$ is redshift, $S_{\rm line}\Delta \nu$ is the line flux in \jykmps{}, $D_{L}$ is luminosity distance in Mpc and $\nu_{\rm obs}$ is the observed line frequency in GHz.
The other is $L'_{\mathrm{\rm line}}$ in the unit of $\rm K\ km/s\ pc^2$:
\begin{equation}
 L'_{\mathrm{\rm line}}  = 3.25 \times 10^{7} S_{\rm line }\Delta \nu \frac{D_{L}^{2}}{(1+z)^{3} \nu_{\rm obs}^{2}}\ [\rm \rm K\ km/s\ pc^2]. 
\end{equation}
% The relation between the two luminosities is
% \begin{equation}
%  L_{\mathrm{\rm line}}  = 3 \times 10^{-11}  \nu_{r} L'_{\mathrm{\rm line}}\ [\rm L_\odot],
% \end{equation}
% where $\nu_{r}$ is rest frame frequency of the line in GHz. 
$L'_{\mathrm{\rm line}}$ is directly linked to the molecular gas mass, as we will explain in Section\,\ref{molecualrmass}.

We detect the dust continuum emission in all the QSOs and five of the companion galaxies in our sample with ALMA. For all QSOs and the companion galaxies that are detected in dust continuum emission, we derive FIR luminosities from star-formation that are higher than $10^{12}$ \lsun{} (see below), which suggests that the QSOs and continuum detected companions are Ultra-luminous infrared galaxies (ULIRGs). The line fluxes and luminosities are shown in Tables  \ref{tab-measurements} and \ref{tab-luminosity}.

In the following Sections, we show the results for the individual QSO fields (Section\,\ref{individual}),  the flux excess seen after tapering the data, which hints at the presence of a faint CGM component (Section\,\ref{cgm-result}), and the FIR luminosities from star formation and corresponding star-formation rates (SFR) as derived from our continuum results (Section\,\ref{sfrs}).                     

\subsection{Individual QSOs}\label{individual}
\textbf{\jone{}:} This QSO shows the brightest \ci{} and \co{} emission in our sample.
In one of the two companion galaxies (C1) previously discovered in \co{} emission, we detect the \ci{} emission as well at  $5 \sigma$. 
The $L^{'}_{\rm [C I](1-0)}/L^{'}_{\rm CO(4-3)}$ ratio ($r_{[\rm C\ I]/CO}$) is a probe of the molecular gas density, and a lower ratio is a signature of higher-density gas (see Section \ref{gas condition} for details).
The $r_{[\rm C\ I]/CO}$ values in the QSO and C1 are 0.36 $\pm$ 0.03 and 0.75 $\pm$ 0.18, which may suggest a higher molecular gas density in the QSO compared to its companion galaxy.

\textbf{\jtwo{}:} 
The \ci{} emission is detected in both this QSO and its companion galaxy (C1), which is located $\sim$240 kpc west of the QSO. 
The \ci{} flux in C1 is much higher than that detected in the QSO. Higher \co{} flux and continuum flux density are also found in our \co{} observations (\paper) in C1 compared to the QSO. The velocity gradient of the \co{} line of C1 resembles either a merger system or a large-scale rotating molecular disk (see \paper{} for details).
The $r_{[\rm C\ I]/CO}$ values are 0.46 $\pm$ 0.14 in the QSO and 0.64 $\pm$ 0.13 in  C1, suggesting that the gas density in \jtwo{} is comparable to that in its companion galaxy.

\textbf{\jthree{}:} 
This QSO shows bright \ci{} and \co{} emission lines. The \ci{} spectrum of the QSO reveals a narrow line width of $110 \pm 5$ \kmps{}. This is among the lowest values found for high-$z$ QSOs, which span a range of $\sim100-1000$ \kmps{} (See Figure 5 in \citealt{carrili13} for details), but is comparable to that found for our \co{} observations (\paper). 
%We recover an increase of 15$\%$ in the \ci{} flux in the 7 m + 12 m arrays data after uv-tapering compared to the 12 m array data at a significance value of $1.3 \sigma$.

\textbf{\jfour{}:} 
The \co{} spectrum of this quasar shows an asymmetric double-peak profile and a similar asymmetric feature is also found in our \ci{} data. Our ALMA data did not cover companion galaxy C1 within the field of view, while for companion galaxy C2, the \ci{} emission fell outside our frequency setup (\paper{}).
%We detect widespread \ci{} emission within the central 5\farcs3 $\times$ 4\farcs9, PA = $-86^{\circ}$ with a total flux of 0.16 \jykmps{} at $1.2 \sigma$ level.

\textbf{\jfive{}:} 
We detect the \ci{} emission in the QSO host galaxy at 5$\sigma$ level in our 12 m array data. However, the \ci{} emission is barely detected in the 7 m + 12 m data after uv-tapering due to the low signal-to-noise of the tapered data. Thus, we here consider the \ci{} emission in the tapered 7 m + 12 m array data as a non-detection. 

\textbf{\jsix{}:} 
This QSO reveals an outflowing \co{} gas component in the QSO spectrum with a line width of 1160 \kmps{} in FWHM offset in velocity by -346 $\pm$ 130 \kmps{} compared to the quasar \citep{li21}. 
In the \ci{} spectrum, we also find a tentative flux excess at the outflow frequency, however, our current SNR is insufficient to confirm this feature. In addition, we find that the \ci{} line width increases from 414 $\pm$ 65 \kmps{} for the 12 m data to 553 $\pm$ 87 \kmps{} for the 7 m + 12 m array data after uv-tapering, which might also be a possible signature of \ci{} outflow.  
An extended \co{} emission in the CGM spreading over 100 kpc has been found in this QSO field \citep{li21}. However, we fail to detect any \ci{} emission in either the 12 m or the 7 m + 12 m arrays data after uv-tapering at the position of the extended \co{} component. 
This leads to an upper limit for the $r_{[\rm C\ I]/CO}$ value in the CGM of $<$ 0.35. The non-detection of the \ci{} in the extended CGM could be due to the relatively lower SNR expected for the \ci{} observations compared to that of the \co{} line, because the \ci{} line is generally a few times weaker than the \co{} line, as can be seen from the \ci{}/\co{} luminosity ratios in the QSOs and companion galaxies (Table \ref{tab-luminosity}).  
This upper limit is lower than the $r_{[\rm C\ I]/CO}$ value in the CGM of the Spiderweb nebulae of 0.6 $\pm$ 0.1 \citep{emonts18}.
Two companion galaxies are detected in our \co{} observations, however we do not detect the \ci{} emission in the companions.

\jsix{} is a very bright radio source with flux densities of 297.0 $\pm$ 0.15 mJy at 1.44 GHz and 314.3 $\pm$ 0.30 mJy at 3.00 GHz, observed with the FIRST and VLASS survey, respectively \citep{becker95,lacy20}.
For QSOs with significant radio emission, the contribution of the radio synchrotron emission to the ALMA continuum emission can be substantial \citep[e.g.,][]{falkendal19}. This needs to be subtracted from the observational value for reliable measurement of the FIR luminosity from star formation, SFR, and dust mass, which trace the cold dust properties of the host galaxy. To subtract the radio synchrotron contribution from the ALMA continuum detection, we first assume the radio emission has a power-law spectrum with the shape of $f_{\nu} \sim \nu^{\alpha}$, where $\alpha$ is the radio power-law index. Secondly, we assume that the radio emission at 1.44 GHz and 3.00 GHz are purely from the radio synchrotron emission, and then derive $\alpha$. 
However, the derived $\alpha$ of 0.08 predicts a radio emission at the ALMA continuum frequency that is even much higher than our detected value, which prevents us from deriving the $L_{\rm FIR(SF)}$ of this QSO. This could be a result of a steepening of the radio power-law index from 3 GHz to the observed ALMA frequency.  In this case, we could not use the radio power-law index with measurements of radio emission at 1.44 GHz and 3.00 GHz to constrain the synchrotron emission at the ALMA observing frequency (148.42 GHz). This means that our assumption of a fixed power-law index between 1.44 and 3.00 GHz is invalid. For this source, we do not subtract the synchrotron emission contribution to the ALMA continuum.

\textbf{\jseven{}:} 
This QSO is bright in \co{}, and it is also among the \ci{} brightest QSOs in our sample. In the companion galaxy C1, which shows the brightest \co{} emission among all the companion galaxies, we detect the brightest \ci{} emission as well. The $r_{[\rm C\ I]/CO}$ values in the QSOs and its companion galaxy are 0.24 $\pm$ 0.03 and 0.37 $\pm$ 0.05, respectively. 

\textbf{\jeight{}:} 
Our \co{} data reveal widespread \co{} emission connecting the QSO and its three close companion galaxies (\paper). 
We detect the \ci{} emission in the QSO, and fail to detect the \ci{} emission in any of the three companion galaxies.  
\jeight{} is covered in both the FIRST and the VLASS survey, with flux densities of  2.69 $\pm$ 0.15 mJy at 1.44 GHz and 1.79 $\pm$ 0.13 mJy at  3.00 GHz in the observed frame. We follow a similar procedure as for \jsix{} to subtract the radio synchrotron contribution from the observed ALMA continuum emission. We obtain $\alpha = -0.55$ and a contribution of 0.21 mJy from the radio source to the observed continuum emission at 144.75 GHz. This leads to a continuum flux density of 0.18 mJy purely from the cold dust emission and a resulting FIR luminosity of 4.3$\times 10^{12}$ \lsun{} from star formation after subtracting the contribution of the radio emission.\footnote{We consider this as a rough estimate, because there might be a contribution of the cold dust emission to the 3 GHz band, and the radio power-law index might steepen between 3 GHz and the ALMA continuum frequency.} Utilizing detections of 2MASS \citep{2mass} and VLASS and adopting the estimated $\alpha = -0.55$, we derived the radio loudness of $R_{f_{\rm 5GHz} / f_{4400 \text{\AA }}}$ = 9, which classifies this source as a radio-detected QSO \citep{liu21}.

\textbf{\jnine{}:} 
This QSO shows extended \co{} emission, however, we do not find any concrete evidence for either extended \ci{} emission or significant flux increase in the 7 m + 12 m array data after uv-tapering.
This QSO has a narrow \ci{} line width of 218 $\pm$ 39 and 164 $\pm$ 35 \kmps{} in the 12 m and 7 m + 12 m array data after uv-tapering, respectively. 

\textbf{\jten{}:} 
This QSO shows widespread \co{} emission connecting the QSO and its two close companion galaxies C1 and C2. 
We detect the \ci{} emission in the QSO and the companion galaxies C1 and C2 at 6.3, 3.3, and 8.3$\sigma$ levels, respectively. Interestingly, we find an offset of 50 $\pm$ 13 \kmps{} between the \co{} and \ci{} line centers for the QSO in the 12 m array spectrum, although the signal is broad (FWHM\,$\sim$500 km\,s$^{-1}$) and the \ci{} line falls at the very edge of the band, so it needs to be verified if this offset is real. The $r_{[\rm C\ I]/CO}$ values are comparable in the QSO and its companions, suggesting similar gas properties in the QSO and its companion galaxies.

\subsection{Extended \ci\ emission}\label{cgm-result}
We study the molecular CGM emission by comparing the \ci{} flux measured in the 12 m array data and the 7 m + 12 m array data after uv-tapering. 
The \ci{} spectra of the 12 m array data are extracted from peak pixels with beam sizes of  1\farcs6$-$2\farcs7 $\times$ 1\farcs2$-$1\farcs8. 
Recent ALMA surveys of the cold ISM (generally traced by \cii{} or CO lines) in the host galaxies of QSOs at $z$=2.4--7.6 reveal typical sizes of less than 5 kpc for the cold ISM (e.g., \citealt{fujimoto20}; \citealt{venemans20}; \citealt{bischetti21}; \citealt{ikeda24}; \citealt{wang24}), with very limited cases showing cold ISM sizes approaching or extending 10 kpc (e.g., \citealt{banerji17}; \citealt{bischetti21}; \citealt{meyer22}). 
The \ci{} source sizes we measure from the 12 m array data (see Appendix for details) suggest that the cold ISM in the host galaxies of the QSOs in our sample tend to be spatially unresolved, with sizes of $\lesssim$ 2\arcsec{} ($\lesssim$\,16 kpc). This is in agreement with the above mentioned studies of the cold ISM in other high-$z$ QSOs. We also do not see any evidence for rotating structures in our 12\,m data that could hint to the presence of large-scale disks on these scales, as has been observed in a few other high-$z$ galaxies \citep[][]{hodge12,dannerbauer17}. Accordingly, we consider that the $\lesssim$ 2\arcsec{} ($\lesssim$\,16 kpc) beam of our 12 m array data fully covers the cold ISM emission of the QSO hosts in our sample.
In this paper, we adopt the definition `CGM' for all the \ci{} emission on scales 
$\gtrsim$\,16 kpc. 
 While this definition is somewhat arbitrary, it nevertheless agrees with findings of other high-$z$ studies \citep[e.g.,][]{fujimoto19,ginolfi20}.

As for the  7 m + 12 m array data,  the \ci{} spectra are extracted from peak pixels within beam sizes of 5\farcs2$-$ 5\farcs4 $\times$ 4\farcs9$-$5\farcs1 after uv-tapering.
We consider the \ci{} emission measured from the 7 m + 12 m array data after uv-tapering to originate from both the ISM emission of the QSO host galaxy and the CGM emission surrounding the QSO.
Accordingly, the \ci{} line flux difference measured from 7 m + 12 m array data after uv-tapering and the 12 m array data represents the CGM emission  on scales of $\sim$2$-$5$^{\prime\prime}$ (16$-$40 kpc).  The uncertainty of the \ci{} flux for the CGM is calculated through $\sqrt{\sigma_{12m}^{2} + \sigma_{7m+12m}^{2}}$, where $\sigma_{12m}$ and $\sigma_{7m+12m}$ are the uncertainties in the Gaussian profile fit to the \ci{} line for the 12 m and 7 m + 12 m data after uv-tapering, as shown in columns 6 and 8 of Table \ref{tab-measurements}, respectively. We consider our estimated uncertainty for the CGM conservative because the noise in the data from the 12 m array and the 7 m + 12 m array is not entirely independent.

\begin{figure*}
\centering
\includegraphics[width=0.5\linewidth]{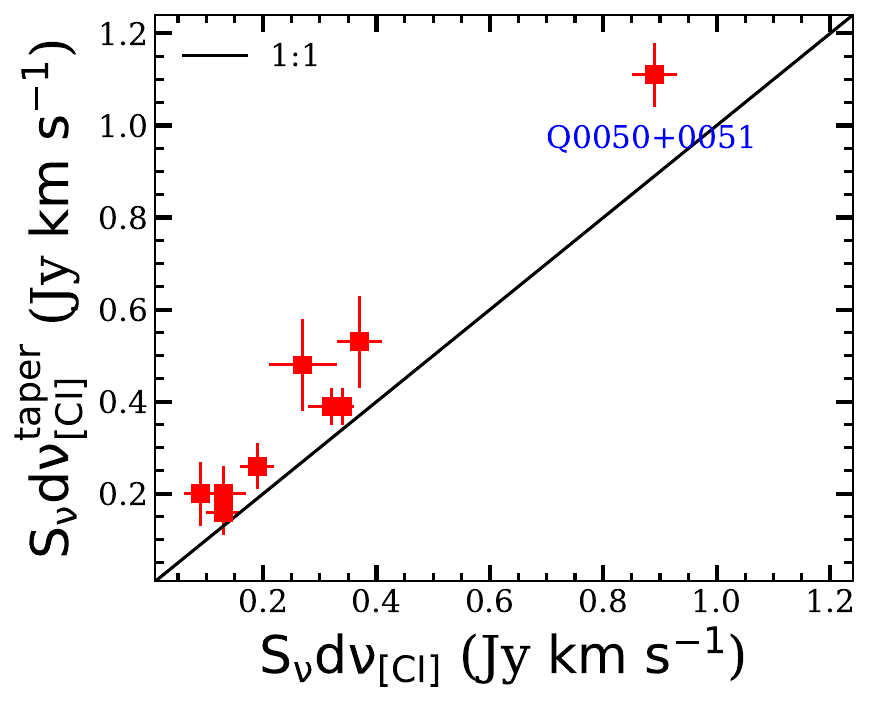}
\caption{\label{flux-comparison}  A comparison between the \ci{} flux measured from the 12 m array data ($S_{\nu} d\nu_{[\rm C\ I]}$) and the 7 m + 12 m array data after uv-tapering ($S_{\nu} d\nu_{[\rm C\ I]}^{taper}$). The red squares represent the measured \ci{} fluxes, and the $S_{\nu} d\nu_{[\rm C\ I]}$ = $S_{\nu} d\nu_{[\rm C\ I]}^{taper}$ relation is shown as a solid black line.
}
\end{figure*}

We detect the \ci{} emission in all of our 10 QSOs (including a 3$\sigma$ detection in \jseven{}) in the 12 m array data and nine of them in the 7 m + 12 m array data after uv-tapering.
 In Figure \ref{flux-comparison}, we show the \ci{} flux measured in the data of the 12 m array and the 7 m + 12 m array after uv-tapering. 
Interestingly, all the nine QSOs show higher flux in the 7 m + 12 m array data compared to the 12 m array data, with a flux difference of $0.03 - 0.22$ \jykmps{}, hinting at the possibility of cold CGM in the majority of our targets.
However, the putative CGM emission in \ci{}, which is measured as a difference between the tapered data and the 12 m array data, is at too low significance  to claim a reliable detection for the individual sources.  The exception is the CGM in \jone{}, where we detect a \ci{} flux (luminosity) of 0.22 \jykmps{} ((2.5 $\pm$ 0.9) $\times 10^{7}$ \lsun{}) at an SNR of 2.7. We also find a \co{} flux increase for \jone{} after uv-tapering (\paper). But the \co{} luminosity in the CGM of \jone{} is less prominent with a luminosity of (1.4 $\pm$ 0.9) $\times 10^{7}$ \lsun{} (1.6$\sigma$), which suggest a 3$\sigma$ upper limit of 2.7 $\times 10^{7}$ \lsun{}. 
We note that in the \jone{} field, there is a companion galaxy at a projected distance of $\sim 45$ kpc from the QSO. To check if there is any contamination from the companion galaxy C1 in the tapered data to the CGM, we  use the fact that the QSO is more luminous in \ci{} than companion C1, and check whether there is any contamination of the QSO emission in the spectra of C1. Comparable fluxes and line peaks are found between the 12 m and 7 m + 12 m array data after uv-tapering for C1  (Figure \,\ref{spectraco}), which is likely an indication of negligible contamination of the QSO flux to C1.  Conversely, we therefore also assume a negligible contamination of the \ci{} emission from C1 to the spectra of the QSO, and hence also to the CGM in \jone{}. 
Any potential CGM emission in the remaining eight QSOs is less prominent with an SNR in the range of $0.5 - 1.8$. 

We also explore the possibility that the extended emission is arising from individual companion galaxies that are undetected in the proximity of the QSOs. Such companion galaxies would be below the detection limit of our \ci{} data, but they would still amount to 15$-$78\,$\%$ of the \ci{} emission of the QSO in each field when added together (see Table \ref{tab-measurements}). Given the rms noise of our data, for most of our QSOs, even 1$-$2 of such companions just below the 3$\sigma$ limit would be sufficient to produce the excessive flux after tapering. 
However, if the \ci{} emission from companion galaxies is hidden just below the noise limit, \co{} emission from the companion galaxies should likely be visible, given that \co{} is typically brighter than \ci{} in the ISM of high-$z$ galaxies \citep[e.g.,][]{israel15,emonts18}.  We looked into the \co{} emission (\paper) from the companion galaxies but could not find a significant emission either.
However, low-metallicity companions would show relatively low \co{} emission with respect to the \ci{} emission \citep[e.g.,][]{papadopoulos04}. 
Therefore, argue that the contribution from companion galaxies to the extended emission is likely not significant, but cannot be completely excluded.

As can be seen from the  $r_{[\rm C\ I]/CO}$ values measured for our QSOs and companion galaxies, the \ci{} emission is generally a few times weaker than the \co{} line. Considering the same observing time and thus similar rms for the \ci{} and \co{} lines, the low detection rate of the CGM emission in the majority of our QSO fields is possibly a result of our limited SNR for detecting the \ci{} emission. 
A detailed study about the \ci{} emission in the CGM through spectral stacking will be presented in a separate paper (Li. et al. in preparation).

\subsection{Star formation rates}\label{sfrs}

We derive the FIR luminosity from star formation (integrated between rest wavelengths 42.5 and 122.5 $\mu m$) for radio-quiet QSOs and companion galaxies based on their continuum flux densities that we detect with ALMA, by means of a normalization and adopting a Modified Black Body (MBB) dust spectral energy distribution (SED) model under the optically thin approximation regime. We adopt a dust temperature of 47 K and emissivity index of 1.6 typical for the dust continuum emission of the  QSO host galaxies \citep{bianchi13}.
In the MBB dust SED model, the dust flux density $S_{\nu}$ is related to the dust emissivity index ($\beta$) and the dust temperature ($T_{\rm dust}$) through:
\begin{equation}
    S_{\nu} \propto \nu^{\beta} B_{\nu}[T_{\rm dust}],
\end{equation}
where $B_{\nu}[T_{\rm dust}]$ is the Planck function for $T = T_{\rm dust}$, and $\nu$ is the dust continuum frequency in the rest frame. 
The $L_{\rm FIR(SF)}$ we derive only considers the dust continuum emission from the host galaxy of the QSO that is heated by star formation, thus we can directly derive the star-formation rate (SFR) from the FIR luminosity ($L_{\rm FIR(SF)}$) through 
\begin{equation}
   {\rm SFR}  =  2.1 \times 10^{-10}\ L_{\rm FIR(SF)}\ [\rm M_{\sun}\ yr^{-1} ] \label{sfr}
\end{equation}
\citep{murphy11}. 

The derived SFR and FIR luminosities from star-formation for the QSOs and companions are shown in Table \ref{tab-fir}.

\section{discussion}\label{discussions}
\subsection{Line to FIR ratio}\label{sec-line-ir-ratio}

\begin{figure*}
\centering
\includegraphics[width=0.49\linewidth]{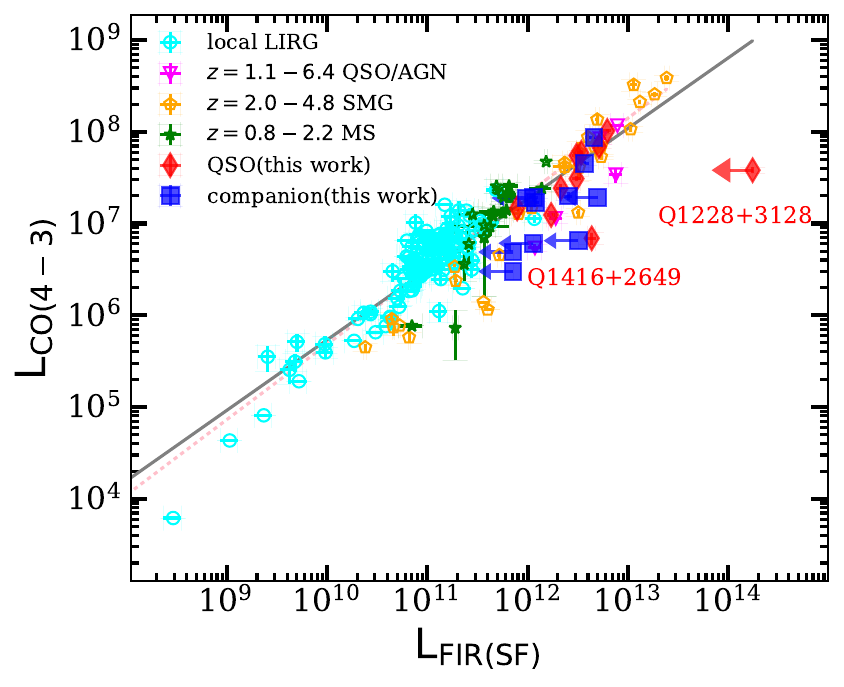}
\includegraphics[width=0.49\linewidth]{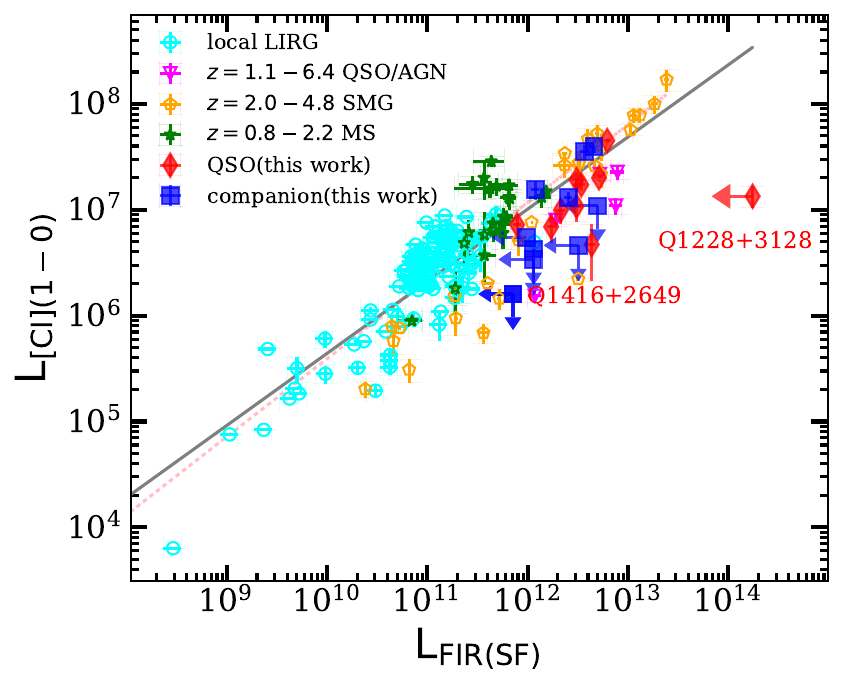}
\caption{\label{line-ir}  $L_{\rm CO(4-3)}$/$L_{\rm FIR(SF)}$ and $L_{\rm [C\ I](1-0)}$/$L_{\rm FIR(SF)}$ ratios for the QSOs and companion galaxies in our sample and samples of galaxies in the literature. 
Literature galaxy samples are local (U)LIRGs (cyan circles), $z= 1.1 - 6.4$ AGN or QSOs (magenta down-triangles), $z= 2.0 - 4.8$ submillimeter galaxies (SMGs, orange pentagons), and  $z= 0.8 - 2.2$ main-sequence (MS) galaxies (green stars) from \citet{valentino20} and \citet{gururajan23}. The $z= 1.1 - 6.4$ AGN or QSOs are from \citet{walter11}, but only the CO$(3-2)$ line is observed. We convert the CO$(3-2)$ luminosity to CO$(4-3)$ luminosity adopting a $L^{'}_{\rm CO(4-3)}/L^{'}_{\rm CO(3-2)} $ ratio of 0.9 typical of high-$z$ QSOs \citep{carrili13}.
 Pink dotted and grey solid lines represent linear fits to the data without and with the QSOs and companion galaxies of this work included. 
} 
\end{figure*}

In Figure \ref{line-ir}, we show the \co{} luminosity against FIR luminosity from star formation (left panel) and the \ci{} luminosity against FIR luminosity from star formation (right panel) for our sample and comparison samples.\footnote{ We note that the cosmology and the method to calculate $L_{\rm FIR(SF)}$ are consistent in all samples.} The local (U)LIRGs, high-$z$ QSOs, SMGs, and MS galaxies follow a linear relation between \co{}, or \ci{}, luminosity and the FIR luminosity from star formation. 
We fit the \ci{} or \co{} $vs$ FIR relations using the polyfit function in the python numpy package in the form of log($L_{\rm line}$) = $\alpha$ + $\beta \ \times$ log($L_{\rm FIR(SF)}$). 
The $\alpha$ and $\beta$ values for the \co{} line are $-$2.34 ($-$2.53) and 0.80 (0.82) with (without) our sample included. 
As for the linear fit to the \ci{} vs. FIR relation, the $\alpha$ and $\beta$ values are $-$1.49 ($-$1.78) and  0.71 (0.74) with (without) our sample included. In short, the inclusion or exclusion of our sample to the fit of the linear relation between \co{}, or \ci{}, and FIR luminosity from star formation does not significantly alter the relation. In addition, we find that the $L_{\rm [C\ I](1-0)}$/$L_{\rm FIR(SF)}$ and  $L_{\rm CO(4-3)}$/$L_{\rm FIR(SF)}$ ratios in \jeight{} are an order of magnitude lower than expected for its $L_{\rm FIR(SF)}$ on the linear relation. The \ci{} and \co{} lines are the molecular gas mass tracers. 
As we will discuss in Section\,\ref{gas condition}, one possible explanation could be due to a lower molecular gas mass relative to its FIR luminosity from star formation compared to the other QSOs and companion galaxies in our sample, as well as other local/high-$z$ galaxies. An alternative explanation could be that the observed ALMA continuum flux density is heavily contaminated by the synchrotron emission even after subtracting the contribution of synchrotron emission to the ALMA flux due to the large uncertainties in deriving the power-law index and extrapolating the radio flux to the observed ALMA frequency. %index $\alpha$.
Another outlier is the radio-loud QSO \jsix{}; the FIR luminosity from star formation of this QSO is not well constrained due to the change of the power-law index for the radio emission. As can be seen from Figure \ref{line-ir}, the FIR luminosity contributed by heating from the star formation should be at least an order of magnitude lower than what we observe to bring it on the relation.

\subsection{Molecular gas properties revealed by line ratios}\label{gas condition}
\begin{figure*}
\centering
\includegraphics[width=1.0\linewidth]{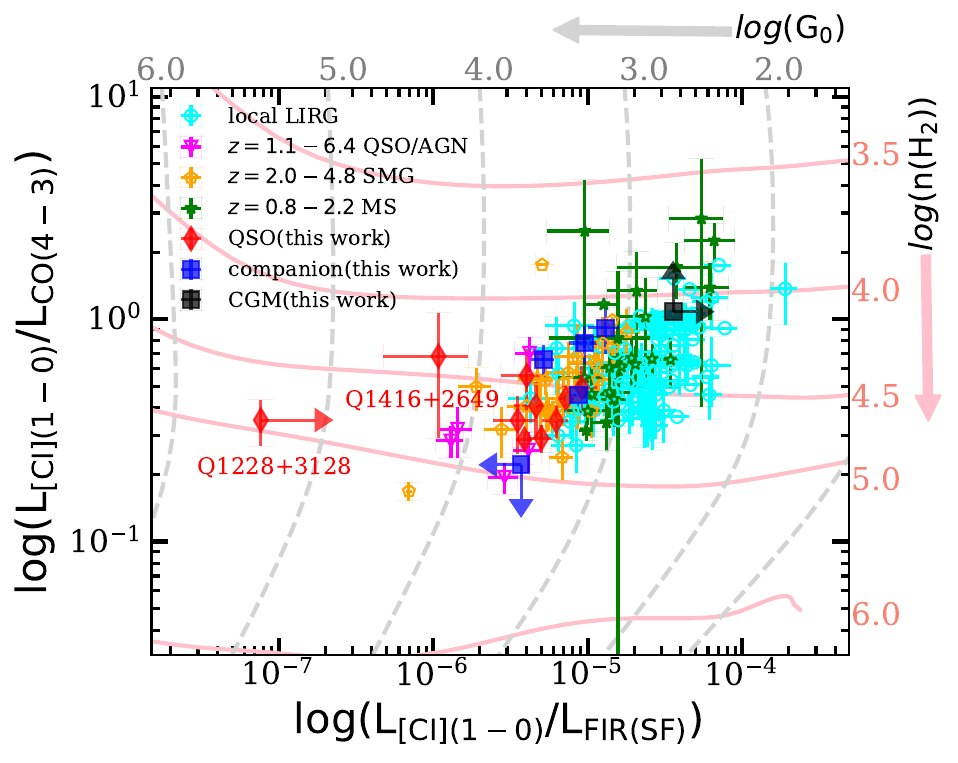}
\caption{\label{line-ratio} $L_{\rm CO(4-3)}$/$L_{\rm [C\ I](1-0)}$ $vs$ $L_{\rm [C\ I](1-0)}$/$L_{\rm FIR(SF)}$ ratios for the QSOs (red filled diamonds), and companion galaxies (blue filled squares), and the tentative detection of the CGM (black filled square) in our sample. 
The comparison samples in the literature are the same as those presented in Figure \ref{line-ir}.
The isocontours are from the photo-dissociation region (PDR) models generated by the PhotoDissociation Region Toolbox adopting the ``wk2000" model (\citealt{kaufman06}; \citealt{wolfire10}; \citealt{neufeld16}). The pink solid contours represent the modeled \ci{}/\co{} luminosity ratio with different molecular hydrogen densities, and the grey dashed contours represent the modeled \ci{}/$L_{\rm FIR(SF)}$ ratio for different ionization parameters.
}
\end{figure*}

Ratios between different molecular emission lines serve as diagnostics of the molecular gas properties in galaxies (e.g., \citealt{papadopoulos04}; \citealt{bothwell17}; \citealt{gururajan23}).  
The \ci{}/\co{} ratio traces the molecular hydrogen density. 
Whereas \co{} can be biased towards tracing denser gas (e.g., in starburst and AGN regions), the \ci{} emission is a reliable tracer for the total molecular gas mass, while $L_{\rm FIR(SF)}$ is directly linked to star formation through Eq. \ref{sfr}. Thus, the ratio between $L_{\rm [C\ I](1-0)}$ and $L_{\rm FIR(SF)}$ is sensitive to the radiation field strength, with a lower ratio probing a more intense ionization field. 
We use the  $L_{\rm [C\ I](1-0)}$/$L_{\rm CO(4-3)}$ and $L_{\rm [C\ I](1-0)}$/$L_{\rm FIR(SF)}$ ratios to study the physical conditions of the molecular gas in QSOs and their companion galaxies by comparing the observed ratios with those predicted in the photo-dissociation region (PDR) model PhotoDissociation Region Toolbox (Figure \ref{line-ratio}). The radiation field strength is described by the ionization parameter $\rm G_{0}$ in Habing field units (1 Habing field corresponds to $\rm 1.6 \times 10^{-3}\ erg\ s^{-1}\ cm^{-2}$; \citealt{habing68}), and the molecular gas density $\rm n(H_{2})$ is in the unit of $\rm cm^{ -3} $.

We show the results in Figure \ref{line-ratio}. Local (U)LIRGs and high-$z$ main-sequence (MS) galaxies have systematically higher  $L_{\rm [C\ I](1-0)}$/$L_{\rm CO(4-3)}$ ratios and higher $L_{\rm [C\ I](1-0)}$/$L_{\rm FIR(SF)}$ ratios compared to high-$z$ SMGs, QSOs and AGN. These suggest that the local (U)LIRGs and MS galaxies tend to have lower gas density and less intense radiation field compared to high-$z$ SMGs, QSOs, and AGN. 
Our QSOs reside in similar regions and thus have similar gas densities and radiation fields as that of high-$z$ SMGs, QSOs and AGN (Figure \ref{line-ratio}). This reveals that the gas within the QSOs in our sample experiences a more intense radiation field and and has higher molecular densities compared to high-$z$ MS galaxies, with values that are at the high end for local (U)LIRGs.
Except for one upper limit, the companion galaxies show similar line ratios as high-$z$ SMGs and are comparable to the lowest ratios found for local (U)LIRGs and high-$z$ MS galaxies. This indicates that, similar to high-$z$ SMGs, the companion galaxies reveal a higher radiation field strength and higher gas density compared to the mean of local (U)LIRGs and high-$z$ MS galaxies. By comparing our detections with models, we estimate a gas density in the range of $10^{4.4}-10^{4.8}\,{\rm cm^{ -3}}$ and radiation field of $10^{3.3}-10^{3.7}$ for the majority of the QSOs (except for the two QSOs \jsix{} and \jeight{} that are detected in radio emission) and density of $10^{4.1}-10^{4.6}\,{\rm cm^{ -3}}$ and radiation field of $10^{3.1}-10^{3.6}$ for the companion galaxies in our sample.  
As for a comparison between the QSOs and the companion galaxies in the QSO fields, the companions reveal slightly higher $L_{\rm [C\ I](1-0)}$/$L_{\rm CO(4-3)}$ and $L_{\rm [C\ I](1-0)}$/$L_{\rm FIR(SF)}$  ratios compared to the QSOs. By comparing the line ratios with models (Figure \ref{line-ratio}), we propose that the QSOs are likely to have slightly higher gas densities and higher radiation fields compared to their companion galaxies. 

We note that the QSO \jeight{} shows an extremely low $L_{\rm [C\ I](1-0)}$/$L_{\rm FIR(SF)}$ ratio of $1.1 \times 10^{-6}$. The $L_{\rm [C\ I](1-0)}$/$L_{\rm CO(4-3)}$ ratio of \jeight{} is not very different from other QSOs and companion galaxies in our sample. 
One explanation could be that the molecular gas in \jeight{} has similar densities as the other QSOs and companion galaxies in our sample, but is exposed to a more intense radiation field. By comparing the line ratios with models, we obtain a gas density of $10^{4.4}\,{\rm cm^{ -3}}$ and a radiation field of $10^{4.3}$ for \jeight{}.
Another explanation for the low $L_{\rm [C\ I](1-0)}$/$L_{\rm FIR(SF)}$ ratio could be a lower molecular gas content of this QSO compared to other sources in our sample. Indeed, this quasar shows the lowest derived molecular gas mass in our sample, based either on \ci{} or \co{} (see Section \ref{molecualrmass}). 
\jeight{} is the only radio-detected source shown on Figure \ref{line-ratio} of our sample with well constrained $L_{\rm FIR(SF)}$  (for the other radio-loud QSO \jsix{}, we cannot constrain $L_{\rm FIR(SF)}$).  
A recent work by \citet{kolwa23} reports systematically fainter \ci{} emission relative to FIR luminosity from star formation in a sample of seven radio-loud AGN. The $L_{\rm [C\ I](1-0)}$/$L_{\rm FIR(SF)}$ ratio in \jeight{} is even lower than the lowest value of $2.9 \times 10^{-6}$ found for the radio-loud AGN in \citet{kolwa23}.
Using the ALMA continuum observations, \citet{falkendal19} interpret the significant offset of a sample of radio galaxies below the main sequence as a result of gas removed from galaxies by the radio source, which leads to quenching. 
Thus, it is possible that similar to the radio-loud AGN in \citet{kolwa23} and \citet{falkendal19}, \jeight{} has \a lower molecular gas mass compared to other QSOs in our sample. This is also supported by the results reported in Section \ref{sec-line-ir-ratio} that the low ratio is found not only in  $L_{\rm [C\ I](1-0)}$/$L_{\rm FIR(SF)}$ but also in $L_{\rm CO(4-3)}$/$L_{\rm FIR(SF)}$, considering both \co{} and \ci{} emission trace the cold molecular gas content. 
In addition, it is the QSO with the lowest derived molecular gas mass among all QSOs in our sample based on either \co{} or \ci{}, although the infrared luminosity due to star formation is an intermediate value that provides an inconsistently high molecular gas mass (Figure \ref{mdust}(c)-(e)).  However, we cannot rule out the possibility that the radio-detected \jeight{} experiences a burst of star formation, as often seen in high-$z$ radio galaxies \citep{drouart14}, possibly as a result of either hierarchical merging or radio-jet-induced star formation \citep[e.g.,][]{miley08}. This will lead to a high $L_{\rm FIR(SF)}$ and thus a low $L_{\rm [C\ I](1-0)}$/$L_{\rm FIR(SF)}$ ratio. 
As for \jsix{}, the $L_{\rm [C\ I](1-0)}$/$L_{\rm CO(4-3)}$ ratio is within the range for other QSOs in our sample, while the $L_{\rm [C\ I](1-0)}$/$L_{\rm FIR(SF)}$ ratio is not well constrained.
More \co{}, \ci{} and dust continuum detections towards radio-loud QSOs will be critical to test if a low $L_{\rm [C\ I](1-0)}$/$L_{\rm FIR(SF)}$ is universal for radio-loud QSOs. 
%\textcolor{blue}{Is it possible that Q1416 has a higher fraction of L(IR) coming from the AGN, and that this drives the low ratio? The low H2 mass makes this scenario less likely, but is it worth mentioning?} \textcolor{red}{we should consider this, references are needed to see if in radio-loud AGNs such contribution is high or low... }

We also show our tentative detection of the \ci{} line in the CGM in the field of \jone{}.
The lower limit for the $L_{\rm [C\ I](1-0)}$/$L_{\rm CO(4-3)}$ ratio of 1.0, which translates into a $r_{[\rm C\ I]/CO}$ value of $>$ 0.8 in the CGM of \jone{} is slightly higher than that detected in the CGM of the Spiderweb galaxy of 0.6 $\pm$ 0.1 \citep{emonts18}. Such high $r_{[\rm C\ I]/CO}$ value in the CGM is higher than the values found in any of the QSOs or companion galaxies in our sample, indicating a lower gas density and less intense radiation field in the CGM compared to the ISM in the QSOs and the companion galaxies.  
The derived gas conditions in the CGM of \jone{} are comparable to the lowest gas densities and radiation fields found for local (U)LIRGs and high-$z$ MS galaxies.

\subsection{Molecular gas mass}\label{molecualrmass}
The \ci{} emission and the low-$J$ CO rotational transition lines trace the bulk of the cold molecular gas and thus serve as tracers for the molecular gas mass. 
In addition, studies of the gas and dust content in local and high-$z$ galaxies suggest a relation between the dust mass and molecular gas mass which is only dependent on gas-phase metallicity (e.g., \citealt{devis19}). 
We here use the \ci{} and \co{} lines, as well as the dust continuum emission, as three independent tracers for the molecular gas mass in the QSOs, companion galaxies, and the CGM.  In Sections \ref{molecularmassco}, \ref{cimasssection}, and \ref{molecularmassdust}, we introduce the methods we adopt for deriving molecular gas masses from the \co{}, \ci{} and dust continuum, respectively.  In Section \ref{comparemass}, we derive the molecular gas masses of the QSOs and companion galaxies in our sample using the three independent molecular gas traces. 
In Section \ref{cgmmass}, we estimate the molecular gas mass locked up in the cold CGM from the \ci{} and \co{} lines.

\subsubsection{CO based molecular gas mass}\label{molecularmassco}
The total molecular gas mass can be directly converted from the CO$(1-0)$ luminosity through
\begin{equation}
M_{\rm mol}^{\rm CO}  = \alpha_{\rm CO} \times L'_{\mathrm{\rm CO(1-0)}}\,[\rm M_\odot], \label{comass}
\end{equation}
where $ \alpha_{\rm CO}$ is the CO-to-H2 conversion factor in unit of \msun{}${[\rm K\ km/s\ pc^2]}^{-1}$ \citep{bolatto13}, and $L'_{\mathrm{\rm CO(1-0)}}$ is the CO($1-0$) luminosity in the unit of $\rm K\ km/s\ pc^2$.

In \paper{}, We have measured the \co{} luminosities for the QSOs and companion galaxies in our sample (Table \ref{tab-luminosity}). This enables us to derive the molecular gas mass from the \co{} line by assuming a $L'_{\mathrm{\rm CO(4-3)}}$-to-$L'_{\mathrm{\rm CO(1-0)}}$ ratio (hereafter $ \rm R_{41}$) value. As is shown in Eq. \ref{comass}, the derived molecular gas mass is dependent on both $\alpha_{\rm CO}$ and $ R_{\rm 41}$. These two parameters vary between different galaxy types, e.g., typical $\alpha_{\rm CO}$ values are 0.8 and 3.6 \msun{}${[\rm K\ km/s\ pc^2]}^{-1}$ for starbursts and normal star-forming galaxies, respectively (e.g., \citealt{downes98}; \citealt{daddi10}; \citealt{genzel10}). Typical $R_{\rm 41}$ values are 0.87, 0.85, and 0.17 for QSOs, starbursts, and star-forming galaxies (e.g., \citealt{carrili13}). 

\subsubsection{\ci{} based molecular gas mass}
\label{cimasssection}
We follow the formula presented in \citet{dunne21} to estimate the molecular gas mass from the \ci{} luminosity. We calculate $M(\rm H_{2})$ as:
\begin{equation}
 M_{\rm mol}^{\rm{[C\,\scriptscriptstyle{I}\scriptstyle{]}}}  = 1.36\frac{(9.51\times10^{-5})}{X_{\mathrm{[C\,\scriptscriptstyle{I}\scriptstyle{]}}}Q_{\rm 10}}L'_{\rm \mathrm{[C\,\scriptscriptstyle{I}\scriptstyle{]}}^3P_1\,-\, ^3P_0}\,[\rm M_\odot], \label{cimass}
\end{equation}
where $X_{\mathrm{[C\,\scriptscriptstyle{I}\scriptstyle{]}}}$ is the carbon abundance relative to $\rm H_{2}$, namely {\bf C/$\rm H_{2}$}. 
$Q_{\rm 10}$ is the excitation factor which is expressed as:
\begin{equation}
Q_{\rm 10} = \frac{3e^{-T_{\rm 1}/T_{\rm ex}}}{1+3e^{-T_{\rm 1}/T_{\rm ex}}+5e^{-T_{\rm 2}/T_{\rm ex}}}, \label{q10}
\end{equation}
where $T_{\rm 1}$ = 23.6 K and $T_{\rm 2}$ = 62.5 K are the excitation energy levels of atomic carbon and $T_{\rm ex}$ is the excitation temperature of the ISM. 

As is shown in Eq. \ref{cimass} and \ref{q10}, $X_{\mathrm{[C\,\scriptscriptstyle{I}\scriptstyle{]}}}$ and $T_{\rm ex}$ are two parameters that influence the resulting molecular gas mass estimated from the \ci{} line. The excitation temperature does not vary significantly between different galaxies, with a median value of 19.4 K for local galaxies \citep{jiao19} and 29.1 K for high-$z$ SMGs and QSOs \citep{walter11}. Such differences in local and high-$z$ galaxies translate into minor differences in $Q_{\rm 10}$. 
We adopt an excitation temperature of 29.1 K, and the resulting $Q_{\rm 10}$ is 0.457. On the other hand, $X_{\mathrm{[C\,\scriptscriptstyle{I}\scriptstyle{]}}}$ is very different in starburst systems (e.g., QSOs and SMGs) compared to normal-starforming galaxies, with commonly adopted values of $8.4\times 10^{-5}$ for high-$z$ SMGs and QSOs and $3.0 \times 10^{-5}$ for normal-starforming galaxies (e.g., \citealt{weiss03}; \citealt{papadopoulos04b};  \citealt{walter11}).
These values result in a molecular gas mass of
\begin{equation}
M_{\rm mol}^{\rm{[C\,\scriptscriptstyle{I}\scriptstyle{]}}}  = 3.37\times  L'_{\mathrm{[C\,\scriptscriptstyle{I}\scriptstyle{]}}^3P_1\,-\, ^3P_0}\, [\rm M_\odot]
\end{equation}
%= 942.3\times L_{\mathrm{[C\,\scriptscriptstyle{I}\scriptstyle{]}}^3P_1\,-\, ^3P_0}\,[M_\odot]
for starbursts and 
\begin{equation}
M_{\rm mol}^{\rm{[C\,\scriptscriptstyle{I}\scriptstyle{]}}}  = 1.20\times  L'_{\mathrm{[C\,\scriptscriptstyle{I}\scriptstyle{]}}^3P_1\,-\, ^3P_0}\, [\rm M_\odot]
\end{equation}
for normal-starforming galaxies.

\subsubsection{Dust-based molecular gas mass}
\label{molecularmassdust}
The dust mass can be derived from dust-related flux density through 
\begin{equation}
M_{\rm dust}=\frac{D_{L}^{2}S_{\nu/(1+z)}}{ (1+\textit{z})k_{\nu}B_{\nu}(T_{\rm dust})}\ [\rm M_\odot],
\end{equation}
where $D_L$ is the luminosity distance, $S_{\nu/(1+z)}$ is the continuum flux density at an observed frequency of $\nu_{\rm rest}/(1+z)$,  $k_{\rm \nu}$ is the mass absorption coefficient (cross-section per unit mass), and $B_{\rm \nu}(T_{\rm  dust})$ is the Planck function at a temperature of $T_{\rm dust}$. 

The molecular gas mass is  converted from the dust mass through 
\begin{equation}
 M_{\rm mol}^{\rm dust}  =  M_{\rm dust}/\delta_{\rm DGR}\  [\rm M_\odot],
\end{equation}
where $\rm \delta_{\rm DGR}$ is the dust-to-gas ratio. $\rm \delta_{\rm DGR}$ is dependent on the gas-phase metallicity (Z) in galaxies through 
\begin{equation}
\rm log(\delta_{\rm DGR})  = (2.45\pm 0.12)log(\frac{Z}{Z_{\odot}})-(2.0\pm 1.4),
\end{equation}
\citep{devis19}. 

\begin{figure*}
\gridline{\fig{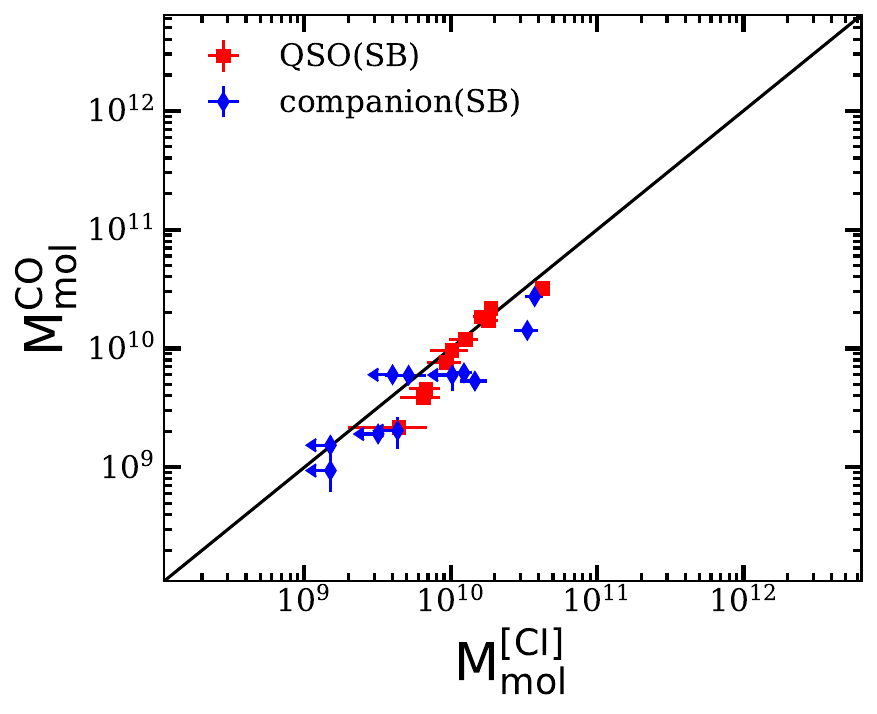}{0.33\textwidth}{(a)}
          \fig{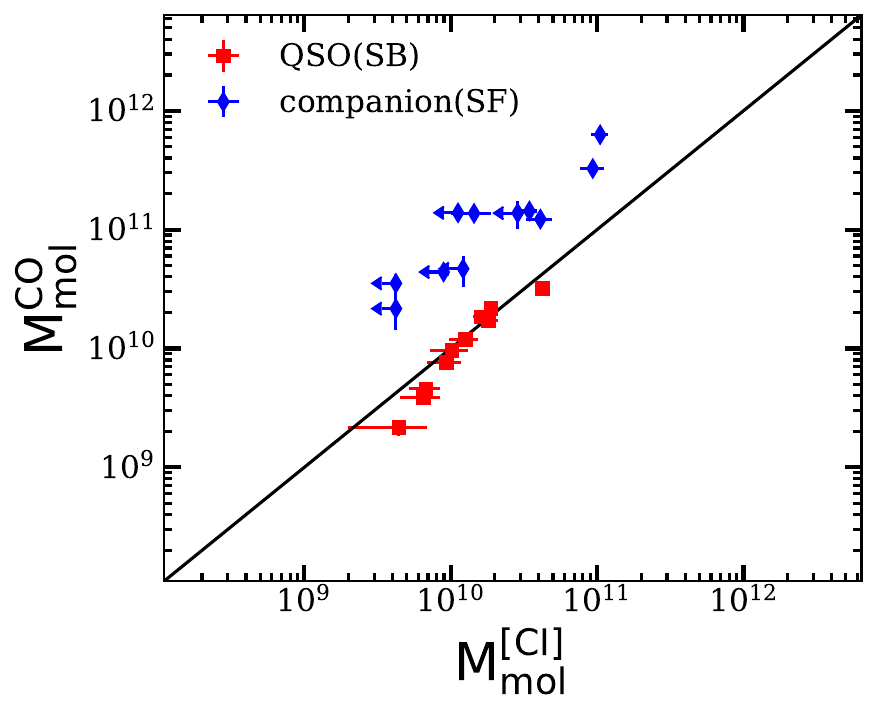}{0.33\textwidth}{(b)}
          }
\gridline{\fig{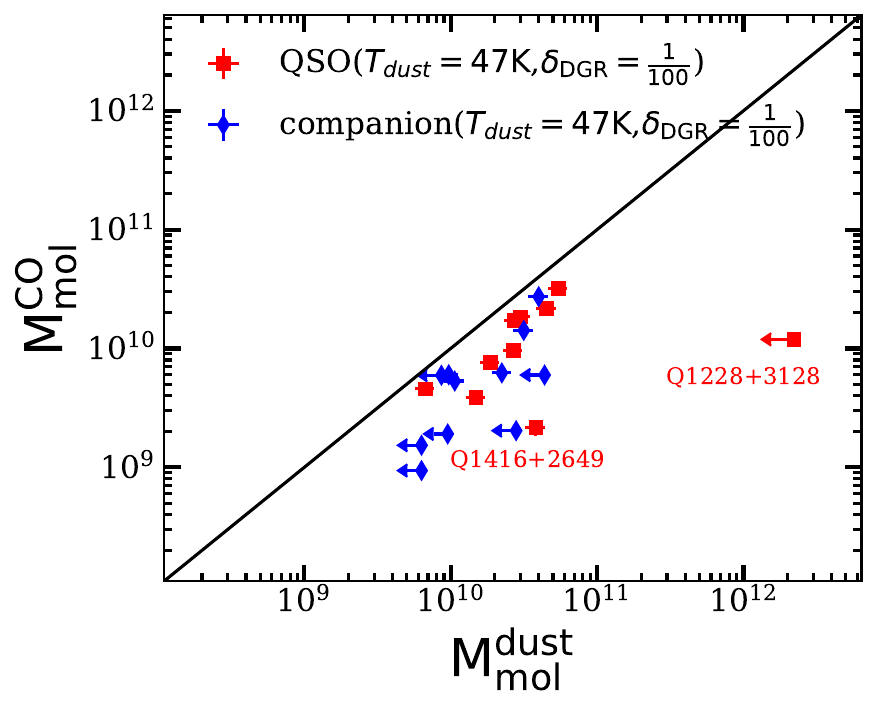}{0.33\textwidth}{(c)}
          \fig{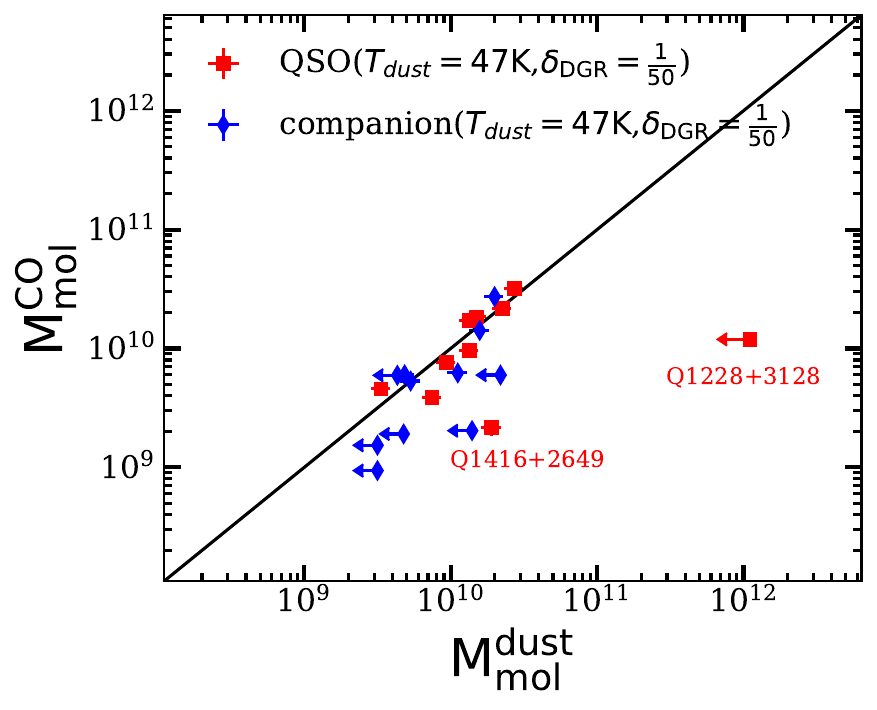}{0.33\textwidth}{(d)}
          \fig{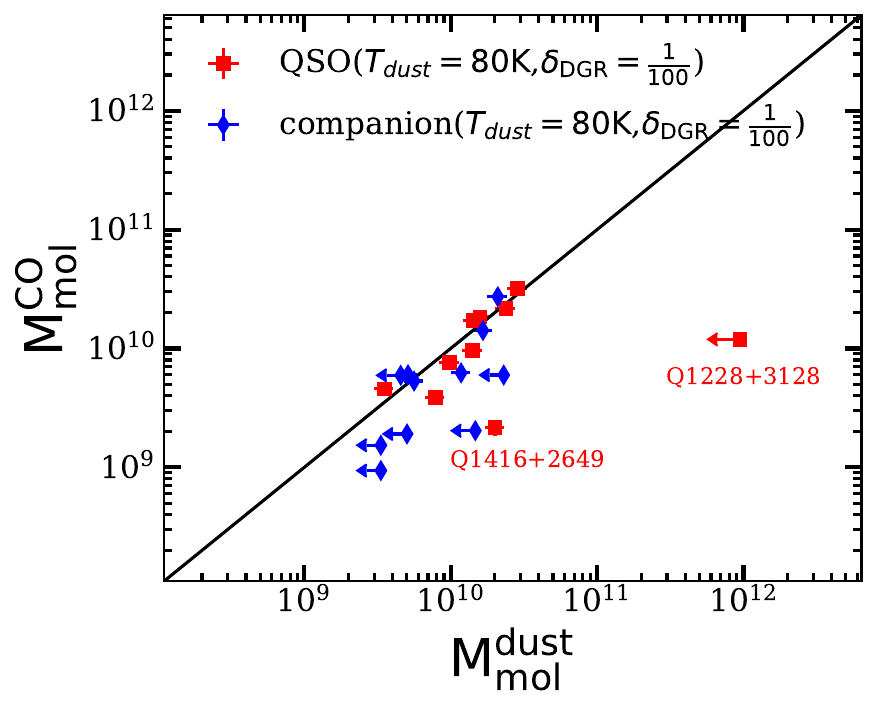}{0.33\textwidth}{(e)}
          }
\caption{\label{mdust}  Molecular gas masses derived based on the \co{}, \ci{} and dust continuum emission. We use $\alpha_{\rm CO}$ = 0.8 \msun{}${[\rm K\ km/s\ pc^2]}^{-1}$, $ R_{\rm 41}$ = 0.87, and $X_{\mathrm{[C\,\scriptscriptstyle{I}\scriptstyle{]}}}$ = $8.4\times 10^{-5}$ to estimate the molecular gas mass using the \co{} and \ci{} lines for the QSOs. We consider all the companion galaxies as either starbursts (a) or star-forming galaxies (b), and the parameters adopted are $\alpha_{\rm CO}$ = 0.8 \msun{}${[\rm K\ km/s\ pc^2]}^{-1}$, $ R_{\rm 41}$ = 0.85, and $X_{\mathrm{[C\,\scriptscriptstyle{I}\scriptstyle{]}}}$ = $8.4\times 10^{-5}$ (for starbursts), and $\alpha_{\rm CO}$ = 3.6 \msun{}${[\rm K\ km/s\ pc^2]}^{-1}$, $ R_{\rm 41}$ =  0.17, and $X_{\mathrm{[C\,\scriptscriptstyle{I}\scriptstyle{]}}}$ = $3.0\times 10^{-5}$ (for star-forming galaxies). For the molecular gas masses derived from the ALMA continuum, we adopt $\delta_{\rm DGR} = \frac{1}{100}$ and $T_{\rm dust}$ = 47 K (c), $\delta_{\rm DGR} = \frac{1}{50}$ and $T_{\rm dust}$ = 47 K (d), and $\delta_{\rm DGR} = \frac{1}{100}$ and $T_{\rm dust}$ = 80 K (e) for the QSOs and companions.
} 
\end{figure*}
\subsubsection{Comparing the molecular gas mass in the ISM derived using different tracers}\label{comparemass}

To compare the estimates of the molecular gas mass based on the different tracers described in Section \ref{molecularmassco}\,$-$\,\ref{molecularmassdust}, one has to rely on important parameters that carry a large degree of uncertainty. For example, \citet{yang17} found that high-$z$ lensed galaxies show a large range of $R_{\rm 41}$ values that affect the CO(4-3) luminosity, \citet{bisbas15} revealed that the CO and \ci{} abundance is influenced heavily by both photo- and cosmic ray ionization, and \citet{kennicutt09} showed that infrared luminosity in part arises from a diffuse dusty medium (“cirrus”) that is not directly associated with star formation. Therefore, any comparison between the various tracers will be hampered by large uncertainties. Nevertheless, it is interesting to compare molecular mass estimates based on values that are widely adopted in the literature.

We adopt $\alpha_{\rm CO}$ = 0.8 \msun{}${[\rm K\ km/s\ pc^2]}^{-1}$ and $ R_{\rm 41}$ = 0.87 to estimate the molecular gas mass from the \co{} line, and $X_{\mathrm{[C\,\scriptscriptstyle{I}\scriptstyle{]}}}$ = $8.4\times 10^{-5}$ to derive the molecular gas based on \ci{} for the QSOs in our sample (see Sections \ref{molecularmassco} and \ref{cimasssection}; all the adopted parameter values are typical for QSOs). 
As for the companion galaxies, we assume two cases, namely the companion galaxies are either all starbursts or star-forming galaxies, and the adopted parameter values are  $\alpha_{\rm CO}$ = 0.8 \msun{}${[\rm K\ km/s\ pc^2]}^{-1}$, $ R_{\rm 41}$ = 0.85, and $X_{\mathrm{[C\,\scriptscriptstyle{I}\scriptstyle{]}}}$ = $8.4\times 10^{-5}$ for the starburst case, and $\alpha_{\rm CO}$ = 3.6 \msun{}${[\rm K\ km/s\ pc^2]}^{-1}$, $ R_{\rm 41}$ =  0.17, and $X_{\mathrm{[C\,\scriptscriptstyle{I}\scriptstyle{]}}}$ = $3.0\times 10^{-5}$ for the star-forming case.
We show the comparison between the molecular gas masses based on \co{} and \ci{} for the QSOs and companion galaxies in Figure \ref{mdust} (a) and (b). The molecular gas mass derived using the \co{} line for the QSOs agrees with those derived from the \ci{} line within uncertainties. For the companion galaxies, only the starburst case suggests comparable molecular gas mass based on \co{} and \ci{}. This suggests that the QSOs and companions are likely starbursts, and both have a low conversion factor of $\alpha_{\rm CO}$\,$\sim$\,0.8 \msun{}${[\rm K\ km/s\ pc^2]}^{-1}$.
This is also consistent with the high SFR derived from their FIR continuum detection (Table \ref{tab-fir}). 

The molecular gas mass based on the dust continuum emission is dependent on both $\delta_{\rm DGR}$ and $T_{\rm dust}$. $\delta_{\rm DGR}$ is only dependent on metallicity \citep{devis19}. 
We do not expect significant differences between the QSOs and their companions as these systems are likely part of the same galaxy groups or clusters, and therefore we can assume that their ages and mechanism of gas supply are similar. In addition, extended cold CGM emission connecting the companion galaxies and the QSOs is detected in \co{} emission for the majority of our QSO fields, which might indicate gas exchange between the QSOs and their companions (\paper).
We first assume a solar metallicity (Z = 1.0 Z$_\sun{}$) which translates to $\rm \delta_{DGR}=\frac{1}{100}$ and a typical $T_{\rm dust}$ = 47 K to derive the molecular gas mass from the continuum emission (Figure \ref{mdust} (c)). Such an assumption leads to higher molecular gas masses derived from the continuum emission compared to those from the \co{} line for the QSOs and companion galaxies in our sample. To get consistent molecular gas masses estimated from the continuum and the \co{} line, we also consider that the QSOs and companion galaxies in our sample have either warmer dust with $T_{\rm dust}$ = 80 K and $\rm \delta_{DGR}=\frac{1}{100}$ or higher metallicity ( Z = 1.3 Z$_\sun{}$) with $T_{\rm dust}$ = 47 K and $\rm \delta_{DGR}=\frac{1}{50}$.
In calculating the \co{} based molecular gas mass, all the QSOs and companion galaxies are treated as starbursts.
The comparison of the molecular gas estimated based on the \co{} and the dust continuum is shown in Figure \ref{mdust} (c-e).
Both the warmer dust and the higher metallicity scenario reveal comparable molecular gas derived based on the \co{} and the continuum emission. This suggests that the QSOs and companions in our sample have either higher dust temperatures than the typical value of 47 K and/or a higher metallicity compared to the solar value. We note that the QSO \jeight{} with a radio-detected AGN is off the one-to-one relation in Figure \ref{mdust} (c-e). This is likely a result of lower \co{} and \ci{} luminosities for its FIR luminosity from star formation compared to all the other QSOs and companions in our sample.
As for the QSO \jsix{}, the dust-derived molecular gas mass is approaching two orders of magnitude lower than what we derive based on the observed continuum flux density to reside on the one-to-one relation for \co{} and dust-based molecular gas masses. 
Future ALMA continuum observations sampling the dust SED at other frequencies may help to better constrain the FIR luminosity of this QSO to further confirm this.   

\subsubsection{Mass estimation of the extended \ci{} from the CGM \label{cgmmass}}
We estimate the molecular gas masses locked up in the CGM based on the \ci{} luminosities presented in Section \ref{cgm-result}. 
The \ci{} emission in the CGM is only tentatively detected in one of the QSOs \jone{}, which has a \ci{} luminosity of (2.5 $\pm$ 0.9) $\times 10^{7}$ \lsun{}. 
As is discussed in Section \ref{cimasssection}, the molecular gas mass based on the \ci{} line is dependent on the carbon abundance, which is different by a factor of $\sim 3$ in the ISM of normal star-forming and starburst galaxies. Due to the limited detection of the \ci{} in the CGM, we have little knowledge about its carbon abundance. 
We here give a very crude estimation of the carbon abundance in the CGM by assuming an $X_{\mathrm{[C\,\scriptscriptstyle{I}\scriptstyle{]}}}$ range of ($3.0 - 8.4) \times 10^{-5}$, similar as that in the ISM. 
This leads to a molecular gas mass of ($1.0 - 2.8$)$\times 10^{10}$ \msun{} based on the \ci{} line. We estimate the molecular CGM masses for the remaining nine quasars based on the 3$\sigma$ upper limits of the \ci{} line, and the resulting range in limits for the molecular gas mass in the CGM is  $<$($0.2 - 1.4$)$\times 10^{10}$ \msun{}.

Utilizing a similar method as \ci{}, we also measure the molecular gas mass using the \co{} emission in the CGM (\paper). The \co{} luminosity in the CGM for \jone{} is (1.4 $\pm$ 0.9) $\times 10^{7}$ \lsun{} (1.6$\sigma$). 
The resulting 3$\sigma$ upper limit for the molecular CGM mass in \jone{} based on the \co{} line is in the range of ($0.7 - 13.6$)$\times 10^{10}$ \lsun{} assuming similar parameters of the CGM as normal star-forming and starburst galaxies. %for CO QSO*19.5=SF 
The 3$\sigma$ upper limits of the molecular CGM masses for the remaining nine QSOs are in the range of $<$($0.1 - 10.2$)$\times 10^{10}$ \msun{}. 
Therefore, the molecular CGM mass based on the \ci{} and \co{} lines are consistent within the uncertainties. 

In addition, we estimate the cosmic baryon fraction using the molecular gas we derive in the CGM. Assuming that all the reprocessed baryons stay within the virial radius, that the cosmic baryon fraction of 15.6$\%$ derived by Planck \citep{planck20} applies to this reservoir, and adopting a typical halo mass for QSOs of $10^{12.5}$ \msun{} (e.g., \citealt{white12}; \citealt{pizzati24}), the baryonic mass within the halo is roughly $4.9 \times 10^{11}$ \msun{}. Our derived molecular gas masses in the CGM of $<$($0.2 - 1.4$)$\times 10^{10}$ \msun{} for the CGM within the  inner 40 kpc represents $<$0.4-3$\% $ of the baryons, which is in the same range as other estimates in the literature \citep{arrigoni22}.

\section{summary} \label{summary}
We report ALMA/ACA observations of the \ci{} line and dust continuum emission in a sample of 10 ultraluminous Type-I QSOs at $z\sim 2$, each showing extended Ly$\alpha$ emission on scales of $>$100 kpc. Utilizing both the 7 m and the 12 m arrays of ALMA, we can constrain the molecular gas properties in both the ISM and the extended CGM. We summarize our main results below.

(1) We detect the \ci{} emission in all of our 10 QSOs in the 12 m array data and nine QSOs in the 7 m + 12 m array data after uv-tapering. Interestingly, all the nine QSOs show higher flux in the 7 m + 12 m array data compared to the 12 m array data, with a flux difference of $0.03 - 0.22$ \jykmps{}, hinting at the possibility of cold CGM emission in the majority of our targets.
However, the \ci{} line is too weak to determine this on a case-by-case basis, except for \jone{} where we detect a \ci{} flux of 0.22 \jykmps{} at an SNR of 2.7. In addition, we detect the dust continuum emission in all the 10 QSOs. 

(2) 13 companion galaxies are identified in our previous \co{} observations in $70\%$  of our QSO fields. We detect the \ci{} emission in $ 40\%$ of the companion galaxies and detect the dust continuum emission in $ 40\%$ of the companion galaxies. 

(3) Utilizing the  $L_{\rm [C\ I](1-0)}$/$L_{\rm CO(4-3)}$ and $L_{\rm [C\ I](1-0)}$/$L_{\rm FIR(SF)}$ ratios as diagnostics of the physical properties of the molecular gas, we find that, similar to other high-$z$ AGN and SMGs, the QSOs and companion galaxies in our sample have higher gas density and more intense radiation field compared to high-$z$ main-sequence galaxies, with values that are at the high-end of the distribution for local (U)LIRGs. In addition, we find a slightly higher gas density and slightly higher radiation field in the QSOs compared to their companion galaxies. 

(4) In one of the radio-detected QSOs in our sample, \jeight{}, we find that the  $L_{\rm CO(4-3)}$ and $L_{\rm [C\ I](1-0)}$ luminosities are an order of magnitude fainter for its $L_{\rm FIR(SF)}$. This could result from a lower molecular gas mass or a more intense radiation field compared to other QSOs and companion galaxies in our sample. 

(5) We use \co{}, \ci{}, and the ALMA continuum emission as independent molecular gas mass tracers for the ISM in the QSOs and their companions. We estimate the molecular gas mass in the QSOs in the range of $(0.4-4.2) \times 10^{10}$ \msun{} and $(0.2-3.7) \times 10^{10}$ \msun{} for companion galaxies where either \co{}, \ci{}, or the ALMA continuum is detected. 
The QSOs and companions also display the same low conversion factor of $\alpha_{\rm CO}$\,$\sim$\,0.8 \msun{}${[\rm K\ km/s\ pc^2]}^{-1}$.
In addition, we constrain the cold molecular CGM mass using the \ci{} emission and obtained a tentative detection of (1.0 $-$ 2.8)$\times 10^{10}$ \msun{} in \jone{} and 3$\sigma$ upper limits of  $<$($0.2 - 1.4$)$\times 10^{10}$ \msun{} for the rest of the nine QSO fields. This translates into a baryon fraction of $<$0.4-3$\% $ in the molecular CGM relative to the total baryonic mass within the halo, which is consistent with values estimated from the literature.

%\section{acknowledgment} 
\begin{acknowledgments}
We acknowledge support from the National Key R\&D Program of China (grant no.~2023YFA1605600), the National Science Foundation of China (grant no.~12073014), the science research grants from the China Manned Space Project with No.~CMS-CSST-2021-A05, and Tsinghua University Initiative Scientific Research Program (No.~20223080023).
We gratefully acknowledge the support from the Shuimu Tsinghua Scholar Program of Tsinghua University. 
This paper makes use of the following ALMA data: ADS/JAO.ALMA$\#$2019.1.01251.S. ALMA is a partnership of ESO (representing its member states), NSF (USA) and NINS (Japan), together with NRC (Canada), MOST and ASIAA (Taiwan), and KASI (Republic of Korea), in cooperation with the Republic of Chile. The Joint ALMA Observatory is operated by ESO, AUI/NRAO and NAOJ.
The National Radio Astronomy Observatory is a facility of the National Science Foundation operated under cooperative agreement by Associated Universities, Inc. MVM acknowledges support from grant PID2021-124665NB-I00 by the Spanish Ministry of Science and Innovation (MCIN) / State Agency of Research (AEI) / 10.13039/501100011033 and by the European Regional Development Fund (ERDF) ``A way of making Europe''. 
RW acknowledges support from the National Natural Science Foundation of China (NSFC) with grant Nos. 12173002, 11991052, and the science research grant from the China Manned Space Project with No. CMS-CSST-2021-A06.
This publication makes use of data products from the Two Micron All Sky Survey, which is a joint project of the University of Massachusetts and the Infrared Processing and Analysis Center/California Institute of Technology, funded by the National Aeronautics and Space Administration and the National Science Foundation. 
Based on observations made with the NASA/ESA Hubble Space Telescope, obtained from the data archive at the Space Telescope Science Institute, which is operated by the Association of Universities for Research in Astronomy, Incorporated, under NASA contract NAS5-26555. Support for Program number HST-GO-16891.002-A was provided through a grant from the STScI under NASA contract NAS526555.
\end{acknowledgments}

\vspace{5mm}
\facilities{ALMA}


\software{CASA \citep{casa22}}

\appendix \label{appendix}
\section*{The extent of the cold ISM in the host galaxies of the QSOs}
\begin{figure*}
\includegraphics[width=0.5\linewidth]{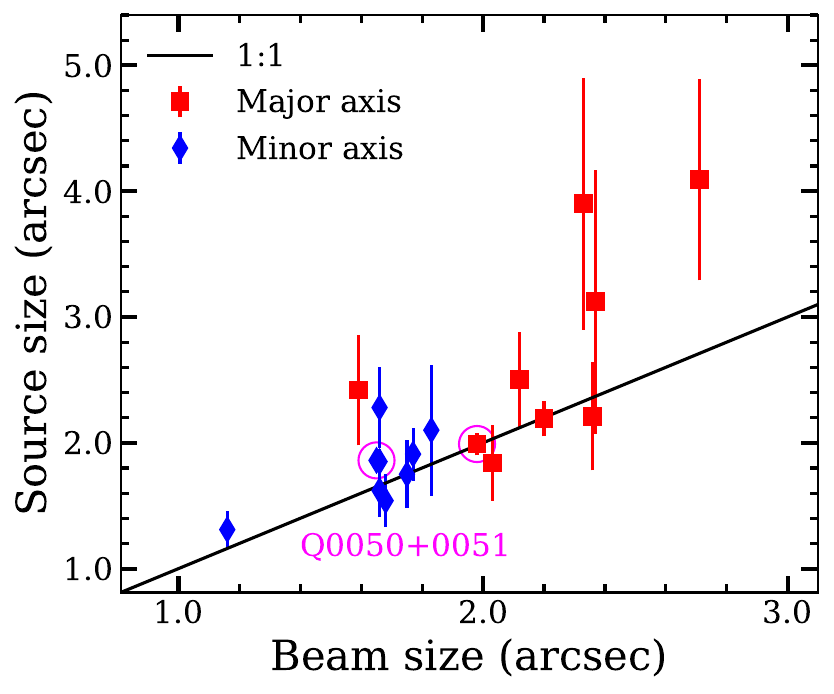}
\caption{\label{size-comparison} A comparison between the \ci{} source sizes convolved with the beam and the beam sizes for the 12 m array data. The red squares represent the sizes of the major axis, and the blue diamonds represent the sizes of the minor axis. 
We show the source size equal to the beam size as a solid black line. We highlight the source with significant CGM emission (\jone{}) in magenta circles.
}
\end{figure*}

We measure the source sizes using the 2D Gaussian fit icon in the CASA viewer. The 2D Gaussian fit returns both the source size convolved with the beam, and the source size deconvolved with beam. We show the derived source sizes and beam sizes in Table \ref{tab:appendix}. 
For 70$\%$ of the sources, the 2D Gaussian fit fails to measure the deconvolved source size and returns a point source. 
For a better visualization, if the host galaxies of the QSOs are spatially resolved in the 12 m array data, we also show the comparison between the source size convolved with the beam and the beam size in Figure \ref{size-comparison}. 
We consider sources where source sizes convolved with the beam comparable to the beam sizes as spatially unresolved. 
%The sources where the source sizes convolved with the beam are comparable to the beam sizes are spatially unresolved. 
From Figure \ref{size-comparison}, the source sizes agree with the beam sizes within 2$\sigma$ uncertainties (the error bars on Figure \ref{size-comparison} represent 1$\sigma$ uncertainties). 
In addition, for the three sources where deconvolved source sizes are returned in the 2D Gaussian fit, the measured deconvolved source sizes are poorly constrained with large uncertainties (SNR less than 3 for the major axis and less than 2 for the minor axis). 
Accordingly, we consider the \ci{} emission from the cold ISM of the QSOs to be spatially unresolved under our beam sizes of 1\farcs6$-$2\farcs7 $\times$ 1\farcs2$-$1\farcs8.

\begin{deluxetable*}{llll}
\tabletypesize{\scriptsize}
\tablenum{A1}
\tablewidth{0pt} 
\tablecaption{Measurements of \ci{} sizes for the QSOs \label{tab:appendix}}
\tablehead{\colhead{Source} & 
\colhead{\ci{} size (convolved with beam)} &\colhead{\ci{} size (deconvolved from beam)} &\colhead{Beam size}\\
 &$\theta_{\rm major}  \times \theta_{\rm minor}; \rm PA$  & $\theta_{\rm major} \times \theta_{\rm minor}; \rm PA$  & $\theta_{\rm major} \times \theta_{\rm minor}; \rm PA$
}
\colnumbers
\startdata 
\jone{}& $(1\farcs99 \pm 0\farcs09) \times (1\farcs86 \pm  0\farcs08); (95\degree \pm 24\degree)$ &\nodata & $1\farcs98 \times 1\farcs65; 78\degree$ \\
\jtwo{}& $(1\farcs84 \pm 0\farcs30) \times (1\farcs54 \pm 0\farcs21); (42\degree \pm 31\degree)$ &\nodata & $2\farcs03 \times 1\farcs68; -81\degree $\\
\jthree{}& $(2\farcs19 \pm 0\farcs14) \times (1\farcs85 \pm 0\farcs10); (75\degree \pm 13\degree) $& \nodata & $2\farcs20 \times 1\farcs66; 70\degree$ \\
\jfour{}& $(2\farcs50 \pm 0\farcs38) \times (2\farcs28 \pm 0\farcs32) ; (125\degree \pm 60\degree) $&$(1\farcs64 \pm 0\farcs71) \times (1\farcs25 \pm 0\farcs96); (176\degree \pm 60\degree) $& $2\farcs12 \times 1\farcs66; -75\degree$ \\
\jfive{}& $(2\farcs21 \pm 0\farcs43) \times (1\farcs75 \pm 0\farcs27); (17\degree \pm 26\degree)$& \nodata&$ 2\farcs36 \times 1\farcs75; 11\degree$\\
\jsix{}& $(3\farcs90 \pm 1\farcs00) \times (1\farcs62 \pm 0\farcs21); (154\degree \pm 5\degree) $&\nodata &$ 2\farcs33 \times 1\farcs66; 12\degree$\\
\jseven{}&$(4\farcs09 \pm 0\farcs80) \times (1\farcs91 \pm 0\farcs21); (15\degree \pm 5\degree) $&$(3\farcs11 \pm 1\farcs18) \times (0\farcs44 \pm 0\farcs68); (21\degree \pm 20\degree) $&$2\farcs71 \times 1\farcs77; 1\degree $\\
\jeight{}& $(3\farcs12 \pm 1\farcs05) \times (2\farcs10 \pm 0\farcs52); (48\degree \pm25 \degree)$& \nodata&$ 2\farcs37 \times 1\farcs83; -22\degree $\\
\jnine{}&$(2\farcs42 \pm 0\farcs44) \times (1\farcs31 \pm 0\farcs15); (121\degree \pm 7\degree) $& $(1\farcs84 \pm 0\farcs66) \times (0\farcs55 \pm 0\farcs33); (126\degree \pm 23\degree)$ &$1\farcs59 \times 1\farcs16; -70\degree$\\
\jten{}&\nodata & \nodata &$ 2\farcs05 \times 1\farcs70; 72\degree$ \\
\enddata
\tablecomments{We show the \ci{} source sizes of the QSOs convolved with the beam, and deconvolved from the beam for the 12 m array data. We also show the beam sizes for the 12 m array data. We do not show the fitting result for \jten{}, because the S/N is too low (see also Figure \ref{intensity_qso}) to give a meaningful fitting result. 
}
\end{deluxetable*}

%\bibliography{sample631}{}
%\bibliographystyle{aasjournal}



\end{document}